\documentclass[10pt,twocolumn,showpacs,preprintnumbers,amsmath,amssymb,aps,prx,longbibliography,superscriptaddress]{revtex4-1}
\usepackage{mathrsfs}
\usepackage{graphicx}
\usepackage{dcolumn}
\usepackage{bm}
\usepackage{amsmath}
\usepackage{amsfonts}
\usepackage{color}
\usepackage{hyperref}

\begin{document}

\title{Open Momentum Space Method for Hofstadter Butterfly and the Quantized Lorentz Susceptibility
}

\author{Biao Lian}
\affiliation{Department of Physics, Princeton University, Princeton, New Jersey 08544, USA}
\author{Fang Xie}
\affiliation{Department of Physics, Princeton University, Princeton, New Jersey 08544, USA}
\author{B. Andrei Bernevig}
\affiliation{Department of Physics, Princeton University, Princeton, New Jersey 08544, USA}

\begin{abstract}
We develop a generic $\mathbf{k}\cdot \mathbf{p}$ open momentum space method for calculating the Hofstadter butterfly of both continuum (moir\'e) models and tight-binding models, where the quasimomentum is directly substituted by the Landau level (LL) operators. By taking a LL cutoff (and a reciprocal lattice cutoff for continuum models), one obtains the Hofstadter butterfly with in-gap spectral flows. For continuum models such as the moir\'e model for twisted bilayer graphene, our method gives a sparse Hamiltonian, making it much more efficient than existing methods. The spectral flows in the Hofstadter gaps can be understood as edge states on a momentum space boundary, from which one can determine the two integers ($t_\nu,s_\nu$) of a gap $\nu$ satisfying the Diophantine equation. The spectral flows can also be removed to obtain a clear Hofstadter butterfly. While $t_\nu$ is known as the Chern number, our theory identifies $s_\nu$ as a dual Chern number for the momentum space, which corresponds to a quantized Lorentz susceptibility $\gamma_{xy}=eBs_\nu$.
\end{abstract}

\date{\today}

\maketitle

Two-dimensional (2D) lattice electrons in large magnetic fields are known to exhibit Hofstadter butterfly spectra \cite{hofstadter1976}. Conventionally, the Hofstadter butterfly is calculated at rational fluxes per unit cell $\varphi=2\pi p/q$ in a basis with translation symmetry of $q$ unit cells, where $p$ and $q$ are coprime integers. The calculation often involves a complicated construction of the matrix elements. In particular, for continuum $\mathbf{k}\cdot \mathbf{p}$ models obtained from plane wave expansions such as the moir\'e model for twisted bilayer graphene (TBG) \cite{bistritzer2011,santos2007,mele2010}, the Hofstadter Hamiltonian matrix is infinite dimensional and dense \cite{bistritzer2011a,janecek2013,gumbs2014,hejazi2019,zhangy2019,tknn1982}, which requires a large cutoff for the spectrum to converge.

In contrast, the Landau levels (LLs) of a $\mathbf{k}\cdot\mathbf{p}$ Hamiltonian at small magnetic fields can be calculated by simply substituting the quasimomentum $\mathbf{k}=(k_x,k_y)$ with $(\frac{a+a^\dag}{\sqrt{2}\ell},\frac{a-a^\dag}{i\sqrt{2}\ell})$, where $a$ and $a^\dag$ are the LL lowering and raising operators, and $\ell$ is the magnetic length \cite{winkler2003}. In this letter, we demonstrate that such a substitution with a LL cutoff (and a reciprocal lattice cutoff for continuum models) provides an efficient method for calculating the Hofstadter butterfly in large magnetic fields, which greatly simplifies the Hamiltonian matrix elements \cite{lian2019b}. In particular, for continuum models, this method yields a sparse Hamiltonian, whose spectrum can be efficiently calculated by the shift-and-invert Lanczos method.

The method can be understood as an open momentum space calculation, where the smaller of the momentum-space LL wavefunction radius cutoff and reciprocal lattice radius cutoff plays the role of a momentum space boundary. As a result, the spectrum contains not only the Hofstadter butterfly, but also in-gap spectral flow levels \cite{asboth2017,lian2019b} which can be understood as ``momentum space edge states". We show that the spectral flows of these edges allow us to determine the two integers $(t_\nu,s_\nu)$ in a Hofstadter gap $\nu$ satisfying the Diophantine equation \cite{claro1979,dana1985,satija2016}, where $t_\nu$ is the Chern number of the gap. Moreover, we show that $s_\nu$ can be interpreted as a dual Chern number for the momentum space, which yields a quantized Lorentz susceptibility (Eq. (\ref{eq-lorentz})). Furthermore, by identifying and removing the momentum space edge states, one can obtain the Hofstadter butterfly without spectral flows. We demonstrate our method for both continuum models and tight binding models in a 2D periodic lattice. We shall denote the lattice Bravais vectors as $\bm{d}_1$ and $\bm{d}_2$, and the reciprocal vectors as $\bm{g}_1$ and $\bm{g}_2$, which satisfy $\bm{g}_i\cdot\bm{d}_{j}=2\pi\delta_{ij}$ ($i,j=1,2$).

\emph{Continuum models.} At zero magnetic field, a continuum model can be written in the real space basis $|\mathbf{r},\alpha\rangle$ as \cite{santos2007,mele2010,bistritzer2011a,janecek2013,gumbs2014,kariyado2019}
\begin{equation}\label{eq-Cmodel}
H^{\alpha\beta}(\mathbf{r})=\epsilon^{\alpha\beta}(-i\nabla)+\sum_{j} V^{\alpha\beta}_{j}e^{i\mathbf{q}_j^{\alpha\beta}\cdot\mathbf{r}}\ ,
\end{equation}
where $\mathbf{r}=(x,y)$ is the real space position, $-i\nabla=-i(\partial_x,\partial_y)$ is the canonical momentum, and we assume there are $M$ intrinsic orbitals labeled by $\alpha,\beta$. $\epsilon^{\alpha\beta}(-i\nabla)$ and $V_j^{\alpha\beta}e^{i\mathbf{q}_j^{\alpha\beta}\cdot\mathbf{r}}$ are the electron kinetic term in free space and the periodic lattice potential between orbitals $\beta$ and $\alpha$, respectively. If one denote $\mathbf{Q}\in\bm{g}_1\mathbb{Z}+\bm{g}_2\mathbb{Z}$ as the reciprocal lattice, and choose the momentum origin of orbital $\alpha$ at $\mathbf{p}_\alpha$, one can define a momentum lattice $\mathbf{Q}_\alpha=\mathbf{p}_\alpha+\mathbf{Q}$ for orbital $\alpha$, and $\mathbf{q}_j^{\alpha\beta}$ in Eq. (\ref{eq-Cmodel}) must be the difference $\mathbf{Q}_\alpha'-\mathbf{Q}_\beta$ between some sites $\mathbf{Q}_\alpha'$ and $\mathbf{Q}_\beta$ (see supplementary material (SM) \cite{suppl} Sec. S2A). Generically, one can always fix all $\mathbf{p}_\alpha=\mathbf{0}$; however, in certain models (e.g., the TBG model \cite{bistritzer2011}) nonzero $\mathbf{p}_\alpha$ choices are preferred.

One can transform the zero-magnetic-field Hamiltonian (\ref{eq-Cmodel}) into the momentum eigenbasis $|\mathbf{k},\mathbf{Q}_\alpha,\alpha\rangle=\int \text{d}^2\mathbf{r} e^{i(\mathbf{k}+\mathbf{Q}_\alpha)\cdot\mathbf{r}}|\mathbf{r},\alpha\rangle$, where $\mathbf{k}$ is in the first Brillouin zone (BZ). The momentum space Hamiltonian under basis $|\mathbf{k},\mathbf{Q}_\alpha,\alpha\rangle$ then takes the form \cite{bistritzer2011,suppl}
\begin{equation}\label{eq-HQQ}
H^{\alpha\beta}_{\mathbf{Q}_\alpha'\mathbf{Q}_\beta}(\mathbf{k})=\epsilon^{\alpha\beta}(\mathbf{k}+\mathbf{Q}_\beta)\delta_{\mathbf{Q}_\alpha'\mathbf{Q}_\beta}+  \sum_{j} V^{\alpha\beta}_{j}\delta_{\mathbf{Q}_\alpha',\mathbf{Q}_\beta+\mathbf{q}_j^{\alpha\beta}}.
\end{equation}

When a uniform out-of-plane magnetic field $\mathbf{B}=B\hat{\mathbf{z}}$ is added, $-i\nabla$ in Eq. (\ref{eq-Cmodel}) is replaced by the kinematic momentum $\bm{\Pi}=-i\nabla-\mathbf{A}(\mathbf{r})$, where $\mathbf{A}(\mathbf{r})$ is the vector potential satisfying $\partial_xA_y-\partial_yA_x=B$. The kinetic momentum satisfies $[\Pi_x,\Pi_y]=i/\ell^2$, where $\ell=1/\sqrt{B}$ is the magnetic length. We also define the guiding center $\mathbf{R}=\mathbf{r}-\ell^2\hat{\mathbf{z}}\times\bm{\Pi}$, which satisfies $[R_x,R_y]=-i\ell^2$, and $[\mathbf{R},\bm{\Pi}]=0$.

The usual Hofstadter method for continuum models employs the Landau basis defined by eigenstates of $R_x$ and $\bm{\Pi}^2$, which has complicated matrix elements \cite{bistritzer2011a,janecek2013,gumbs2014,hejazi2019,zhangy2019,tknn1982}. Here we shall take a different basis, under which we prove the nonzero magnetic field Hamiltonian can be simply obtained by the zero-field momentum-space Hamiltonian (\ref{eq-HQQ}) with the substitution of Eq. (\ref{eq-sub}).

We define $R_{\hat{\bm{\tau}}}=\mathbf{R}\cdot\hat{\bm{\tau}}$ as the guiding center along unit vector $\hat{\bm{\tau}}$, where we choose $\frac{\hat{\bm{\tau}}\cdot(\hat{\mathbf{z}}\times\bm{g}_1)} {\hat{\bm{\tau}}\cdot(\hat{\mathbf{z}}\times\bm{g}_2)}$ irrational. 
We also define a set of (linearly dependent) LL operators $a_{\mathbf{Q}_\alpha}=\frac{\ell}{\sqrt{2}}[\Pi_{x}-Q_{\alpha,x}-k_{0,x}+ i(\Pi_y-Q_{\alpha,y}-k_{0,y})]$ and their conjugates $a_{\mathbf{Q}_\alpha}^\dag$ associated with momentum sites $\mathbf{Q}_\alpha$, where $\mathbf{k}_0=(k_{0,x},k_{0,y})$ is a freely chosen real vector which we call the \emph{center momentum}. We then construct an orthonormal basis $|\lambda,\mathbf{Q}_\alpha,n,\alpha\rangle$ for orbital $\alpha$ and reciprocal site $\mathbf{Q}_\alpha$ by requiring
\begin{equation}\label{eq-Rt}
\begin{split}
&R_{\hat{\bm{\tau}}}|\lambda,\mathbf{Q}_\alpha,n,\alpha \rangle=[\lambda-\ell^2\hat{\bm{\tau}}\cdot(\hat{\mathbf{z}}\times \mathbf{Q}_\alpha)]|\lambda,\mathbf{Q}_\alpha,n,\alpha\rangle,\\
&a_{\mathbf{Q}_\alpha}^\dag a_{\mathbf{Q}_\alpha}|\lambda,\mathbf{Q}_\alpha,n,\alpha \rangle= n|\lambda,\mathbf{Q}_\alpha,n,\alpha \rangle.
\end{split}
\end{equation}
Here $n\ge0$ is an integer LL number, while $\lambda$ is a real number chosen in the set $\lambda+\ell^2\hat{\bm{\tau}}\cdot(\hat{\mathbf{z}}\times\bm{g}_1)\mathbb{Z} +\ell^2\hat{\bm{\tau}}\cdot(\hat{\mathbf{z}}\times\bm{g}_2)\mathbb{Z}$ representing the set, or abstractly, $\lambda\in\mathbb{R}/[\ell^2\hat{\bm{\tau}}\cdot(\hat{\mathbf{z}}\times\bm{g}_1)\mathbb{Z} +\ell^2\hat{\bm{\tau}}\cdot(\hat{\mathbf{z}}\times\bm{g}_2)\mathbb{Z}]$ (see SM \cite{suppl} Sec. S2B). 
It can then be proved that all the states $|\lambda,\mathbf{Q}_\alpha,n,\alpha\rangle$ form a complete basis for the continuum model satisfying $\langle \lambda,\mathbf{Q}_\alpha,n,\alpha|\lambda',\mathbf{Q}_\beta',n',\beta\rangle=\delta_{\lambda\lambda'}\delta_{\mathbf{Q}_\alpha,\mathbf{Q}_\beta'}\delta_{nn'}\delta_{\alpha\beta}$.

The above basis $|\lambda,\mathbf{Q}_\alpha,n,\alpha\rangle$ is advantageous because the nonzero-magnetic-field Hamiltonian is diagonal in $\lambda$ and independent of $\lambda$. 
In SM \cite{suppl} Sec. S2B, we show the Hamiltonian in a fixed $\lambda$ subspace is
\begin{equation}\label{eq-HQ}
H_{\mathbf{Q}_\alpha'\mathbf{Q}_\beta}^{\lambda,\alpha\beta} =\epsilon^{\alpha\beta}(\hat{\bm{\kappa}}_{\mathbf{Q}_\beta}+\mathbf{k}_0+\mathbf{Q}_\beta)\delta_{\mathbf{Q}_\alpha'\mathbf{Q}_\beta}+  \sum_{j} V_{j}\delta_{\mathbf{Q}_\alpha',\mathbf{Q}_\beta+\mathbf{q}_j^{\alpha\beta}},
\end{equation}
where we have defined $\hat{\bm{\kappa}}_{\mathbf{Q}_\alpha}=\frac{1}{{\sqrt{2}\ell}}(a_{\mathbf{Q}_\alpha}+a^\dag_{\mathbf{Q}_\alpha}, -ia_{\mathbf{Q}_\alpha}+ia^\dag_{\mathbf{Q}_\alpha})$. Without ambiguity, we can drop the subindex $\mathbf{Q}_\alpha$ and simplify $(a_{\mathbf{Q}_\alpha},a_{\mathbf{Q}_\alpha}^\dag)$ as $(a,a^\dag)$, which acts as $a|\lambda,\mathbf{Q}_\alpha,n,\alpha\rangle=\sqrt{n}|\lambda,\mathbf{Q}_\alpha,n-1,\alpha\rangle$ and $a^\dag|\lambda,\mathbf{Q}_\alpha,n,\alpha\rangle=\sqrt{n+1}|\lambda,\mathbf{Q}_\alpha,n+1,\alpha\rangle$. The Hamiltonian (\ref{eq-HQ}) is then exactly the zero-field Hamiltonian $H^{\alpha\beta}_{\mathbf{Q}_\alpha'\mathbf{Q}_\beta}(\mathbf{k})$ in Eq. (\ref{eq-HQQ}) with the substitution
\begin{equation}\label{eq-sub}
k_x\rightarrow \frac{a+a^\dag}{\sqrt{2}\ell}+k_{0,x}\ ,\quad k_y\rightarrow\frac{a-a^\dag}{i\sqrt{2}\ell}+k_{0,y}\ ,
\end{equation}
as we claimed earlier. One then only need calculate the spectrum for a fixed $\lambda$. Different $\lambda$ and $\lambda'$ subspaces have identical spectra, but have eigenstates differing by displacement $\lambda-\lambda'$ in the $\hat{\bm{\tau}}$ direction ($R_{\hat{\bm{\tau}}}$ eigenvalue).

To numerically calculate the spectrum of Hamiltonian (\ref{eq-HQ}), one can fix a center momentum $\mathbf{k}_0$, take a LL cutoff $n\le N_L$, and take a cutoff of reciprocal lattice $\mathbf{Q}_\alpha$ at a boundary enclosing $N_Q$ BZs. This yields a Hamiltonian of size $MN_LN_Q$ for $M$ intrinsic orbitals. If $\epsilon(\mathbf{k})$ only contains polynomials up to $\Delta$-th power of $\mathbf{k}$, and the number of $\mathbf{q}_j^{\alpha\beta}$ is finite, $\langle\lambda,\mathbf{Q}_\alpha',m,\alpha|H|\lambda,\mathbf{Q}_\beta,n,\beta\rangle$ will be zero for $|m-n|>\Delta$ or $|\mathbf{Q}_\alpha'-\mathbf{Q}_\beta|>\max(|\mathbf{q}_j^{\alpha\beta}|)$, so
the Hamiltonian $H$ is a sparse matrix. The low-energy eigenstates and spectrum can then be efficiently calculated by the Lanczos algorithm.

\begin{figure}[tbp]
\begin{center}
\includegraphics[width=3.4in]{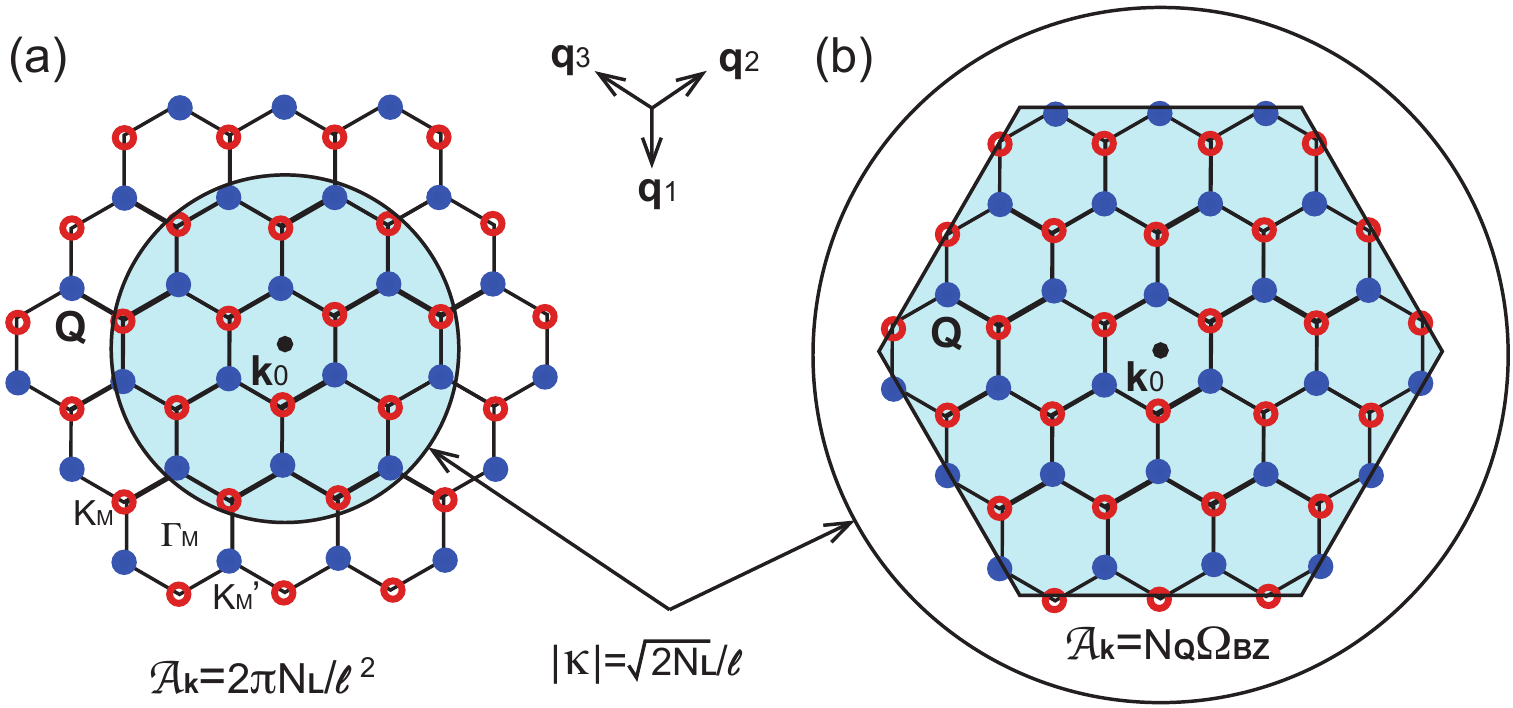}
\end{center}
\caption{(a) When $\varphi/2\pi<N_Q/N_L$, the momentum space (the shaded area) has a circular boundary of radius $\sqrt{2N_L}/\ell$. (b) When $\varphi/2\pi>N_Q/N_L$, the momentum space boundary is the reciprocal lattice boundary enclosing $N_Q$ BZs (the shaded area). 
}
\label{kspace}
\end{figure}

\begin{figure}[tbp]
\begin{center}
\includegraphics[width=3.4in]{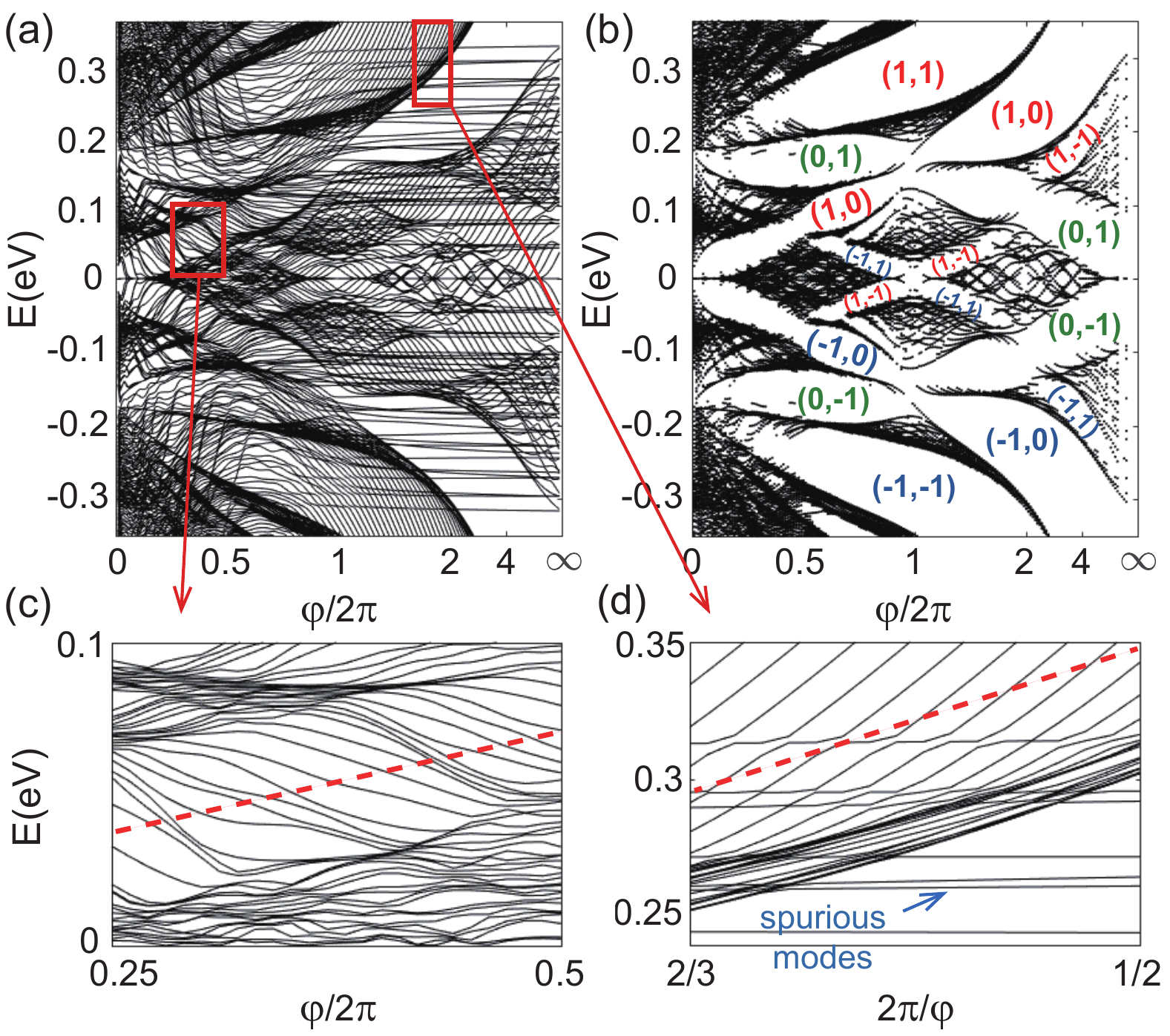}
\end{center}
\caption{(a) Hofstadter butterfly and spectral flow of $\theta=2.2^\circ$ TBG with $N_Q=37$ and $N_L=60$, and $\mathbf{k}_0$ at $\Gamma$ point. The horizontal axis $\varphi/2\pi$ is linearly plotted in $[0,1]$, and deformed into $2-2\pi/\varphi$ in $[1,\infty]$. (b) The Hofstadter butterfly after deleting the edge states (with $w=\min\{\ell^{-1},1.6\sqrt{\Omega_{BZ}}\}$, $P_c=0.5$), and $(t_\nu,s_\nu)$ in the gaps. (c) Zoom-in plot in the regime $\varphi/2\pi<N_Q/N_L$. (d) Zoom-in plot in the regime $\varphi/2\pi>N_Q/N_L$.
}
\label{Ghof}
\end{figure}

The cutoffs $N_Q$ and $N_L$, however, lead to spectral flows in the Hofstadter gaps due to the absence of periodic boundary conditions \cite{asboth2017,lian2019b}. As an example, we calculate the Hofstadter butterfly of the TBG continuum model defined on a honeycomb momentum lattice $\mathbf{Q}_\alpha$  \cite{bistritzer2011}, which has a Dirac kinetic term $\epsilon(\mathbf{k})=v_F\bm{\sigma}^*\cdot\mathbf{k}$, and $2\times2$ hopping matrices $V_j$ between the nearest momentum sites, where $\bm{\sigma}^*=(\sigma_x,-\sigma_y)$ are the Pauli matrices (SM \cite{suppl} Sec. S3). Fig. \ref{Ghof}(a) shows the TBG spectrum at twist angle $\theta=2.2^\circ$ versus the flux per unit cell $\varphi=B|\bm{d}_1\times\bm{d}_2|$, where we take $N_Q=37$ and $N_L=60$. Besides the Hofstadter butterfly, one can see numerous in-gap spectral flow levels.

The in-gap spectral flows are generically due to the presence of boundaries which host edge states \cite{asboth2017,lian2019b}. 
Here, as illustrated in Fig. \ref{kspace}(a) and (b), the cutoff $N_Q$ sets a boundary of momentum radius $\sqrt{\frac{N_Q\Omega_{BZ}}{\pi}}$ enclosing $N_Q$ BZs, where $\Omega_{BZ}=4\pi^2/|\bm{d}_1\times\bm{d}_2|$ is the BZ area; while the LL cutoff $N_L$ yields a boundary $\sqrt{\langle\hat{\bm{\kappa}}_{\mathbf{Q}_\alpha}^2\rangle}\le \frac{\sqrt{2N_L}}{\ell}$ for $\hat{\bm{\kappa}}_{\mathbf{Q}_\alpha}$ in the Hamiltonian (\ref{eq-HQ}). The smaller value of $\sqrt{\frac{N_Q\Omega_{BZ}}{\pi}}$ and $\frac{\sqrt{2N_L}}{\ell}$ then serves as a momentum space boundary radius for Hamiltonian (\ref{eq-HQ}) (SM \cite{suppl} Sec. S2C), which gives rise to edge states. 

The momentum space edge state levels then generate the spectral flows versus the magnetic field $B$. This can be understood from the Diophantine equation \cite{claro1979,dana1985,satija2016,suppl} satisfied by the $\nu$-th Hofstadter gap ($\nu\in\mathbb{Z}$) at $\varphi=2\pi p/q$ flux per unit cell:
\begin{equation}
t_\nu p+s_\nu q=\nu\ ,
\end{equation}
where $(t_\nu,s_\nu)$ are two integer quantum numbers characterizing the gap. $t_\nu$ is the Chern number of the gap, while $s_\nu$ is referred to as the electromechanical quantum number in Ref. \cite{bistritzer2011a}. It is often rewritten as
\begin{equation}\label{eq-rho}
t_\nu\frac{\varphi}{2\pi}+s_\nu=\rho\ ,
\end{equation}
where $\rho=\nu/q$ is the number of occupied bulk states per unit cell in the gap \cite{suppl,wannier1978,streda1982}. 
Here it is more useful to rewrite it in a dual form
\begin{equation}\label{eq-rhoK}
t_\nu+s_\nu\frac{2\pi}{\varphi}=\rho_K\ ,
\end{equation}
where $\rho_K=2\pi\rho/\varphi=\nu/p$. 
In SM \cite{suppl} Sec. S2F, we show that $\rho_K$ gives the number of occupied bulk states \emph{per BZ} in the gap for the Hamiltonian (\ref{eq-HQ}) at a fixed $\lambda$. Furthermore, in Eq. (\ref{eq-tknn-s}), we show that $s_\nu$ plays the role of a dual Chern number for the momentum space.  Eq. (\ref{eq-rhoK}) then determines the in-gap spectral flows (Fig. \ref{Ghof}(a)) in two different regimes as follows.

In the first regime $\varphi/2\pi<N_Q/N_L$, the momentum space boundary is a circle enclosing a $\varphi$-dependent area $\mathcal{A}_K=2\pi N_L/\ell^2$ centered at $\mathbf{k}_0$ (Fig. \ref{kspace}(a)). In a gap, the total number of occupied states in the momentum area $\mathcal{A}_K$ is $\mathcal{N}_K=\rho_K\mathcal{A}_K/\Omega_{BZ}=N_L\rho_K\varphi/2\pi= N_L\rho$. Therefore, by Eq. (\ref{eq-rhoK}) we have
\begin{equation}\label{eq-Nks}
\mathcal{N}_K=N_L(t_\nu\varphi/2\pi+s_\nu)\ .
\end{equation}
In a bulk gap, the number of occupied states $\mathcal{N}_K$ can only change by pumping edge states into (out of) the bulk. Therefore, the edge states necessarily produce in-gap spectral flows, where the rate of flowing levels is $\text{d}\mathcal{N}_K/\text{d}(\varphi/2\pi)=N_Lt_\nu$ by Eq. (\ref{eq-Nks}). Fig. \ref{Ghof}(c) shows a gap in this regime, where the midgap line (dashed line) crosses $16$ levels as $\varphi/2\pi$ increases from $0.25$ to $0.5$, and $N_L=60$. The flow rate is then $\text{d}\mathcal{N}_K/\text{d}(\varphi/2\pi)=16/(0.5-0.25)=64$, so we can identify the Chern number of the gap as the integer closest to $N_L^{-1}\text{d}\mathcal{N}_K/\text{d}(\varphi/2\pi)=1.07$, namely, $t_\nu=1$. Further, at $\varphi/2\pi=0.5$, we counted there are $\mathcal{N}_K=32$ levels from zero energy to the midgap energy (for TBG, $\mathcal{N}_K=0$ is at zero energy (SM \cite{suppl} Sec. S2G)), so we find the gap has $s_\nu=\mathcal{N}_K/N_L-t_\nu\varphi/2\pi=0$ from Eq. (\ref{eq-Nks}).

In the second regime $\varphi/2\pi>N_Q/N_L$, the momentum space boundary is given by cutoff $N_Q$, which encloses a $\varphi$ independent area $\mathcal{A}_K=N_Q\Omega_{BZ}$ (Fig. \ref{kspace}(b)). The number of occupied states $\mathcal{N}_K=\rho_K\mathcal{A}_K/\Omega_{BZ}$ in momentum area $\mathcal{A}_K$ in a gap is then
\begin{equation}\label{eq-Nkg}
\mathcal{N}_K=N_Q(t_\nu+2\pi s_\nu/\varphi)\ .
\end{equation}
This yields a spectral flow rate $\text{d}\mathcal{N}_K/\text{d}(2\pi/\varphi)=N_Qs_\nu$. 
Besides, for TBG which has a Dirac kinetic term, there are $2N_Q$ horizontal levels at $\varphi/2\pi>N_Q/N_L$ in Fig. \ref{Ghof}(a), which are spurious zero modes due to LL cutoff $N_L$ (see SM \cite{suppl} Sec. S3). 
These spurious levels should be excluded when counting $\mathcal{N}_K$. Fig. \ref{Ghof}(d) shows a gap in this regime, where the midgap line crosses $6$ levels (excluding the spurious modes) as $2\pi/\varphi$ decreases from $2/3$ to $1/2$, and $N_Q=37$. The flow rate is then $\text{d}\mathcal{N}_K/\text{d}(2\pi/\varphi)=6/(2/3-1/2)=36$, thus $s_\nu$ can be identified as the integer closest to $N_Q^{-1}\text{d}\mathcal{N}_K/\text{d}(2\pi/\varphi)=0.97$, namely, $s_\nu=1$. Further, we counted there are $\mathcal{N}_K=55$ levels (excluding the spurious modes) between midgap and zero energy at $2\pi/\varphi=1/2$, thus the gap has a Chern number $t_\nu=\mathcal{N}_K/N_Q-2\pi s_\nu/\varphi=1$ from Eq. (\ref{eq-Nkg}).

We note that models with a Dirac kinetic term $\epsilon(\mathbf{k})=v_F\bm{\sigma}^*\cdot\mathbf{k}$ would have $\mathcal{N}_K=0$ defined at half filling (zero energy for TBG), while models with a lower-bounded kinetic term (e.g., $\epsilon(\mathbf{k})=k^2/2m_0$) would have $\mathcal{N}_K=0$ below the lowest band (SM \cite{suppl} Sec. S2G). More generically, if a gap persists below and above $\varphi/2\pi=N_Q/N_L$, one can identify $t_\nu$ and $s_\nu$ separately from the spectral flow rates at small and large $\varphi$, after which one can obtain $\mathcal{N}_K$ of the gap from Eq. (\ref{eq-Nks}) or (\ref{eq-Nkg}).

The edge states and spurious modes can be easily removed from the spectrum. We define a boundary projector $P_{\kappa_b,w}$ onto basis $|\lambda,\mathbf{Q}_\alpha,n,\alpha\rangle$ with $n>(\kappa_b-w)^2\ell^2/2$ for some $w>0$, where $\kappa_b\approx\min\{\frac{\sqrt{2N_L}}{\ell}, \sqrt{\frac{N_Q\Omega_{BZ}}{\pi}}\}$ is the radius of momentum space boundary. 
We can then identify the eigenstates with $\langle P_{\kappa_b,w} \rangle> P_c$ above certain value $P_c\in[0,1]$ as momentum space edge states within distance $w$ to the boundary, and delete them to obtain a bulk Hofstadter spectrum. For example, Fig. \ref{Ghof}(b) is obtained by setting $w=\min\{\ell^{-1},1.6\sqrt{\Omega_{BZ}}\}$ and $P_c=0.5$.

\emph{Tight-binding models.} The substitution (\ref{eq-sub}) can also be employed to calculate the Hofstadter butterfly of tight-binding models. Given the position $\bm{u}_\alpha$ of each Wannier orbital $\alpha$ in a unit cell in the continuum space, we denote orbital $\alpha$ at position $\bm{D}+\bm{u}_\alpha$ as $|\bm{D},\alpha\rangle$, where $\bm{D}\in \bm{d}_1\mathbb{Z}+\bm{d}_2\mathbb{Z}$ is the lattice vector. The Hamiltonian under Peierls substitution \cite{Luttinger1951,kita2005,alex2018} then takes the form
\begin{equation}\label{eq-Htb}
H=\sum_{j,\alpha,\beta}t_{j}^{\alpha\beta} T_{\bm{D}_j+\bm{u}_\alpha- \bm{u}_\beta}\ ,
\end{equation}
where $\bm{D}_j\in\bm{d}_1\mathbb{Z}+\bm{d}_2\mathbb{Z}$, $t_{j}^{\alpha\beta}$ is the hopping from $|\bm{D},\beta\rangle$ to $|\bm{D}+\bm{D}_j,\alpha\rangle$, and
\begin{equation}
T_{\bm{D}_j+\bm{u}_\alpha- \bm{u}_\beta}=\sum_{\bm{D}} e^{i\int_{c_{\alpha\beta}}\mathbf{A}(\mathbf{r})\cdot \text{d}\mathbf{r}}|\bm{D}+\bm{D}_j,\alpha\rangle\langle \bm{D},\beta|
\end{equation}
is the translation operator, with $c_{\alpha\beta}$ being the straight line segment from $\bm{D}+\bm{u}_\beta$ to $\bm{D}+\bm{D}_j+\bm{u}_\alpha$. 
At zero magnetic field, the Hamiltonian can be transformed into the momentum space basis $|\mathbf{k},\alpha\rangle=\sum_{\bm{D}}e^{i\mathbf{k}\cdot(\mathbf{D}+\bm{u}_\alpha)}|\bm{D},\alpha\rangle$ as
\begin{equation}\label{eq-Htbk}
H^{\alpha\beta}(\mathbf{k})=\sum_{j}t_{j}^{\alpha\beta}e^{-i\mathbf{k}\cdot(\bm{D}_j+\bm{u}_\alpha- \bm{u}_\beta)}\ .
\end{equation}

At nonzero magnetic field, we define a basis as $\overline{|\lambda,n,\alpha\rangle}=\sum_{\bm{D}}|\bm{D},\alpha\rangle\langle \bm{D}+\bm{u}_\alpha,\alpha|\lambda,\mathbf{0},n,\alpha\rangle$, where $|\bm{D}+\bm{u}_\alpha,\alpha\rangle$ is the continuum space position eigenstate at position $\mathbf{r}=\bm{D}+\bm{u}_\alpha$, 
$|\lambda,\mathbf{0},n,\alpha\rangle$ is the state defined in Eq. (\ref{eq-Rt}) in the continuum space at reciprocal site $\mathbf{0}$, and $\lambda\in\mathbb{R}/[\ell^2\hat{\bm{\tau}}\cdot(\hat{\mathbf{z}}\times\bm{g}_1)\mathbb{Z} +\ell^2\hat{\bm{\tau}}\cdot(\hat{\mathbf{z}}\times\bm{g}_2)\mathbb{Z}]$. One can then show that $\overline{|\lambda,n,\alpha\rangle}$ forms a complete orthonormal basis of Hamiltonian (\ref{eq-Htb}) satisfying $\overline{\langle\lambda',n',\beta}|\overline{\lambda,n,\alpha\rangle} =\delta_{\lambda\lambda'}\delta_{n'n}\delta_{\beta\alpha}$ (SM \cite{suppl} Sec. S4A). Furthermore, $T_{\bm{D}_j+\bm{u}_\alpha- \bm{u}_\beta}$ is diagonal in $\lambda$ and takes the $\lambda$ independent form
\begin{equation}\label{eq-TD}
T^{\lambda,\alpha\beta}_{\bm{D}_j+\bm{u}_\alpha- \bm{u}_\beta}=e^{-i(\hat{\bm{\kappa}}+\mathbf{k}_0)\cdot(\bm{D}_j+\bm{u}_\alpha- \bm{u}_\beta)}
\end{equation}
in a fixed $\lambda$ subspace between basis $\overline{|\lambda,n,\beta\rangle}$ and $\overline{|\lambda,n',\alpha\rangle}$, where $\hat{\bm{\kappa}}=\frac{1}{{\sqrt{2}\ell}}(a+a^\dag, -ia+ia^\dag)$, with $a\overline{|\lambda,n,\alpha\rangle}=\sqrt{n}\overline{|\lambda,n-1,\alpha\rangle}$ and $a^\dag\overline{|\lambda,n,\alpha\rangle}=\sqrt{n+1}\overline{|\lambda,n+1,\alpha\rangle}$ (SM \cite{suppl} Sec. S4A). Therefore, the nonzero magnetic field tight-binding Hamiltonian (\ref{eq-Htb}) in a fixed $\lambda$ is given by the zero-field momentum space Hamiltonian (\ref{eq-Htbk}) with substitution (\ref{eq-sub}). For nonstandard Peierls substitutions along nonstraight $c_{\alpha\beta}$ paths, $e^{-i(\hat{\bm{\kappa}}+\mathbf{k}_0)\cdot(\bm{D}_j+\bm{u}_\alpha- \bm{u}_\beta)}$ in Eq. (\ref{eq-TD}) becomes the path-ordered integral $\mathcal{P}e^{-i\int_{c_{\alpha\beta}}(\hat{\bm{\kappa}}+\mathbf{k}_0)\cdot \text{d}\mathbf{r}}$ (SM \cite{suppl} Sec. S4B).

\begin{figure}[tbp]
\begin{center}
\includegraphics[width=3.4in]{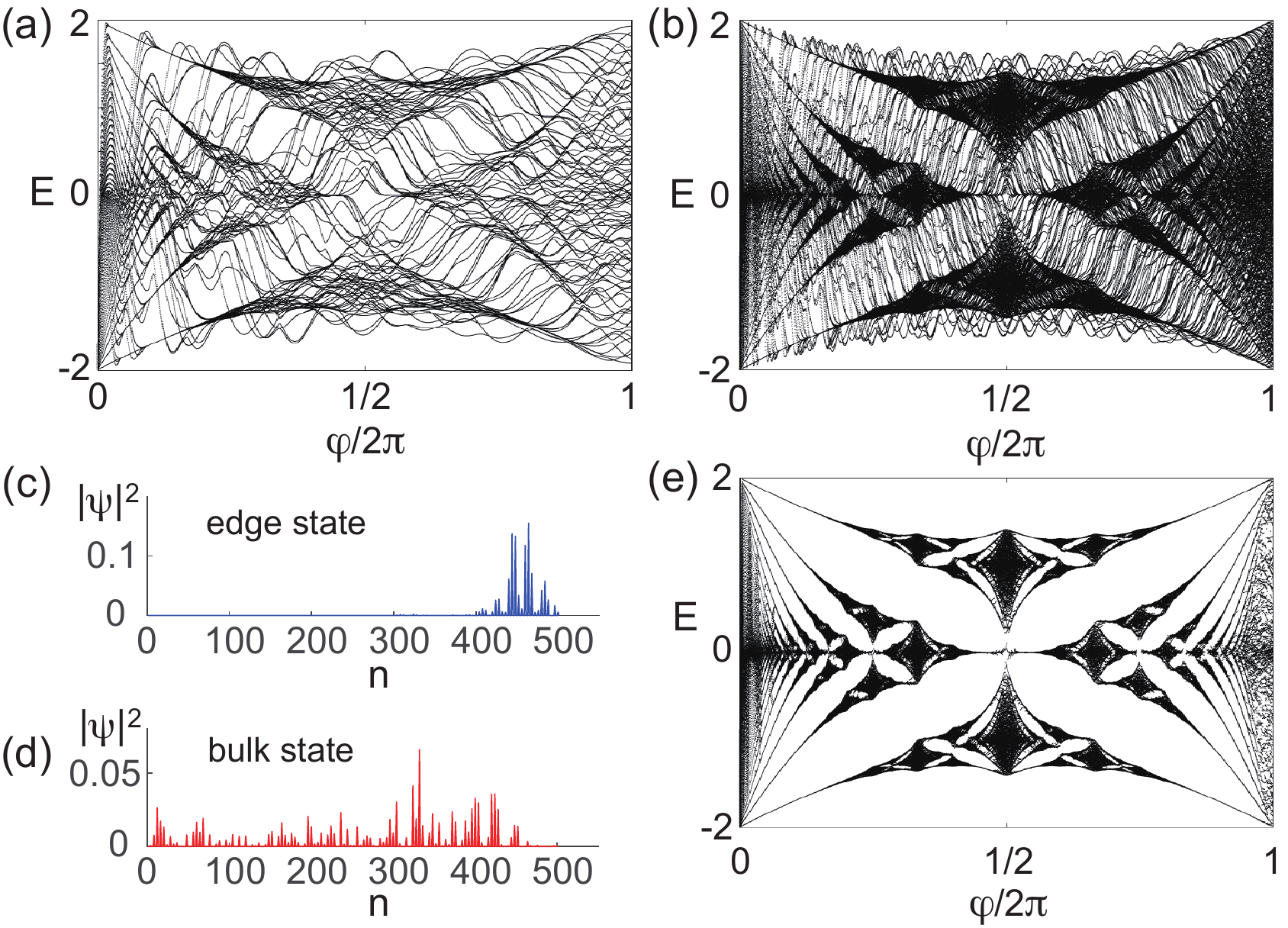}
\end{center}
\caption{The Hofstadter spectrum for tight-binding model $H(\mathbf{k})=-\cos k_x-\cos k_y$ with $\mathbf{k}_0=\mathbf{0}$ and LL cutoff (a) $N_L=100$ and (b) $N_L=500$. (c) Probability distribution of a typical momentum space edge state in (b) versus LL number $n$. (d) Probability distribution of a typical bulk state in (b). (e) The Hofstadter butterfly obtained by deleting the edge states in (b) (with $w=3.5\ell^{-1}\sqrt{\varphi+0.5}$, $P_c=0.5$), which looks identical to that obtained by usual methods.
}
\label{TBhof}
\end{figure}

The Hofstadter butterfly can then be numerically calculated with a LL cutoff, namely, $n\le N_L$. Fig. \ref{TBhof}(a) and (b) show the spectrum of the square lattice tight-binding model $H(\mathbf{k})=-\cos k_x-\cos k_y$ \cite{hofstadter1976} with cutoffs $N_L=100$ and $N_L=500$, respectively, where we set $\mathbf{k}_0=\mathbf{0}$. The spectrum exhibits both the Hofstadter butterfly and the spectral flows, which can again be understood as momentum space edge states. Since tight-binding models have no cutoff in the reciprocal lattice, the momentum space boundary is always at radius $\kappa_b=\frac{\sqrt{2N_L}}{\ell}$ given by $N_L$, and the spectral flows always satisfy Eq. (\ref{eq-Nks}).

One can define a boundary projector $P_{\kappa_b,w}$ onto basis $\overline{|\lambda,n,\alpha\rangle}$ with $n>(\kappa_b-w)^2\ell^2/2$ for certain $w>0$, and identify the eigenstates with $\langle P_{\kappa_b,w}\rangle>P_c$ for some $P_c\in[0,1]$ as momentum space edge states. Fig. \ref{TBhof}(c) and (d) show the probability of a typical edge state and bulk state versus LL number $n$, respectively. By deleting the edge states, one can obtain a high-quality Hofstadter butterfly without spectral flows, as shown in Fig. \ref{TBhof}(e) (where $w=3.5\ell^{-1}\sqrt{\varphi+0.5}$ and $P_c=0.5$).

\emph{Quantized Lorentz susceptibility}. The Chern number $t_\nu$ is known to give a quantized Hall conductance via the Kubo formula $\sigma_{xy}=i\partial_\omega \int \text{d}\omega'\langle G_{\omega+\omega'}\hat{j}_{x} G_{\omega'}\hat{j}_{y}\rangle|_{\omega\rightarrow 0}=t_\nu\frac{e^2}{h}$ \cite{tknn1982}, where $G_\omega$ is the Green's function at energy $\omega$, and $\hat{\bm{j}}=(\hat{j}_x,\hat{j}_y)$ is the uniform current operator. The duality between Eqs. (\ref{eq-rho}) and (\ref{eq-rhoK}) suggests that $s_\nu$ behaves as a dual Chern number for the momentum space, thus $s_\nu$ should also give a quantized response. Indeed, by noting that the natural momentum-space dual of the current operator $\hat{\bm{j}}$ is the force operator $\hat{\bm{F}}=e\mathbf{B}\times\frac{\text{d}\mathbf{R}}{\text{d}t}=(\hat{F}_x,\hat{F}_y)$, we find $s_\nu$ leads to a quantized \emph{Lorentz susceptibility} (SM \cite{suppl} Sec. S6)
\begin{equation}\label{eq-lorentz}
\gamma_{xy}=-i\frac{\partial}{\partial \omega} \int \text{d}\omega'\langle G_{\omega+\omega'}\hat{F}_{x} G_{\omega'}\hat{F}_{y}\rangle\Big|_{\omega\rightarrow 0}=eB s_\nu.
\end{equation}
It yields a Lorentz force per unit cell $F_{x}=\gamma_{xy} v_y$ on the system when the lattice is moving at velocity $v_y$. Furthermore, a formula similar to the Thouless-Kohmoto-Nightingale-den Nijs formula \cite{tknn1982} at flux per unit cell $\varphi=2\pi p/q$ can be derived for $s_\nu$ (SM \cite{suppl} Sec. S6B2):
\begin{equation}\label{eq-tknn-s}
s_\nu=-i\sum_{n\in\text{occ}}\int_{\bm{d}\in\Omega_M} \frac{\text{d}^2\bm{d} }{2\pi}\hat{\mathbf{z}}\cdot\langle\partial_{\bm{d}}w_{n,\bm{d}}|\times|\partial_{\bm{d}}w_{n,\bm{d}}\rangle\ ,
\end{equation}
where $\Omega_M$ is a torus with periods $\bm{d}_1$ and $\bm{d}_2/p$ serving as a ``dual magnetic BZ", $|w_{n,\bm{d}}\rangle=e^{i \ell^{-2}(\hat{\mathbf{z}}\times\mathbf{R})\cdot\bm{d}}|\psi_{n,\ell^{-2}\hat{\mathbf{z}}\times\bm{d}}\rangle$ (see the explicit form in SM \cite{suppl} Sec. S6C) is defined using the Bloch eigenstates $|\psi_{n,\mathbf{k}}\rangle$ of band $n$, and $n$ runs over all occupied bands.

\emph{Discussion}. It is worth noting that the cutoffs in our method affect the resolution but not the shape of the Hofstadter butterfly. Our method greatly simplifies the matrix element construction compared to usual methods \cite{hofstadter1976,bistritzer2011a}, and require neither rational flux per unit cell nor large magnetic unit cells, making it easy to calculate the Hofstadter spectra of complicated models \cite{lian2019b,arbeitman2020,lu2020multiple,burg2020evidence}. Moreover, it leads to a sparse Hamiltonian for continuum models. At small magnetic fields, our method reduces to the LL calculations of $\mathbf{k}\cdot\mathbf{p}$ Hamiltonians expanded at center momentum $\mathbf{k}_0$. The large magnetic field spectrum is insensitive to the choice of $\mathbf{k}_0$.

\begin{acknowledgments}
\emph{Acknowledgments}. We thank Michael Zaletel, Hoi Chun Po and Junyi Zhang for helpful discussions. B.L. acknowledge the support of Princeton Center for Theoretical Science at Princeton University at the early stage of this work. B.A.B was supported by the DOE Grant No. DE-SC0016239, the Schmidt Fund for Innovative Research, Simons Investigator Grant No. 404513, and the Packard Foundation. Further support was provided by the NSF-EAGER No. DMR 1643312, NSF-MRSEC No. DMR-1420541 and DMR-2011750, ONR No. N00014-20-1-2303, Gordon and Betty Moore Foundation through Grant GBMF8685 towards the Princeton theory program, BSF Israel US foundation No. 2018226, and the Princeton Global Network Funds.

\end{acknowledgments}

\bibliography{hof_ref}

\begin{widetext}

\section*{Supplementary Material}

\tableofcontents



\section{The algebra in a magnetic field}\label{sec-algebra}
We first review the algebra obeyed by 2-dimensional (2D) electrons in a uniform static magnetic field $B$ in the continuum real space (in the first quantized language). We denote $\mathbf{r}=(x,y)$ as the position operator, and $-i\nabla=(-i\hbar\partial_x,-i\hbar\partial_y)$ as the canonical momentum operator.

Assume the magnetic field $B$ corresponds to a gauge field $\mathbf{A}(\mathbf{r})=(A_x(\mathbf{r}),A_y(\mathbf{r}))$, which satisfies $\partial_xA_y-\partial_yA_x=B$. We can then define the magnetic length $\ell=\sqrt{\hbar/eB}$, where $e$ is the electron charge, and $\hbar$ is the Planck constant. The kinematic momentum operator of an electron in a magnetic field is given by $\bm{\Pi}=(\Pi_x,\Pi_y)$, which satisfies
\begin{equation}\label{Seq-Pi}
\Pi_x=-i\partial_x-eA_x\ ,\qquad \Pi_y=-i\partial_y-eA_y\ ,\qquad [\Pi_x,\Pi_y]=\frac{i\hbar^2}{\ell^2}\ ,
\end{equation}
or in vector form $\bm{\Pi}=-i\nabla-\mathbf{A}(\mathbf{r})$.
In the absence of the gauge field, $\bm{\Pi}=-i\nabla$ is the same as the canonical momentum.
We also define the real space guiding center coordinates $\mathbf{R}=(R_x,R_y)$, which satisfy
\begin{equation}\label{Sec-Guiding}
R_x=x+\frac{\ell^2}{\hbar}\Pi_y\ ,\qquad R_y=y-\frac{\ell^2}{\hbar}\Pi_x\ ,\qquad [R_x,R_y]=-i\ell^2\ .
\end{equation}
It can be written in the vector form as $\mathbf{R}=\mathbf{r}-\frac{\ell^2}{\hbar}\hat{\mathbf{z}}\times\bm{\Pi}$.
Semiclassically, the guiding center is the central position of the cyclotron motion of an electron in magnetic field $B$. The kinematic momentum operator $\bm{\Pi}$ commutes with the guiding center operator $\mathbf{R}$, namely, $[\Pi_x,R_x]=[\Pi_x,R_y]=[\Pi_y,R_x]=[\Pi_y,R_y]=0$.

For convenience, hereafter we set $e=\hbar=1$, unless recovery of the original units is needed. Besides, we always understand $\mathbf{r}$ as the position operator instead of a vector parameter, except that $|\mathbf{r}\rangle$ stands for a state at position $\mathbf{r}$ (where $\mathbf{r}$ is a parameter). 

\section{Basis Completeness and Matrix Elements for Continuum models}\label{sec-CM}
In this section, we give the detailed derivation of the basis we choose and the Hamiltonian matrix elements for continuum models at zero magnetic field and nonzero magnetic field. 

\subsection{Continuum model in real space}

We consider a continuum model with $M$ intrinsic orbitals per zero-magnetic-field unit cell. For example, in the one-valley one-spin twisted bilayer graphene (TBG) continuum model in Ref. \cite{bistritzer2011} (see also Sec. \ref{sec-tBLG}), the intrinsic orbitals are graphene sublattice and layer indices. In more generic examples with spin-orbit coupling, intrinsic orbitals also include spin, etc. We denote the lattice Bravais vectors as $\bm{d}_1$ and $\bm{d}_2$, and the reciprocal vectors as $\bm{g}_1$ and $\bm{g}_2$, which satisfy 
\begin{equation}\label{Seq-gd}
\bm{g}_i\cdot\bm{d}_{j}=2\pi\delta_{ij}\ ,\qquad (i,j=1,2).
\end{equation}
We denote the reciprocal lattice as 
\begin{equation}\label{Seq-reciprocal-lattice}
\mathbf{Q}=m_1\bm{g}_1+m_2\bm{g_2}\ ,\qquad (m_1,m_2\in\mathbb{Z}).
\end{equation}
\emph{In the absence of magnetic field}, the continuum model Hamiltonian in a continuum space with a lattice potential is of the generic form
\begin{equation}
\mathbf{H}=\int \text{d}^2\mathbf{r} c^\dag_\alpha(\mathbf{r}) \widetilde{H}^{\alpha\beta}(\mathbf{r}) c_\beta(\mathbf{r})\ ,
\end{equation}
where
\begin{equation}\label{Seq-Cmodel0}
\widetilde{H}^{\alpha\beta}(\mathbf{r})=\widetilde{\epsilon}^{\alpha\beta}(-i\nabla)+\sum_{j} \widetilde{V}^{\alpha\beta}_{j}e^{i\mathbf{q}_j\cdot\mathbf{r}}\ .
\end{equation}
Here $\alpha,\beta$ denote the $M$ intrinsic orbitals, $c_\alpha(\mathbf{r})$, $c^\dag_\alpha(\mathbf{r})$ are the electron annihilation and creation operators of orbital $\alpha$ at position $\mathbf{r}$, $\widetilde{\epsilon}^{\alpha\beta}(-i\nabla)$ is the kinetic term in free space, $\widetilde{V}^{\alpha\beta}_{j}$ is the momentum $\mathbf{q}_j$ component of the lattice potential. 
The momentum $\mathbf{q}_j$ of the lattice potential component $\widetilde{V}^{\alpha\beta}_{j}$ satisfy
\begin{equation}
\mathbf{q}_j\in\mathbf{Q}\ ,
\end{equation}
where $\mathbf{Q}$ is the set of reciprocal lattice sites in Eq. (\ref{Seq-reciprocal-lattice}); thus the lattice potential is periodic with lattice Bravais vectors $\bm{d}_1,\bm{d}_2$. Besides, the Hamiltonian is Hermitian, namely, $\widetilde{\epsilon}^{\alpha\beta}(-i\nabla)=\widetilde{\epsilon}^{\beta\alpha}(-i\nabla)^\dag$, and $V^{\alpha\beta}_{j}=V^{\beta\alpha*}_{\overline{j}}$ for momenta $\mathbf{q}_{\overline{j}}=-\mathbf{q}_j$.

Here we note that $\mathbf{H}$ denotes the second quantized Hamiltonian, and $\widetilde{H}^{\alpha\beta}(\mathbf{r})$ denotes the first quantized single-particle Hamiltonian. The basis of the first quantized Hamiltonian $\widetilde{H}^{\alpha\beta}(\mathbf{r})$ is given by 
\begin{equation}\label{Seq-r-basis0}
c^\dag_\alpha(\mathbf{r})|0\rangle\ ,
\end{equation}
with $|0\rangle$ being the vacuum state. Since we do not consider interactions, we can work in the first quantized single-particle Hamiltonian hereafter.

\subsection{Continuum model in real space with orbital-dependent momentum origin shifts}

In writing the Hamiltonian (\ref{Seq-Cmodel0}) above, the momentum origins of all orbitals are chosen at the $\Gamma$ point of the BZ of the lattice. In some continuum models, it is convenient to shift the momentum origins of different orbitals $\alpha$ to some desired momenta $\mathbf{p}_\alpha$. This is done by transforming the single-particle Hamiltonian (\ref{Seq-Cmodel0}) from the real space basis in Eq. (\ref{Seq-r-basis0}) into a new real space basis defined by
\begin{equation}\label{Seq-r-basis}
|\mathbf{r},\alpha\rangle=e^{-i\mathbf{p}_\alpha\cdot\mathbf{r}}c^\dag_\alpha(\mathbf{r})|0\rangle\ ,
\end{equation}
where $\mathbf{p}_\alpha$ is an orbital $\alpha$ dependent momentum vector which can be chosen freely. Here we shall restrict the choices of $\mathbf{p}_\alpha$ so that 
\begin{equation}\label{Seq-palpha-assumption}
\mathbf{p}_\alpha= \mathbf{p}_\beta \qquad \text{if} \qquad \widetilde{\epsilon}^{\alpha\beta}(-i\nabla)\neq0\ ,
\end{equation}
which ensures the kinetic term under the new basis (\ref{Seq-r-basis}) to be a function of $-i\nabla$ only and is independent of $\mathbf{r}$ (see Eq. (\ref{Seq-Cmodel})). 
An example of models with such orbital-dependent momentum origin shifts is the TBG continuum model originally written down in Ref. \cite{bistritzer2011}, where the orbitals $\alpha$ of the upper layer have $\mathbf{p}_\alpha=k_\theta(0,1)$, and the orbitals $\beta$ of the lower layer have $\mathbf{p}_{\beta}=k_\theta(0,-1)$ (see Sec. \ref{sec-tBLG} for definition of $k_\theta$ and more details). Condition (\ref{Seq-palpha-assumption}) is also satisfied for the TBG continuum model, since the kinetic term $\widetilde{\epsilon}^{\alpha\beta}(-i\nabla)$ between an orbital $\alpha$ in the upper layer and an orbital $\beta$ in the lower layer is zero.

Under the assumption (\ref{Seq-palpha-assumption}), we find the first quantized single-particle Hamiltonian transforms under the new basis (\ref{Seq-r-basis}) into
\begin{equation}\label{Seq-Cmodel}
H^{\alpha\beta}(\mathbf{r})=\langle \mathbf{r},\alpha|\mathbf{H}|\mathbf{r},\beta \rangle=e^{i\mathbf{p}_\alpha\cdot\mathbf{r}}\widetilde{H}^{\alpha\beta}(\mathbf{r})e^{-i\mathbf{p}_\beta\cdot\mathbf{r}}=\epsilon^{\alpha\beta}(-i\nabla)+\sum_{j} V^{\alpha\beta}_{j}e^{i\mathbf{q}_j^{\alpha\beta}\cdot\mathbf{r}}\ ,
\end{equation}
where we have defined
\begin{equation}\label{Seq-Cmodel-parameter}
\epsilon^{\alpha\beta}(-i\nabla)=\widetilde{\epsilon}^{\alpha\beta}(-i\nabla-\mathbf{p}_\alpha)\ ,\qquad V^{\alpha\beta}_{j}=\widetilde{V}^{\alpha\beta}_{j}\ ,\qquad \mathbf{q}_j^{\alpha\beta}=\mathbf{q}_j+\mathbf{p}_\alpha-\mathbf{p}_\beta\in \mathbf{Q}_\alpha-\mathbf{Q}_\beta\quad (\mathbf{q}_j\in \mathbf{Q})\ ,
\end{equation}
and the momentum lattice $\mathbf{Q}_\alpha$ for orbital $\alpha$ is defined as
\begin{equation}\label{Seq-palpha}
\mathbf{Q}_\alpha=\mathbf{p}_\alpha+\mathbf{Q}=\mathbf{p}_\alpha+m_1\bm{g}_1+m_2\bm{g}_2\ , \qquad (m_1,m_2\in\mathbb{Z})\ ,
\end{equation}
where $\mathbf{Q}$ is the reciprocal lattice sites in Eq. (\ref{Seq-reciprocal-lattice}). In particular, we note that with the constraint (\ref{Seq-palpha-assumption}), the transformed kinetic term $\epsilon^{\alpha\beta}(-i\nabla)$ in Eq. (\ref{Seq-Cmodel-parameter}) under the new real space basis (\ref{Seq-r-basis}) is still only a function of $-i\nabla$, and does not depend on $\mathbf{r}$. In contrast, if we choose vectors $\mathbf{p}_\alpha$ which do not satisfy the constraint (\ref{Seq-palpha-assumption}), after the transformation of Eq. (\ref{Seq-Cmodel}), we would have a kinetic term $\epsilon^{\alpha\beta}(-i\nabla,\mathbf{r})=\widetilde{\epsilon}^{\alpha\beta}(-i\nabla-\mathbf{p}_\alpha)e^{i(\mathbf{p_\alpha-p_\beta})\cdot\mathbf{r}}$ that is position $\mathbf{r}$ dependent, which brings unnecessary complications. Therefore, we will impose the constraint (\ref{Seq-palpha-assumption}).

We note that the single-particle Hamiltonian in Eq. (\ref{Seq-Cmodel}) is generically invariant up to a unitary transformation under the translation of a Bravais lattice vector $\bm{d}_i$. Under the real space basis $|\mathbf{r},\alpha\rangle$ in Eq. (\ref{Seq-r-basis}), if we define the translation operator of distance $\bm{d}$ which satisfies $T_{\bm{d}}|\mathbf{r},\alpha\rangle=|\mathbf{r}+\bm{d},\alpha\rangle$, its real space representation in the first quantized language is given by $T_{\bm{d}}=e^{-\bm{d}\cdot\nabla}$, and we have
\begin{equation}\label{Seq-Hr+d}
H^{\alpha\beta}(\mathbf{r}+\bm{d}_i)=T_{\bm{d}_i}^{-1} H^{\alpha\beta}(\mathbf{r})T_{\bm{d}_i}=e^{\bm{d}\cdot\nabla}\left[\epsilon^{\alpha\beta}(-i\nabla)+\sum_{j} V^{\alpha\beta}_{j}e^{i\mathbf{q}_j^{\alpha\beta}\cdot\mathbf{r}}\right]e^{-\bm{d}\cdot\nabla}=e^{i\mathbf{p}_\alpha\cdot \bm{d}_i} H^{\alpha\beta}(\mathbf{r})e^{-i\mathbf{p}_\beta\cdot \bm{d}_i}\ ,
\end{equation}
where $\bm{d}_i$ ($i=1,2$) is a Bravais lattice vector, and we have used the assumption (\ref{Seq-palpha-assumption}), and the relation (\ref{Seq-gd}). It is clear that if we set all $\mathbf{p}_\alpha=0$, we would have the usual translational invariance $H^{\alpha\beta}(\mathbf{r}+\bm{d}_i)=H^{\alpha\beta}(\mathbf{r})$ which does not involve a unitary transformation.

For generality, we shall use the first quantized continuum model single-particle Hamiltonian in Eq. (\ref{Seq-Cmodel}), which has orbital-dependent momentum origin shifts $\mathbf{p}_\alpha$, and the real space basis is defined in Eq. (\ref{Seq-r-basis}). We note that one can always choose all $\mathbf{p}_\alpha=0$, in which case the form of the continuum model Hamiltonian reduces back to Eq. (\ref{Seq-Cmodel0}).

\subsection{Transforming the model at zero magnetic field into momentum space}

Eq. (\ref{Seq-Cmodel}) gives the single-particle Hamiltonian in the absence of magnetic field. It can be written in the momentum space by Fourier transformation. To do this, we define the momentum space basis
\begin{equation}
|\mathbf{k},\mathbf{Q}_\alpha,\alpha\rangle=\frac{1}{\sqrt{\Omega_{\text{tot}}}}\int \text{d}^2\mathbf{r} e^{i(\mathbf{k}+\mathbf{Q}_\alpha)\cdot\mathbf{r}}|\mathbf{r},\alpha\rangle\ ,
\end{equation}
where $\mathbf{k}$ is the quasimomentum in the first Brillouin zone (BZ), $\mathbf{Q}_\alpha$ for intrinsic orbital $\alpha$ is defined in Eq. (\ref{Seq-palpha}), and $\Omega_{\text{tot}}$ is the total area of the system. Under this momentum space basis $|\mathbf{k},\mathbf{Q}_\alpha,\alpha\rangle$, the single-particle Hamiltonian $H^{\alpha\beta}(\mathbf{r})$ in Eq. (\ref{Seq-Cmodel}) transforms into the following form diagonal in $\mathbf{k}$ (here $\mathbf{k,k'}$ are in the first BZ):
\begin{equation}\label{Seq-HQQ1}
\begin{split}
&H^{\alpha\beta}_{\mathbf{Q}_\alpha'\mathbf{Q}_\beta}(\mathbf{k',k})=\langle\mathbf{k}', \mathbf{Q}_\alpha',\alpha|\mathbf{H}|\mathbf{k},\mathbf{Q}_\beta,\beta\rangle=\frac{1}{\Omega_{\text{tot}}} \int \text{d}^2\mathbf{r} d^2\mathbf{r}' e^{-i(\mathbf{k}'+\mathbf{Q}_\alpha')\cdot\mathbf{r}} \langle \mathbf{r},\alpha|\mathbf{H}|\mathbf{r}',\beta\rangle e^{i(\mathbf{k}+\mathbf{Q}_\beta)\cdot\mathbf{r}'} \\
=&\frac{1}{\Omega_{\text{tot}}} \int \text{d}^2\mathbf{r} e^{-i(\mathbf{k}'+\mathbf{Q}_\alpha')\cdot\mathbf{r}} H^{\alpha\beta}(\mathbf{r})e^{i(\mathbf{k}+\mathbf{Q}_\beta)\cdot\mathbf{r}}=\frac{1}{\Omega_{\text{tot}}} \int \text{d}^2\mathbf{r} e^{-i(\mathbf{k}'+\mathbf{Q}_\alpha')\cdot\mathbf{r}} \left[\epsilon^{\alpha\beta}(-i\nabla)+\sum_{j} V^{\alpha\beta}_{j}e^{i\mathbf{q}_j^{\alpha\beta}\cdot\mathbf{r}}\right]e^{i(\mathbf{k}+\mathbf{Q}_\beta)\cdot\mathbf{r}} \\
=&\delta_{\mathbf{k,k'}}\left[\epsilon^{\alpha\beta}(\mathbf{k}+\mathbf{Q}_\beta)\delta_{\mathbf{Q}'_\alpha,\mathbf{Q}_\beta}+  \sum_{j} V^{\alpha\beta}_{j}\delta_{\mathbf{Q}_\alpha',\mathbf{Q}_\beta+\mathbf{q}_j^{\alpha\beta}}\right]=\delta_{\mathbf{k,k'}}H^{\alpha\beta}_{\mathbf{Q}_\alpha'\mathbf{Q}_\beta}(\mathbf{k})\ ,
\end{split}
\end{equation}
where we have used the definition of $\mathbf{q}_j^{\alpha\beta}$ in Eq. (\ref{Seq-Cmodel-parameter}), and the condition (\ref{Seq-palpha-assumption}). The Hamiltonian diagonal in $\mathbf{k}$
\begin{equation}\label{Seq-HQQ}
H^{\alpha\beta}_{\mathbf{Q}_\alpha'\mathbf{Q}_\beta}(\mathbf{k})=\epsilon^{\alpha\beta}(\mathbf{k}+\mathbf{Q}_\beta)\delta_{\mathbf{Q}'_\alpha,\mathbf{Q}_\beta}+  \sum_{j} V^{\alpha\beta}_{j}\delta_{\mathbf{Q}_\alpha',\mathbf{Q}_\beta+\mathbf{q}_j^{\alpha\beta}}
\end{equation}
then gives Eq. (2) in the main text.

\subsection{The Basis and Hamiltonian in nonzero magnetic field}

Under \emph{a uniform magnetic field} $\mathbf{B}=B\hat{\mathbf{z}}$, the canonical momentum $-i\nabla$ in Eq. (\ref{Seq-Cmodel}) will be replaced by the kinetic momentum $\bm{\Pi}=-i\nabla-\mathbf{A}(\mathbf{r})$, where $\nabla\times\mathbf{A}(\mathbf{r})=\mathbf{B}$. The (first quantized) single-particle Hamiltonian then reads 
\begin{equation}\label{Seq-CmodelB}
H^{\alpha\beta}(\mathbf{r})=\epsilon^{\alpha\beta}(\bm{\Pi})+\sum_{j} V^{\alpha\beta}_{j}e^{i\mathbf{q}_j^{\alpha\beta}\cdot\mathbf{r}}\ .
\end{equation}
In the below, we will construct a basis under magnetic field, in which we show that the Hamiltonian in magnetic field is block diagonalized into blocks with identical matrix elements (different blocks differ by guiding center translations in the real space, see explanations below Eq. (\ref{Seq-HLQQ})), and each block is simply given by Eq. (\ref{Seq-HQQ}) (the momentum space Hamiltonian at zero magnetic field) with the substitution
\begin{equation}
\mathbf{k}\rightarrow \hat{\bm{\kappa}}+\mathbf{k}_0\ ,\qquad  \hat{\bm{\kappa}}= \frac{1}{{\sqrt{2}\ell}}(a+a^\dag, -ia+ia^\dag)\ ,
\end{equation}
where $\mathbf{k}_0$ is an arbitrarily chosen momentum vector, $a$ and $a^\dag$ are lowering and raising operators satisfying $[a,a^\dag]=1$ which we will define below, and $\ell$ is the magnetic length.

To begin, we would like to find a set of mutually commuting operators to define the quantum numbers of our basis. First, we define a set of Landau level (LL) lowering and raising operators associated with the momentum lattice sites $\mathbf{Q}_\alpha=(Q_{\alpha,x},Q_{\alpha,y})$ of orbital $\alpha$ as
\begin{equation}
a_{\mathbf{Q}_\alpha}=\frac{\ell}{\sqrt{2}}[\Pi_{x}-Q_{\alpha,x}-k_{0,x}+ i(\Pi_y-Q_{\alpha,y}-k_{0,y})]\ ,\quad a_{\mathbf{Q}_\alpha}^\dag=\frac{\ell}{\sqrt{2}}[\Pi_{x}-Q_{\alpha,x}-k_{0,x}- i(\Pi_y-Q_{\alpha,y}-k_{0,y})]\ ,
\end{equation}
where $\mathbf{k}_0=(k_{0,x},k_{0,y})$ is a freely chosen fixed momentum which we call the center momentum. They satisfy the commutation relation $[a_{\mathbf{Q}_\alpha},a_{\mathbf{Q}_\alpha}^\dag]=1$. We note that for two different sites $\mathbf{Q}_\alpha$ and $\mathbf{Q}_\alpha'$, the operators $a_{\mathbf{Q}_\alpha}$ and $a_{\mathbf{Q}_\alpha'}$ only differ by a constant shift, thus are linearly dependent. We also note that $[a_{\mathbf{Q}_\alpha}^\dag a_{\mathbf{Q}_\alpha},a_{\mathbf{Q}_\alpha'}^\dag a_{\mathbf{Q}_\alpha'}]\neq 0$ for $\mathbf{Q}_\alpha\neq \mathbf{Q}_\alpha'$. The eigenvalue of $a_{\mathbf{Q}_\alpha}^\dag a_{\mathbf{Q}_\alpha}$ for a given $\mathbf{Q}$ runs over all the nonnegative integers.

Secondly, we define (recall that $\mathbf{R}=\mathbf{r}-\frac{\ell^2}{\hbar}\hat{\mathbf{z}}\times\bm{\Pi}$)
\begin{equation}
R_{\hat{\bm{\tau}}}=\mathbf{R}\cdot\hat{\bm{\tau}}=R_x\hat{\tau}_x+R_y\hat{\tau}_y
\end{equation}
as the guiding center along the $\hat{\bm{\tau}}$ direction, where $\hat{\bm{\tau}}=(\hat{\tau}_x,\hat{\tau}_y)$ is a unit vector so chosen that $\frac{\hat{\bm{\tau}}\cdot(\hat{\mathbf{z}}\times\bm{g}_1)} {\hat{\bm{\tau}}\cdot(\hat{\mathbf{z}}\times\bm{g}_2)}$ is an irrational number (the reason will be explained below Eq. (\ref{Seq-irrational})). In an infinite real space (which we assume is the case here), the eigenvalue of $R_{\hat{\bm{\tau}}}$ runs over all real numbers $\mathbb{R}$. This can be seen from the fact that one can shift the eigenvalue of $R_{\hat{\bm{\tau}}}$ by any value $b\in\mathbb{R}$ using the operator $e^{ib \mathbf{R}\cdot(\hat{\bm{\tau}}\times\hat{\mathbf{z}})/\ell^2}$, namely, 
\begin{equation}\label{seq-Rshift}
e^{ib \mathbf{R}\cdot(\hat{\bm{\tau}}\times\hat{\mathbf{z}})/\ell^2} R_{\hat{\bm{\tau}}} e^{-ib \mathbf{R}\cdot(\hat{\bm{\tau}}\times\hat{\mathbf{z}})/\ell^2}=R_{\hat{\bm{\tau}}}+b\ ,\qquad (b\in\mathbb{R})
\end{equation}
We note that once $R_{\hat{\bm{\tau}}}$ is diagonalized for a given direction $\hat{\bm{\tau}}$, one cannot further diagonalize the guiding center coordinate $R_{\hat{\bm{\tau}}'}=\mathbf{R}\cdot\hat{\bm{\tau}}'$ along any other direction $\hat{\bm{\tau}}'\neq \pm\hat{\bm{\tau}}$, since $[R_{\hat{\bm{\tau}}'},R_{\hat{\bm{\tau}}}]=-i\ell^2\hat{\mathbf{z}}\cdot(\hat{\bm{\tau}}'\times \hat{\bm{\tau}})\neq0$ unless $\hat{\bm{\tau}}'= \pm\hat{\bm{\tau}}$ (Note that $R_{\hat{\bm{\tau}}}=-R_{-\hat{\bm{\tau}}}$, so diagonalization of $R_{\hat{\bm{\tau}}}$ is sufficient). 

Since $[\mathbf{R},\bm{\Pi}]=0$, it is easy to see that 
\begin{equation}
[R_{\hat{\bm{\tau}}},a_{\mathbf{Q}_\alpha}]=[R_{\hat{\bm{\tau}}},a_{\mathbf{Q}_\alpha}^\dag]=[R_{\hat{\bm{\tau}}},a_{\mathbf{Q}_\alpha}^\dag a_{\mathbf{Q}_\alpha}]=0 
\end{equation}
for any $\mathbf{Q}_\alpha$. Therefore, $R_{\hat{\bm{\tau}}}$ and $a_{\mathbf{Q}_\alpha}^\dag a_{\mathbf{Q}_\alpha}$ (for a given $\mathbf{Q}_\alpha$) can be diagonalized simultaneously. For a given $\mathbf{Q}_\alpha$, there is no other functions of $\mathbf{R},\bm{\Pi}$ independent of $R_{\hat{\bm{\tau}}}$ and $a_{\mathbf{Q}_\alpha}^\dag a_{\mathbf{Q}_\alpha}$ which commute with both $R_{\hat{\bm{\tau}}}$ and $a_{\mathbf{Q}_\alpha}^\dag a_{\mathbf{Q}_\alpha}$, so $R_{\hat{\bm{\tau}}}$ and $a_{\mathbf{Q}_\alpha}^\dag a_{\mathbf{Q}_\alpha}$ form the maximal commuting set of operators $\mathbf{R},\bm{\Pi}$. We shall therefore use $R_{\hat{\bm{\tau}}}$ and $a_{\mathbf{Q}_\alpha}^\dag a_{\mathbf{Q}_\alpha}$ to define the quantum numbers of our basis.

In the previous Hofstadter method for continuum models \cite{bistritzer2011a}, the chosen basis makes the lattice potential term $e^{i\mathbf{q}_j^{\alpha\beta}\cdot \mathbf{r}}$ in Eq. (\ref{Seq-CmodelB}) a complicated dense matrix. We hope to find a different basis where the operator $e^{i\mathbf{q}_j^{\alpha\beta}\cdot \mathbf{r}}$ have simple matrix elements, so that the Hamiltonian takes a much simpler form. Now we describe the construction of such a basis (in which the operator $e^{i\mathbf{q}_j^{\alpha\beta}\cdot \mathbf{r}}$ has very simple matrix elements, see Eq. (\ref{Seq-Rq})). For each momentum site $\mathbf{Q}_\alpha$, we define a set of basis $|\lambda,\mathbf{Q}_\alpha,n,\alpha \rangle$ satisfying
\begin{equation}\label{Seq-Bbasis}
R_{\hat{\bm{\tau}}}|\lambda,\mathbf{Q}_\alpha,n,\alpha \rangle=[\lambda-\ell^2\hat{\bm{\tau}}\cdot(\hat{\mathbf{z}}\times \mathbf{Q}_\alpha)]|\lambda,\mathbf{Q}_\alpha,n,\alpha\rangle\ , \qquad a_{\mathbf{Q}_\alpha}^\dag a_{\mathbf{Q}_\alpha}|\lambda,\mathbf{Q}_\alpha,n,\alpha \rangle= n|\lambda,\mathbf{Q}_\alpha,n,\alpha \rangle\ ,
\end{equation}
where $n\ge0$ is a nonnegative integer (Landau level number), $\lambda$ is a real number, and $\alpha$ is the intrinsic orbital index. Note that we have defined the eigenvalue of $R_{\hat{\bm{\tau}}}$ in a $\mathbf{Q}_\alpha$ dependent way (we could have defined the $R_{\hat{\bm{\tau}}}$ eigenvalue without $\ell^2\hat{\bm{\tau}}\cdot(\hat{\mathbf{z}}\times \mathbf{Q}_\alpha)$, which would be equivalent to the basis used in usual method \cite{bistritzer2011a}). One advantage of such a definition is that, the operator $e^{i\mathbf{q}_j^{\alpha\beta}\cdot \mathbf{r}}$ will not change the number $\lambda$ when acting on the basis $|\lambda,\mathbf{Q}_\alpha,n,\alpha \rangle$, as we will prove in Eq. (\ref{Seq-Rq}). Another advantage of such a definition in Eq. (\ref{Seq-Bbasis}) is that, for a fixed number $\lambda$, the basis $|\lambda,\mathbf{Q}_\alpha,n,\alpha \rangle$ of all quantum numbers $\alpha$, $\mathbf{Q}_\alpha$ and $n$ are orthonormal. To see this, consider two momentum sites $\mathbf{Q}_\alpha,\mathbf{Q}_\alpha'\in\mathbf{p}_\alpha+\bm{g}_1\mathbb{Z}+\bm{g}_2\mathbb{Z}$ (as defined in Eq. (\ref{Seq-palpha})), and assume $\mathbf{Q}_\alpha-\mathbf{Q}_\alpha'=m_1\bm{g}_1+m_2\bm{g}_2$. If two states $|\lambda,\mathbf{Q}_\alpha,n,\alpha \rangle$ and $|\lambda,\mathbf{Q}_\alpha',n',\alpha \rangle$ have equal $R_{\hat{\bm{\tau}}}$ eigenvalues, namely,
\begin{equation}\label{Seq-irrational}
\lambda-\ell^2\hat{\bm{\tau}}\cdot(\hat{\mathbf{z}}\times \mathbf{Q}_\alpha)=\lambda-\ell^2\hat{\bm{\tau}}\cdot(\hat{\mathbf{z}}\times \mathbf{Q}_\alpha') \quad \rightarrow \quad m_1\hat{\bm{\tau}}\cdot(\hat{\mathbf{z}}\times\bm{g}_1)+m_2\hat{\bm{\tau}}\cdot(\hat{\mathbf{z}}\times\bm{g}_2)=0\ ,
\end{equation} 
we must have $m_1=m_2=0$, namely, $\mathbf{Q}_\alpha'=\mathbf{Q}_\alpha$, since $\frac{\hat{\bm{\tau}}\cdot(\hat{\mathbf{z}}\times\bm{g}_1)} {\hat{\bm{\tau}}\cdot(\hat{\mathbf{z}}\times\bm{g}_2)}$ is an irrational number. Vice versa, if $\mathbf{Q}_\alpha'\neq\mathbf{Q}_\alpha$, the two states $|\lambda,\mathbf{Q}_\alpha,n,\alpha \rangle$ and $|\lambda,\mathbf{Q}_\alpha',n',\alpha \rangle$ will have different $R_{\hat{\bm{\tau}}}$ eigenvalues and will be orthogonal to each other. Therefore, the basis with a fixed number $\lambda$ satisfies the orthonormal relation:
\begin{equation}\label{Seq-lambda-ortho}
\langle\lambda,\mathbf{Q}_\beta',n',\beta|\lambda,\mathbf{Q}_\alpha,n,\alpha\rangle =\delta_{\beta\alpha}  \delta_{\mathbf{Q}_\beta'\mathbf{Q}_\alpha}\delta_{n'n}\ .
\end{equation}
We note that Eq. (\ref{Seq-lambda-ortho}) would not hold if we define $R_{\hat{\bm{\tau}}}|\lambda,\mathbf{Q}_\alpha,n,\alpha \rangle=\lambda|\lambda,\mathbf{Q}_\alpha,n,\alpha\rangle$ instead, as two different $\mathbf{Q}$'s would have the same eigenvalue $\lambda$.

As we will show later below Eq. (\ref{Seq-Rq}), the way of defining the basis in Eq. (\ref{Seq-Bbasis}) will greatly simplify the matrix elements of Hamiltonian (\ref{Seq-CmodelB}). However, before we move on, we note that the set of basis with a fixed number $\lambda$ (satisfying Eq. (\ref{Seq-lambda-ortho})) is not a complete basis for the Hamiltonian (\ref{Seq-CmodelB}), since for a fixed $\lambda$, the $R_{\hat{\bm{\tau}}}$ eigenvalue $\lambda-\ell^2\hat{\bm{\tau}}\cdot(\hat{\mathbf{z}}\times \mathbf{Q}_\alpha)$ (of all momentum sites $\mathbf{Q}_\alpha$) does not run over the entire real number set $\mathbb{R}$ (the complete set of $R_{\hat{\bm{\tau}}}$ eigenvalues is $\mathbb{R}$, see the argument above Eq. (\ref{seq-Rshift})). Therefore, we need to allow $\lambda$ to run over a certain set to make the basis $|\lambda,\mathbf{Q}_\alpha,n,\alpha\rangle$ a complete basis. We now prove this can be done by assuming $\lambda$ runs over the following quotient set:
\begin{equation}\label{Seq-lam}
\lambda \in \Lambda_{\hat{\bm{\tau}}}=\mathbb{R}/[\ell^2\hat{\bm{\tau}}\cdot(\hat{\mathbf{z}}\times\bm{g}_1)\mathbb{Z} +\ell^2\hat{\bm{\tau}}\cdot(\hat{\mathbf{z}}\times\bm{g}_2)\mathbb{Z}]\ ,
\end{equation}
namely, each number $\lambda$ labels a coset $\{\lambda + \ell^2\hat{\bm{\tau}}\cdot(\hat{\mathbf{z}}\times\bm{g}_1)\mathbb{Z} +\ell^2\hat{\bm{\tau}}\cdot(\hat{\mathbf{z}}\times\bm{g}_2)\mathbb{Z}\}$ of the subgroup $\ell^2\hat{\bm{\tau}}\cdot(\hat{\mathbf{z}}\times\bm{g}_1)\mathbb{Z} +\ell^2\hat{\bm{\tau}}\cdot(\hat{\mathbf{z}}\times\bm{g}_2)\mathbb{Z}$ in the real number group $\mathbb{R}$. Two numbers $\lambda$ and $\lambda'$ label the same coset if they belongs to the same coset. For example, assume we have $\ell^2\hat{\bm{\tau}}\cdot(\hat{\mathbf{z}}\times\bm{g}_1)=1$ and $\ell^2\hat{\bm{\tau}}\cdot(\hat{\mathbf{z}}\times\bm{g}_2)=\sqrt{2}$, then the quantum number $\lambda=0$ will label the coset $\{\mathbb{Z}+\sqrt{2}\mathbb{Z}\}$. Accordingly, $\lambda=0$ is identical to $\lambda=1$ and $\lambda=\sqrt{2}$, etc. In contrast, $\lambda=0$ and $\lambda=\sqrt{3}$ are not identical, which belong to different cosets. For definiteness, we will pick a fixed element $\lambda$ in each coset to represent the coset, so each coset is represented by a definite number $\lambda$. For example, we can choose to use the definite number $\lambda=0$ (instead of $1$, $\sqrt{2}$, etc) to represent the coset $\{\mathbb{Z}+\sqrt{2}\mathbb{Z}\}$. In this way, we can represent each element (coset) in the set $\Lambda_{\hat{\bm{\tau}}}$ in Eq. (\ref{Seq-lam}) by a definite number $\lambda$.

With the set of representative numbers $\lambda$ given in Eq. (\ref{Seq-lam}), we now show that the basis $|\lambda,\mathbf{Q}_\alpha,n,\alpha\rangle$ forms a complete orthonormal basis. Note that for each $\lambda$ in Eq. (\ref{Seq-lam}), the $R_{\hat{\bm{\tau}}}$ eigenvalue $\lambda-\ell^2\hat{\bm{\tau}}\cdot(\hat{\mathbf{z}}\times \mathbf{Q}_\alpha)$ of all $\mathbf{Q}_\alpha$ runs over all the numbers in the coset $\{\lambda-\ell^2\hat{\bm{\tau}}\cdot(\hat{\mathbf{z}}\times\mathbf{p}_\alpha) + \ell^2\hat{\bm{\tau}}\cdot(\hat{\mathbf{z}}\times\bm{g}_1)\mathbb{Z} +\ell^2\hat{\bm{\tau}}\cdot(\hat{\mathbf{z}}\times\bm{g}_2)\mathbb{Z}\}$, where $\mathbf{p}_\alpha$ is the momentum origin shift of orbital $\alpha$ defined in Eq. (\ref{Seq-r-basis}), and recall that $\mathbf{Q}_\alpha=\mathbf{p}_\alpha+\bm{g}_1\mathbb{Z}+\bm{g}_2\mathbb{Z}$ as defined in Eq. (\ref{Seq-palpha}). Therefore, the $R_{\hat{\bm{\tau}}}$ eigenvalue of $|\lambda,\mathbf{Q}_\alpha,n,\alpha\rangle$ of all $\lambda$ in Eq. (\ref{Seq-lam}) and all $\mathbf{Q}_\alpha$ runs over all the real numbers $\mathbb{R}$ (since $\lambda$ is a continuous variable in the quotient set (\ref{Seq-lam})). In particular, two orbital $\alpha$ states $|\lambda,\mathbf{Q}_\alpha,n,\alpha\rangle$ and $|\lambda',\mathbf{Q}_\alpha',n',\alpha\rangle$ will have their $R_{\hat{\bm{\tau}}}$ eigenvalues in different cosets if $\lambda\neq\lambda'$ ($\lambda, \lambda'\in \Lambda_{\hat{\bm{\tau}}}$ in Eq. (\ref{Seq-lam}), note that we have chosen a unique definite number $\lambda$ in each coset to represent the coset in Eq. (\ref{Seq-lam})). Therefore, we have the orthonormal relation
\begin{equation}\label{Seq-ortho}
\langle\lambda',\mathbf{Q}_\beta',n',\beta|\lambda,\mathbf{Q}_\alpha,n,\alpha\rangle =\delta_{\beta\alpha} \delta_{\lambda\lambda'}  \delta_{\mathbf{Q}_\beta'\mathbf{Q}_\alpha}\delta_{n'n}\ .
\end{equation}
Thus, the basis $|\lambda,\mathbf{Q}_\alpha,n,\alpha\rangle$ with $\lambda$ defined in Eq. (\ref{Seq-lam}) forms a complete orthonormal basis of Hamiltonian (\ref{Seq-Bbasis}).

We now show that the most important advantage of the basis $|\lambda,\mathbf{Q}_\alpha,n,\alpha\rangle$ satisfying Eq. (\ref{Seq-ortho}) is, the Hamiltonian (\ref{Seq-CmodelB}) is diagonal in $\lambda$, and its matrix elements are independent of $\lambda$. To see this, we first note that from the commutation relations in Sec. (\ref{sec-algebra}), we have
\begin{equation}\label{Seq-Rq}
e^{-i\mathbf{q}\cdot\mathbf{r}}R_{\hat{\bm{\tau}}}e^{i\mathbf{q}\cdot\mathbf{r}}= R_{\hat{\bm{\tau}}}-\ell^2\hat{\bm{\tau}}\cdot(\hat{\mathbf{z}}\times \mathbf{q})\ ,\qquad e^{-i\mathbf{q}\cdot\mathbf{r}}a_{\mathbf{Q}_\alpha}e^{i\mathbf{q}\cdot\mathbf{r}}= a_{\mathbf{Q}_\alpha}+\frac{\ell}{\sqrt{2}}(q_x+iq_y)=a_{\mathbf{Q}_\alpha-\mathbf{q}}\ ,
\end{equation}
where $\mathbf{r}$ is the position operator, and $\mathbf{q}$ is an arbitrary momentum vector. Therefore, we have
\begin{equation}
R_{\hat{\bm{\tau}}}e^{i\mathbf{q}\cdot\mathbf{r}}|\lambda,\mathbf{Q}_\alpha,n,\alpha\rangle = e^{i\mathbf{q}\cdot\mathbf{r}}(R_{\hat{\bm{\tau}}}-\ell^2\hat{\bm{\tau}}\cdot(\hat{\mathbf{z}}\times \mathbf{q}))|\lambda,\mathbf{Q}_\alpha,n,\alpha\rangle =[\lambda-\ell^2\hat{\bm{\tau}}\cdot(\hat{\mathbf{z}}\times (\mathbf{Q}_\alpha+\mathbf{q}))] e^{i\mathbf{q}\cdot\mathbf{r}} |\lambda,\mathbf{Q}_\alpha,n,\alpha\rangle 
\end{equation}
\begin{equation}
a_{\mathbf{Q}_\alpha+\mathbf{q}}^\dag a_{\mathbf{Q}_\alpha+\mathbf{q}} e^{i\mathbf{q}\cdot\mathbf{r}}|\lambda,\mathbf{Q}_\alpha,n,\alpha\rangle = e^{i\mathbf{q}\cdot\mathbf{r}}a_{\mathbf{Q}_\alpha}^\dag a_{\mathbf{Q}_\alpha}|\lambda,\mathbf{Q}_\alpha,n,\alpha\rangle =n e^{i\mathbf{q}\cdot\mathbf{r}} |\lambda,\mathbf{Q}_\alpha,n,\alpha\rangle\ ,
\end{equation}
which indicates
\begin{equation}\label{Seq-Rq}
e^{i\mathbf{q}\cdot\mathbf{r}}|\lambda,\mathbf{Q}_\alpha,n,\alpha\rangle=|\lambda,\mathbf{Q}_\alpha+\mathbf{q},n,\alpha\rangle\ ,
\end{equation}
where $\mathbf{r}$ is the position operator (understood as $\sum_\alpha\int \mathbf{r}|\mathbf{r},\alpha\rangle\langle\mathbf{r},\alpha|\text{d}^2\mathbf{r}$ when expanded in position basis), and it also plays the role of the generator of momentum translations. 

We then examine the matrix elements of Hamiltonian (\ref{Seq-CmodelB}). Using Eq. (\ref{Seq-Rq}), we find the lattice potential term $V^{\alpha\beta}_{j}e^{i\mathbf{q}_j^{\alpha\beta}\cdot\mathbf{r}}$ in the Hamiltonian (\ref{Seq-CmodelB}) has matrix elements
\begin{equation}\label{Seq-Vj}
\langle\lambda',\mathbf{Q}_\alpha',m,\alpha|V_{j}^{\alpha\beta}e^{i\mathbf{q}_j^{\alpha\beta}\cdot \mathbf{r}}|\lambda,\mathbf{Q}_\beta,n,\beta\rangle= V_{j}^{\alpha\beta}\langle\lambda',\mathbf{Q}_\alpha',m,\alpha |\lambda,\mathbf{Q}_\beta+\mathbf{q}_j^{\alpha\beta},n,\beta\rangle = V^{\alpha\beta}_j \delta_{\lambda'\lambda}\delta_{\mathbf{Q}_\alpha',\mathbf{Q}_\beta+\mathbf{q}_j^{\alpha\beta}}\delta_{mn}\ ,
\end{equation}
where $\mathbf{r}$ on the left hand side is understood as the position operator (instead of a number). Therefore, the lattice hopping potential term $V^{\alpha\beta}_{j}e^{i\mathbf{q}_j^{\alpha\beta}\cdot\mathbf{r}}$ is diagonal in quantum numbers $\lambda$ and $n$ when acting on basis $|\lambda,\mathbf{Q}_\alpha,n,\alpha\rangle$, while changes the orbital from $\beta$ to $\alpha$, and shifts the reciprocal momentum $\mathbf{Q}_\beta$ to $\mathbf{Q}_\beta+\mathbf{q}_j^{\alpha\beta}$.

As for the kinetic term $\epsilon(\mathbf{\Pi})$, we first note that the kinetic momentum $\bm{\Pi}$ commutes with $R_{\hat{\bm{\tau}}}$ and does not change the $R_{\hat{\bm{\tau}}}$ eigenvalue (which has one-to-one correspondence with the pair of quantum numbers $\lambda$ and $\mathbf{Q}_\alpha$). Therefore, the matrix elements of $\epsilon(\bm{\Pi})$ has to be diagonal in both $\lambda$ and $\mathbf{Q}_\alpha$. Besides, for each orbital $\alpha$, we can rewrite the kinetic momentum as $\bm{\Pi}=\hat{\bm{\kappa}}_{\mathbf{Q}_\alpha}+\mathbf{k}_0+\mathbf{Q}_\alpha$, where we have defined a $\mathbf{Q}_\alpha$ dependent operator 
\begin{equation}
\hat{\bm{\kappa}}_{\mathbf{Q}_\alpha}=\frac{1}{{\sqrt{2}\ell}}(a_{\mathbf{Q}_\alpha}+a^\dag_{\mathbf{Q}_\alpha}, -ia_{\mathbf{Q}_\alpha}+ia^\dag_{\mathbf{Q}_\alpha})\ .
\end{equation}
Note that $\hat{\bm{\kappa}}_{\mathbf{Q}_\alpha}$ only acts on the quantum number $n$ of basis $|\lambda,\mathbf{Q}_\alpha,n,\alpha\rangle$, which obeys the following rules:
\begin{equation}
a_{\mathbf{Q}_\alpha}|\lambda,\mathbf{Q}_\alpha,n,\alpha\rangle=\sqrt{n}|\lambda,\mathbf{Q}_\alpha,n-1,\alpha\rangle\ ,\qquad  a^\dag_{\mathbf{Q}_\alpha}|\lambda,\mathbf{Q}_\alpha,n,\alpha\rangle=\sqrt{n+1}|\lambda,\mathbf{Q}_\alpha,n+1,\alpha\rangle\ .
\end{equation}
Therefore, it is easy to find that
\begin{equation}\label{Seq-K1}
\begin{split}
\langle\lambda',\mathbf{Q}_\beta',m,\beta|\epsilon(\bm{\Pi})|\lambda,\mathbf{Q}_\alpha,n,\alpha\rangle =& \delta_{\lambda'\lambda}\delta_{\mathbf{Q}_\beta',\mathbf{Q}_\alpha} \langle\lambda',\mathbf{Q}_\beta',m,\beta| \epsilon(\hat{\bm{\kappa}}_{\mathbf{Q}_\alpha}+\mathbf{k}_0+\mathbf{Q}_\alpha)|\lambda,\mathbf{Q}_\alpha,n,\alpha\rangle \\
=&\delta_{\lambda'\lambda}\delta_{\mathbf{Q}_\beta',\mathbf{Q}_\alpha}  [\epsilon^{\beta\alpha}(\hat{\bm{\kappa}}_{\mathbf{Q}_\alpha}+\mathbf{k}_0+\mathbf{Q}_\alpha)]_{mn}\ .
\end{split}
\end{equation}
Mathematically, we can define an operator 
\begin{equation}\label{Seq-kappa}
\hat{\bm{\kappa}}= \frac{1}{{\sqrt{2}\ell}}(a+a^\dag, -ia+ia^\dag)
\end{equation}
without the $\mathbf{Q}_\alpha$ subindex, where $a$ and $a^\dag$ are some lowering and raising operators satisfying $[a,a^\dag]=1$ (which need not have any relation with $a_{\mathbf{Q}_\alpha}$ and $a_{\mathbf{Q}_\alpha}^\dag$). It is then easy to see that the matrix element in Eq. (\ref{Seq-K1}) is mathematically equal to
\begin{equation}\label{Seq-K2}
[\epsilon^{\beta\alpha}(\hat{\bm{\kappa}}_{\mathbf{Q}_\alpha}+\mathbf{k}_0+\mathbf{Q}_\alpha)]_{mn}=[\epsilon^{\beta\alpha}(\hat{\bm{\kappa}}+\mathbf{k}_0+\mathbf{Q}_\alpha)]_{mn} =\langle m|\epsilon^{\beta\alpha}(\hat{\bm{\kappa}}+\mathbf{k}_0+\mathbf{Q}_\alpha) |n\rangle\ ,
\end{equation}
where $|n\rangle$ is the basis of $a$ and $a^\dag$ defined by $a|n\rangle=\sqrt{n}|n-1\rangle$ and $a^\dag|n\rangle=\sqrt{n+1}|n+1\rangle$. Therefore, mathematically one could replace $\hat{\bm{\kappa}}_{\mathbf{Q}_\alpha}$ by $\hat{\bm{\kappa}}$ without ambiguity, provided that one remembers that $\hat{\bm{\kappa}}$ acts on the quantum number $n$.

From Eqs. (\ref{Seq-Vj}), (\ref{Seq-K1}) and (\ref{Seq-K2}), it is easy to see that the matrix elements of both the kinetic term and the lattice potential term of Hamiltonian (\ref{Seq-CmodelB}) are diagonal in $\lambda$ and independent of $\lambda$. Therefore, we can divide the entire Hilbert space into subspaces with different (continuous) quantum number $\lambda$; the energy spectra of all the $\lambda$ subspaces are the same. Within the subspace of a fixed $\lambda$, the Hamiltonian matrix element from basis $|\lambda,\mathbf{Q}_\beta,n,\beta\rangle$ to $|\lambda,\mathbf{Q}_\alpha',m,\alpha\rangle$ can be written as
\begin{equation}\label{Seq-HLQQ}
H^{\lambda,\alpha\beta}_{\mathbf{Q}_\alpha'\mathbf{Q}_\beta,mn}=\left[\epsilon^{\alpha\beta}(\hat{\bm{\kappa}}+\mathbf{k}_0+\mathbf{Q}_\beta)\right]_{mn} \delta_{\mathbf{Q}'_\alpha,\mathbf{Q}_\beta}+  \sum_{j} V^{\alpha\beta}_{j}\delta_{\mathbf{Q}_\alpha',\mathbf{Q}_\beta+\mathbf{q}_j^{\alpha\beta}}\delta_{mn}\ ,
\end{equation}
where $[\epsilon^{\alpha\beta}(\hat{\bm{\kappa}}+\mathbf{k}_0+\mathbf{Q}_\beta)]_{mn}$ is defined in Eq. (\ref{Seq-K2}). This is nothing but the zero magnetic field momentum space Hamiltonian (\ref{Seq-HQQ}) with the substitution $\mathbf{k}\rightarrow \hat{\bm{\kappa}}+\mathbf{k}_0$ (given also in Eq. (5) of the main text). It is then sufficient to compute the spectrum within just one fixed $\lambda$ subspace.

We note that different $\lambda$ sectors have different eigenstate wave functions, although they have identical Hamiltonian matrix elements independent of $\lambda$ as given by Eq. (\ref{Seq-HLQQ}). To see this explicitly, if we take a Landau gauge perpendicular to the $\hat{\bm{\tau}}$ direction, $\mathbf{A}=B(\mathbf{r}\cdot\hat{\bm{\tau}})(\hat{\mathbf{z}}\times\hat{\bm{\tau}})$, we have
\begin{equation}
\langle \mathbf{r},\alpha |\lambda,\mathbf{Q}_\alpha,n,\alpha\rangle =e^{i[ \lambda-\ell^2\hat{\bm{\tau}}\cdot(\hat{\mathbf{z}}\times \mathbf{Q}_\alpha)][\mathbf{r}\cdot(\hat{\mathbf{z}}\times\hat{\bm{\tau}})]/\ell -[\mathbf{r}\cdot\hat{\bm{\tau}}- \lambda+\ell^2\hat{\bm{\tau}}\cdot(\hat{\mathbf{z}}\times \mathbf{Q}_\alpha)]^2/2\ell^2}h_n\Big(\ell^{-1}[\mathbf{r}\cdot\hat{\bm{\tau}}- \lambda+\ell^2\hat{\bm{\tau}}\cdot(\hat{\mathbf{z}}\times \mathbf{Q}_\alpha)]\Big),
\end{equation}
where $h_n(x)$ is the $n$-th Hermite polynomial. Therefore, one can see explicitly that the wave functions $|\lambda,\mathbf{Q}_\alpha,n,\alpha\rangle$ at different $\lambda$ have different guiding centers. 
Assume the subspace Hamiltonian $H^{\lambda,\alpha\beta}_{\mathbf{Q}_\alpha'\mathbf{Q}_\beta,mn}$ in sector $\lambda$ has an eigenstate $|\lambda,u\rangle=\sum_{\alpha,\mathbf{Q}_\alpha,n}u_{\alpha,\mathbf{Q}_\alpha,n}|\lambda,\mathbf{Q}_\alpha,n,\alpha\rangle$ at energy energy $E_{u}$, where the coefficients $u_{\alpha,\mathbf{Q}_\alpha,n}$ are independent of $\lambda$. Then the subspace Hamiltonian $H^{\lambda',\alpha\beta}_{\mathbf{Q}_\alpha'\mathbf{Q}_\beta,mn}$ in sector $\lambda'$ will have an eigenstate $|\lambda',u\rangle=\sum_{\alpha,\mathbf{Q}_\alpha,n}u_{\alpha,\mathbf{Q}_\alpha,n}|\lambda',\mathbf{Q}_\alpha,n,\alpha\rangle$ at the same energy $E_{u}$. However, the $\hat{\bm{\tau}}$ direction guiding center coordinate $R_{\hat{\bm{\tau}}}$ of the two wave functions $|\lambda,u\rangle$ and $|\lambda',u\rangle$ will differ by $\lambda'-\lambda$, namely, the central position of the two wave functions are different. 

We also note that, if there is a disorder potential that breaks the periodicity of the lattice model, e.g., a potential term $\delta \widetilde{V}^{\alpha\beta}e^{i\widetilde{\mathbf{q}}\cdot\mathbf{r}}$ from orbitals $\beta$ to $\alpha$ with $\widetilde{\mathbf{q}}\neq \mathbf{Q}+\mathbf{p}_\alpha-\mathbf{p}_\beta$ for any reciprocal vector $\mathbf{Q}$, this term will couple different $\lambda$ sectors, and the Hamiltonian will no longer be diagonal in $\lambda$ and independent of $\lambda$. By Eq. (\ref{Seq-Rq}) and Eq. (\ref{Seq-Bbasis}), such a term $\delta \widetilde{V}^{\alpha\beta}e^{i\widetilde{\mathbf{q}}\cdot\mathbf{r}}$ will couple the sector of coset labeled by $\lambda$ (defined in Eq. (\ref{Seq-lam})) with the sector of coset of $\lambda-\ell^2\hat{\bm{\tau}}\cdot(\hat{\mathbf{z}}\times [\widetilde{\mathbf{q}}+\mathbf{p}_\beta-\mathbf{p}_\alpha)]$, which does not live in the same coset. In this paper, we shall not consider any disorder potential breaking the periodicity of the lattice model.

In particular, if the kinetic energy $\epsilon(\mathbf{k})$ is a polynomial of $\mathbf{k}$ up to power $\Delta$ ($\Delta\in\mathbb{Z}_+$), the matrix element in Eq. (\ref{Seq-HLQQ}) has to be zero for $|m-n|>\Delta$. In this case, the Hamiltonian (\ref{Seq-HLQQ}) is sparse.

\subsection{Numerical Hofstadter calculations: the cutoffs and the momentum space boundary}\label{Sec-numerical-cutoff}

In numerical calculations, one needs to take a cutoff in the reciprocal lattice at a boundary enclosing an area of $N_Q$ BZs, and a cutoff in the LL quantum number $n\le N_L$. The Hamiltonian (\ref{Seq-HLQQ}) is then a matrix of size $MN_L N_Q$. We now explain how the cutoffs $N_Q$ and $N_L$ set a momentum space boundary for Hamiltonian (\ref{Seq-HLQQ}).

For concreteness, assume the cutoff of the reciprocal lattice encloses a circular momentum space area $N_Q\Omega_{BZ}$ centered at the center momentum $\mathbf{k}_0$, where $\Omega_{BZ}=|\bm{g}_1\times\bm{g}_2|=4\pi^2/|\bm{d}_1\times\bm{d}_2|$ is the BZ area. This restricts the reciprocal momentum sites within a momentum radius 
\begin{equation}
|\mathbf{k}_0+\mathbf{Q}_\alpha|\le \sqrt{N_Q\Omega_{BZ}/\pi}\ . 
\end{equation}
In addition, by Eq. (\ref{Seq-kappa}), with the LL cutoff $N_L$, for any states we have
\begin{equation}
|\hat{\bm{\kappa}}|^2=\langle \hat{\bm{\kappa}}^2\rangle=\frac{1}{\ell^2}\langle(2a^\dag a+1) \rangle \le \frac{2N_L+1}{\ell^2}\approx \frac{2N_L}{\ell^2}\ ,
\end{equation}
where we have defined $|\hat{\bm{\kappa}}|=\sqrt{\langle \hat{\bm{\kappa}}^2\rangle}$ as the norm of the operator $\hat{\bm{\kappa}}$. Therefore, $\hat{\bm{\kappa}}$ is restricted within a radius $| \hat{\bm{\kappa}} |\le \sqrt{2N_L}/\ell$.

The Hamiltonian (\ref{Seq-HLQQ}) can be viewed as a lattice model in the momentum space with ``hoppings" $V_{j}$ between nearby reciprocal sites, and an ``on-site potential energy" $\epsilon(\hat{\bm{\kappa}}+\mathbf{k}_0+\mathbf{Q}_\alpha)$ on each site $\mathbf{Q}_\alpha$, and $\hat{\bm{\kappa}}$ plays the role of the ``position" operator in the momentum space (although its $x$ and $y$ components are noncommuting). The momentum space probability (norm square of amplitude) of the basis wave function $|\lambda,\mathbf{Q}_\alpha,n,\alpha\rangle$ is concentrated circularly near a ring of radius $\sqrt{\langle \hat{\bm{\kappa}}^2\rangle}\approx \sqrt{2n}/\ell$.

The radius cutoff of $\mathbf{k}_0+\mathbf{Q}_\alpha$ and the radius cutoff of $\hat{\bm{\kappa}}$ become equal when
\begin{equation}
\sqrt{2N_L}/\ell=\sqrt{N_Q\Omega_{BZ}/\pi}\ ,\qquad \rightarrow \qquad \frac{\varphi}{2\pi}=\frac{B|\bm{d}_1\times\bm{d}_2|}{2\pi}=\frac{2\pi}{\ell^2\Omega_{BZ}}=\frac{N_Q}{N_L}\ ,
\end{equation}
where $\varphi=B|\bm{d}_1\times\bm{d}_2|$ is the magnetic flux per unit cell, and we have used the Brillouin zone area $\Omega_{BZ}=4\pi^2/|\bm{d}_1\times\bm{d}_2|$ and $B=1/\ell^2$. Given $N_Q$, $N_L$ which we pick in our calculation, this flux $\varphi/2\pi=N_Q/N_L$ then separates the system into a small $B$ regime and a large $B$ regime as follows.

If $\sqrt{2N_L}/\ell<\sqrt{N_Q\Omega_{BZ}/\pi}$, or equivalently $\varphi/2\pi<N_Q/N_L$ (the small $B$ field regime), the momentum space ``position" $\hat{\bm{\kappa}}$ has a hard cutoff $|\hat{\bm{\kappa}}|\le \sqrt{2N_L}/\ell$ which serves as the momentum space boundary.

On the contrary, if $\sqrt{2N_L}/\ell>\sqrt{N_Q\Omega_{BZ}/\pi}$, or equivalently $\varphi/2\pi>N_Q/N_L$ (the large $B$ field regime), we have the expectation value of $|\hat{\bm{\kappa}}+\mathbf{k}_0+\mathbf{Q}_\alpha|>\sqrt{2N_L}/\ell-\sqrt{N_Q\Omega_{BZ}/\pi}$ for all sites $\mathbf{Q}_\alpha$ if a state has expectation value $|\hat{\bm{\kappa}}|>\sqrt{N_Q\Omega_{BZ}/\pi}$. In general, the kinetic energy, or ``on-site potential energy" in the momentum space $\epsilon(\hat{\bm{\kappa}}+\mathbf{k}_0+\mathbf{Q}_\alpha)$, is an increasing function of  $|\hat{\bm{\kappa}}+\mathbf{k}_0+\mathbf{Q}_\alpha|$. If an eigenstate has a large expectation value of $|\hat{\bm{\kappa}}+\mathbf{k}_0+\mathbf{Q}_\alpha|$ for any $Q_\alpha$ (because it has expectation value $|\hat{\bm{\kappa}}|>\sqrt{N_Q\Omega_{BZ}/\pi}$), it will also have a large expectation value of kinetic energy $\epsilon(\hat{\bm{\kappa}}+\mathbf{k}_0+\mathbf{Q}_\alpha)$ for any $\mathbf{Q}_\alpha$, thus its eigenenergy is expected to be large, and
cannot be a reliable eigenstate of the low energy Hofstadter bulk bands. 

Therefore, from the reasoning the above, we can define a momentum boundary radius given by the cutoffs $N_L$ and $N_Q$ as
\begin{equation}\label{Seq-kappa-b}
\kappa_b\approx\min\{\frac{\sqrt{2N_L}}{\ell}, \sqrt{\frac{N_Q\Omega_{BZ}}{\pi}}\}\ ,
\end{equation}
and a trustable low-energy eigenstate in the Hofstadter bulk bands should have its expectation value $|\hat{\bm{\kappa}}|<\kappa_b$. Any state with expectation value $|\hat{\bm{\kappa}}|\gtrsim \kappa_b$ are effectively localized on the momentum space boundary at radius $\kappa_b$ (if we view $\hat{\bm{\kappa}}$ as a momentum space coordinate), which we will call the \emph{momentum space edge states}. Accordingly, we call the states with $|\hat{\bm{\kappa}}|<\kappa_b$ the \emph{momentum space bulk states}. The momentum space bulk states give the Hofstadter bulk band spectra we want to calculate.

As we discussed in the main text (below main text Eq. (10)), the momentum space edge states can be identified by a boundary projection operator $P_{\kappa_b,w}$, the matrix elements of which are defined by
\begin{equation}\label{Seq-Pb}
\langle\lambda',\mathbf{Q}_\alpha',m,\beta| P_{\kappa_b,w} |\lambda,\mathbf{Q}_\beta,n,\alpha\rangle=\Theta\Big(n-(\kappa_b-w)^2\ell^2/2\Big) \delta_{\lambda'\lambda}\delta_{mn} \delta_{\alpha\beta}\delta_{\mathbf{Q}_\alpha',\mathbf{Q}_\beta}\ ,
\end{equation}
where $\Theta(x)$ is the Heaviside unit step function, $\kappa_b\approx\min\{\frac{\sqrt{2N_L}}{\ell}, \sqrt{\frac{N_Q\Omega_{BZ}}{\pi}}\}$ as defined in Eq. (\ref{Seq-kappa-b}), and $w>0$ is a parameter one can vary representing the defining width of the edge states. An eigenstate with large expectation value $\langle P_{\kappa_b,w}\rangle$ will be mainly concentrated at $|\hat{\bm{\kappa}}|>\kappa_b-w$ in the momentum space, namely, within distance $w$ inside the boundary radius $\kappa_b$, thus are momentum space edge states. These edge eigenstates with large $\langle P_{\kappa_b,w}\rangle>P_c$ above a certain chosen threshold $P_c\in[0,1]$ can then be deleted in the energy spectrum, so that only the bulk spectrum Hofstadter butterfly is kept (see main text Fig. 2(b), which is calculated with cutoffs $N_Q=37$ and $N_L=60$). In practice, one may properly adjust $w$ and $\langle P_{\kappa_b,w}\rangle$ for different magnetic fields $B$ and different energy range to reach a cleaner Hofstadter butterfly. Generically, the momentum space edge states in a smaller (larger) Hofstadter gap will be less (more) localized at the momentum boundary, thus requires a larger (smaller) edge width $w$ and a smaller (larger) projection threshold $P_c$. In calculating main text Fig. 2(b), we have chosen $w=\min\{\ell^{-1},1.6\sqrt{\Omega_{BZ}}\}$, and the edge projection threshold $P_c=0.5$. Here the factor $1.6$ in front of $\sqrt{\Omega_{BZ}}$ is numerically tested to be a good choice for eliminating most edge states in the main Hofstadter gaps of the TBG model calculated here. Generically the optimal order 1 factor in front of $\sqrt{\Omega_{BZ}}$ depends on the models calculated (and can even be chosen to be dependent on the magnetic field $B$ and the energy of the eigenstate). Generically, we suggest to choose $w$ to be around the order of the smaller one of $\ell^{-1}$ and $\sqrt{\Omega_{BZ}}$.

Besides, we note that the Hofstadter butterfly with edge states deleted may have remaining edge state ``hairs" near the edges of the Hofstadter gaps (i.e., not a clean Hofstadter butterfly), as shown in Fig. 2(b). This is because the momentum space edge states approaching the Hofstadter gap edges are more and more delocalized from the momentum space boundary, thus some of such edge states cannot satisfy the criteria $\langle P_{\kappa_b,w}\rangle>P_c$ and thus cannot be deleted. By making the momentum boundary radius $\kappa_b$ larger (which is computationally more expensive) and choose a larger edge width $w$, one can reduce such edge state ``hairs" and improve the clearness of the Hofstadter butterfly. 

\subsection{Number of states per Brillouin zone in a fixed $\lambda$ sector}\label{Sec-lambda-states}
Here we discuss the number of occupied states per Brillouin zone (i.e., per reciprocal lattice ``unit cell") $\rho_K$ of the Hamiltonian $H^{\lambda,\alpha\beta}_{\mathbf{Q}_\alpha'\mathbf{Q}_\beta,mn}$ in a fixed $\lambda$ sector (coset) given a Fermi energy, which we used in the main text Eq. (8).

Consider a magnetic field $B$ corresponding to flux per unit cell $\varphi=B\Omega=\ell^{-2}\Omega$, where $\Omega=|\bm{d}_1\times \bm{d}_2|$ is the zero-magnetic-field unit cell area. Assume the Fermi energy is $\epsilon_F$, and the number of occupied states (below the Fermi energy $\epsilon_F$) per zero-magnetic-field unit cell area $\Omega$ in the real space is $\rho$. 

To find out the number of occupied states in each $\lambda$ sector, we first count how many $\lambda$ sectors there are. In an infinite real space and with an infinite reciprocal lattice $\mathbf{Q}$ (i.e., without reciprocal lattice cutoff), the set of $\lambda$ in Eq. (\ref{Seq-lam}) is an infinite set, the size of which cannot be perceived easily. To make the set of $\lambda$ finite, we assume the system has a large but finite real space area $\Omega_{\text{tot}}=N_{\Omega}\Omega$ (where $N_\Omega$ is the number of zero-magnetic-field unit cells), and take a finite reciprocal lattice number cutoff $N_Q$ (as we did in the numerical calculation described in Sec. \ref{Sec-numerical-cutoff}), but we keep the Landau level number cutoff $N_L\rightarrow\infty$. By Eq. (\ref{Seq-Bbasis}), for each orbital $\alpha$, each $\lambda$ sector consists of the sub-Hilbert spaces with $R_{\hat{\bm{\tau}}}$ eigenvalues $\lambda-\ell^2\hat{\bm{\tau}}\cdot(\hat{\mathbf{z}}\times \mathbf{Q}_\alpha)$ with $\mathbf{Q}_\alpha=\mathbf{Q}+\mathbf{p}_\alpha$ (defined in Eq. (\ref{Seq-palpha})) running over all $\mathbf{Q}$ within the reciprocal lattice cutoff $N_Q$. This means each $\lambda$ sector consists of the sub-Hilbert spaces of $N_Q$ different (discrete) eigenvalues of $R_{\hat{\bm{\tau}}}$. On the other hand, for each orbital $\alpha$, the total number of (discrete) $R_{\hat{\bm{\tau}}}$ eigenvalues in the system is equal to the degeneracy of a single Landau level in the free space, which is
\begin{equation}
N_R=\frac{\Omega_{\text{tot}}}{2\pi\ell^2}\ .
\end{equation}
This is because in the free space, the Landau level states of a definite LL band can be completely labeled by the eigenvalue of the guiding center along a certain direction (here chosen to be $R_{\hat{\bm{\tau}}}$). Since each $\lambda$ sector allows $N_Q$ different eigenvalues of $R_{\hat{\bm{\tau}}}$, and different $\lambda$ sectors are orthogonal to each other in the Hilbert space, we conclude the number of $\lambda$ sectors (cosets) in our method are related by
\begin{equation}
N_\lambda=\frac{N_R}{N_Q}=\frac{\Omega_{\text{tot}}}{2\pi\ell^2 N_Q}\ .
\end{equation}

Meanwhile, since there are $\rho$ occupied states per zero-magnetic-field unit cell area $\Omega=|\bm{d}_1\times \bm{d}_2|$, the total number of occupied states is given by
\begin{equation}
N_{\text{occ}}=\rho\frac{\Omega_{\text{tot}}}{\Omega}\ .
\end{equation}
Since the number of $R_{\hat{\bm{\tau}}}$ eigenvalues is $N_R$, we can define the number of occupied states in each $R_{\hat{\bm{\tau}}}$ eigenvalue sector as
\begin{equation}
\rho_K=\frac{N_{\text{occ}}}{N_R}=\rho\frac{2\pi\ell^2}{\Omega}=\frac{2\pi}{\varphi}\rho\ .
\end{equation}
Note that $\rho_K$ is nothing but the LL filling fraction. On the other hand, the occupied states should evenly belong to each $\lambda$ sector, since different $\lambda$ sectors have the same spectrum. Therefore, we conclude the number of states occupied in each $\lambda$ sector (in the $N_L\rightarrow\infty$ and finite $N_Q$ case considered here, i.e., $N_Q/N_L\rightarrow0$) is
\begin{equation}\label{seq-NQfinite}
\mathcal{N}_{K}^{(\lambda)}=\frac{N_{\text{occ}}}{N_{\lambda}}=\frac{N_{\text{occ}}}{N_R}N_Q=\frac{2\pi}{\varphi}\rho N_Q=\rho_KN_Q\ ,
\end{equation}
which is independent of $\lambda$. Since $N_Q$ is the reciprocal lattice cutoff, or the number of Brillouin zones (BZs) we keep in the reciprocal lattice (note that we assumed $N_L\rightarrow\infty$ which does not give a momentum space cutoff), we see that $\rho_K=(2\pi/\varphi)\rho$ can be understood as the number of occupied states per BZ in a fixed $\lambda$ sector. 

At rational flux $\varphi=2\pi p/q$, according to the Diophantine equation (\ref{Seq-rho}), we have $\rho=\nu/q$ where $\nu$ is the number of occupied magnetic bands in the magnetic unit cell, and thus we find the number of occupied states per BZ (with $\lambda$ fixed) is $\rho_K=(2\pi/\varphi)\rho=\nu/p$. Accordingly, the Diophantine equation (\ref{Seq-rho}) can be rewritten as 
\begin{equation}
t_\nu+s_\nu\frac{2\pi}{\varphi}=\rho_K\ ,
\end{equation}
i.e., main text Eq. (8).

Since the total number of occupied states $\mathcal{N}_{K}^{(\lambda)}=N_Q\rho_K$ in a $\lambda$ sector is independent of $\lambda$, hereafter we simply denote it as $\mathcal{N}_{K}$, as appears in the main text Eqs. (9) and (10).


\subsection{Numerical determination of number of occupied states}\label{Seq-NK}
This subsection discusses how to determine the number of occupied states $\mathcal{N}_K$ in a gap in numerical calculations. We note that $\mathcal{N}_K$ here is counted below the mid-gap energy of a gap, which may disperse as a function of magnetic field $B$ (e.g., the inclined dashed lines in the main text Fig. 2(c)-(d)). Generically, $\mathcal{N}_K$ counted in this way will slightly depend on where the mid-gap energy is chosen, which determines how many in-gap edge states are included in $\mathcal{N}_K$ in addition to the bulk states. However, the relative error in counting $\mathcal{N}_K$ will tend to zero as $N_L$ and $N_Q$ increase, since the ratio between the number of in-gap edge states and the number of bulk band states will tend to zero when the momentum space area increases.

In the main text, we have shown that in the calculation with LL cutoff $N_L$ and reciprocal cutoff $N_Q$ (in a fixed $\lambda$ sector), there are two regimes: 

i) The regime $\varphi/2\pi<N_Q/N_L$, for which the momentum space boundary is at radius $\kappa_b=\sqrt{2N_L}/\ell$, and the momentum space bulk area is $\mathcal{A}_K=2\pi N_L/\ell^2$, so the number of occupied states $\mathcal{N}_K=\rho_K\mathcal{A}_K/\Omega_{BZ}$ in the fixed $\lambda$ sector (note that this is different from Eq. (\ref{seq-NQfinite}) where $\varphi/2\pi>N_Q/N_L$) satisfies Eq. (9) in the main text, namely,
\begin{equation}\label{Seq-Nks}
\mathcal{N}_K=N_L(t_\nu\varphi/2\pi+s_\nu)\ ,
\end{equation}
where $t_\nu$ is the Chern number of the gap, and $s_\nu$ is another integer which we show in Sec. \ref{sec-LS} can be understood as a dual Chern number for the momentum space. Accordingly, the in-gap spectral flow rate in a gap is given by 
\begin{equation}\label{Seq-Nks1}
\frac{\text{d}\mathcal{N}_K}{\text{d}(\varphi/2\pi)}=N_Lt_\nu, 
\end{equation}
which allows us to determine $t_\nu$ of a gap in this regime by counting the number of states flowing across the midgap energy per flux number $\varphi/2\pi$, as described in the example given below main text Eq. (9) (shown in main text Fig. 2c). Further, if the number of occupied states $\mathcal{N}_K$ in the gap at some flux $\varphi$ is known (counted relative to some reference point, which will be discussed below), we could also derive $s_\nu$ of the gap from Eq. (\ref{Seq-Nks}).

ii) The regime $\varphi/2\pi>N_Q/N_L$, for which the momentum space boundary is at radius $\kappa_b=\sqrt{N_Q\Omega_{BZ}/\pi}$, and the momentum space bulk area is $\mathcal{A}_K=N_Q\Omega_{BZ}$, so the number of occupied states $\mathcal{N}_K=\rho_K\mathcal{A}_K/\Omega_{BZ}$ in the fixed $\lambda$ sector satisfies main text Eq. (10), namely,
\begin{equation}\label{Seq-Nkg}
\mathcal{N}_K=N_Q(t_\nu+2\pi s_\nu/\varphi)\ .
\end{equation}
Accordingly, the in-gap spectral flow rate in a gap is given by 
\begin{equation}\label{Seq-Nkg1}
\frac{\text{d}\mathcal{N}_K}{\text{d}(2\pi/\varphi)}=N_Qs_\nu, 
\end{equation}
which allows us to determine $s_\nu$ of a gap in this regime by counting the number of states (without deleting momentum space edge states by the projector method in Sec. \ref{Sec-numerical-cutoff}) flowing across the midgap energy per inverse flux number $2\pi/\varphi$, as described in the example given below main text Eq. (10) (shown in main text Fig. 2d). Further, if the number of occupied states $\mathcal{N}_K$ in the gap at some flux $\varphi$ is known, we could also derive $t_\nu$ of the gap from Eq. (\ref{Seq-Nkg}).

In either regime, finding out $\mathcal{N}_K$ at some $\varphi$ in the numerical results allows us to derive both $t_\nu$ and $s_\nu$ of a gap. In numerical calculations, however, the number of occupied states $\mathcal{N}_K$ below a gap should be counted relative to some reference energy level where $\mathcal{N}_K=0$ is defined. Therefore, we need to find out where the $\mathcal{N}_K=0$ energy level is defined, which depends on models. This is discussed below.

\subsubsection{Determination of $\mathcal{N}_K=0$}

A generic method to find out $\mathcal{N}_K=0$ is the following: first, find a gap which extends over both the small $\varphi$ regime $\varphi/2\pi<N_Q/N_L$ and the large $\varphi$ regime $\varphi/2\pi>N_Q/N_L$ (this is the condition for this generic method to work), which we call the reference gap. By extracting out the spectral flow rate in the small and large $\varphi$ regimes, one can derive $t_\nu$ and $s_\nu$ of the reference gap from Eqs. (\ref{Seq-Nks1}) and (\ref{Seq-Nkg1}), respectively. Then, $\mathcal{N}_K$ of this reference gap at any $\varphi$ can be determined from Eqs. (\ref{Seq-Nks}) and (\ref{Seq-Nkg}). Then, the number of occupied states $\mathcal{N}_K$ in any other gap $j$ at a given $\varphi$ can be determined in the numerical calculation by counting the number of states from the midgap energy of the reference gap to the midgap energy of gap $j$, plus the number of occupied states in the reference gap which is known. From the reference gap, one could determine which energy level corresponds to $\mathcal{N}_K=0$.

As an example, in the Hofstadter butterfly of the TBG model in the main text Fig. 2(a) and 2(b) (see Sec. \ref{sec-tBLG} for details), we can take the $(1,0)$ gap (labeled in main text Fig. 2(b)) as such a reference gap. This Hofstadter spectrum is calculated by setting $N_Q=37$ and $N_L=60$. In the regime $\varphi/2\pi<N_Q/N_L$, the spectral flow rate across the mid-gap energy of gap $(1,0)$ can be counted along the dashed line in Fig. 2(c) to be $d\mathcal{N}_K/d(\varphi/2\pi)\approx 16/0.25=64=N_Lt_\nu$ ($16$ levels when $\varphi/2\pi$ increases from 0.25 to 0.5), so we find $t_\nu=1$ being the integer closest to $64/N_L$. Then in the regime $\varphi/2\pi>N_Q/N_L$, there are no levels flowing in the $(1,0)$ gap except for some horizontal lines, which are spurious Dirac zero modes of the TBG model and should not be counted in $\mathcal{N}_K$ (see Sec. \ref{Sec-spurious} for detailed explanation). Therefore, the spectral flow rate in this regime is $d\mathcal{N}_K/d(2\pi/\varphi)=0=N_Qs_\nu$, which leads to $s_\nu=0$. Thus the gap is labeled by quantum numbers $(t_\nu,s_\nu)=(1,0)$. Then, by Eq. (\ref{Seq-Nks}), we can find, for instance, $\mathcal{N}_K=N_L(1\times 0.5+0)=30$ at flux $\varphi/2\pi=0.5<N_Q/N_L$. Therefore, the $\mathcal{N}_K=0$ level can be found by counting $30$ levels downwards from the midgap energy of the $(1,0)$ gap at flux $\varphi/2\pi=0.5$, which turns out to be approximately the level at zero energy (see the main text Fig. 2(c)).

We further discuss the following two special cases, where the reference point $\mathcal{N}_K=0$ can be determined more easily:

i) Models with a kinetic energy bounded from below, e.g., a single-orbital model with a Hamiltonian $\widetilde{H}(\mathbf{r})=\epsilon(-i\nabla)+\sum_{j} V_{j}e^{i\mathbf{q}_j\cdot\mathbf{r}}$ with a quadratic kinetic energy $\epsilon(\mathbf{k})=k^2/2m_0\ge0$, where $m_0$ is the electron effective mass. In this case, one must have Chern number $t_\nu=0$ below the lowest energy band of the entire spectrum, i.e., when no states are occupied at all. By Eq. (\ref{Seq-Nkg}), one then finds $\mathcal{N}_K=0$ below the lowest energy band in the limit $\varphi\rightarrow\infty$. Since there are no states at lower energies, there should be no spectra flows below the lowest band with respect to $\varphi$, so one should have $\mathcal{N}_K=0$ below the lowest energy band at any flux $\varphi$.

ii) Models with a Dirac kinetic energy (which has no lower bound), such as the TBG model in Eq. (\ref{Seq-HQQTBG}) which has a kinetic term $\epsilon(\mathbf{k})=v_F\bm{\sigma}^*\cdot\mathbf{k}$ (where $\bm{\sigma}^*=(\sigma_x,-\sigma_y)$). In this case, if there is no LL cutoff and reciprocal lattice cutoff, the energy spectrum of the system does not have a lower bound. With a LL cutoff $N_L$ and a reciprocal lattice cutoff $N_Q$, the Hamiltonian size is $MN_LN_Q$ for $M$ intrinsic orbitals (each Dirac kinetic term $\epsilon(\mathbf{k})=v_F\bm{\sigma}^*\cdot\mathbf{k}$  has two intrinsic orbitals). In the $\varphi\rightarrow\infty$ limit, the energies of the eigenstates are dominated by the kinetic term $\epsilon(\mathbf{k})=v_F\bm{\sigma}^*\cdot\mathbf{k}$ (since $\mathbf{k}$ is replaced by $\hat{\bm{\kappa}}+\mathbf{k}_0$ and $\hat{\bm{\kappa}}\propto \ell^{-1}=\sqrt{B}$ which goes to infinity as $\varphi\rightarrow\infty$), which should give a (nearly) particle-hole symmetric spectrum because of the particle-hole symmetry of the Dirac kinetic term. Accordingly, all the spectral flows should be (nearly) particle-hole symmetric about the half filling, which fixes $\mathcal{N}_K=0$ at the half filling point. Therefore, in the limit $\varphi\rightarrow\infty$, one has $\mathcal{N}_K=0$ at the half filling of the Hamiltonian with cutoffs $N_L$ and $N_Q$, e.g., the filling between the $MN_LN_Q/2$-th level and the $(MN_LN_Q/2)+1$-th level (energetically sorted). Since the total number of levels $MN_LN_Q$ does not change with respect to $\varphi$ (for fixed $N_L$ and $N_Q$), and $\mathcal{N}_K=0$ is a reference filling independent of $\varphi$, we conclude that $\mathcal{N}_K=0$ is between the $MN_LN_Q/2$-th level and the $(MN_LN_Q/2)+1$-th level for any $\varphi$. In particular, for the TBG model in Eq. (\ref{Seq-HQQTBG}), the energy spectrum is particle-hole symmetric at all $\varphi$, so the half filling point $\mathcal{N}_K=0$ is at zero energy at any $\varphi$. We note that when further counting the number of occupied states $\mathcal{N}_K$ of certain gaps relative to this half filling $\mathcal{N}_K=0$ point, one needs to exclude the unphysical spurious modes due to LL cutoff $N_L$ as discussed in Sec. \ref{Sec-spurious}.


\section{The example of the TBG continuum model}\label{sec-tBLG}

In this section, we explain the Hofstadter spectrum calculation of the one-valley TBG continuum model in Ref. \cite{bistritzer2011}, the results of which are given in the main text Fig. 2.

\subsection{Description of the model}

The model consists of the Dirac electrons of the same valley of two graphene layers, which are relatively twisted by angle $\theta$. Besides, we only consider one spin, namely, the model is spinless. The model can be written in real space as a $4\times4$ matrix
\begin{equation}\label{Seq-HtBLG}
H_{\text{TBG}}=\left(
\begin{array}{cc}
-iv_F\bm{\sigma}^*\cdot\nabla & \sum_{j=1}^3V_je^{i\mathbf{q}_j\cdot\mathbf{r}}\\
\sum_{j=1}^3V_j^\dag e^{-i\mathbf{q}_j\cdot\mathbf{r}} & -iv_F\bm{\sigma}^*\cdot\nabla \\
\end{array}
\right)\ ,
\end{equation}
where the upper (lower) two basis are the A and B sublattices of the upper (lower) monolayer graphene, $\bm{\sigma}^*=(\sigma_x,-\sigma_y)$ are the Pauli matrices acting on A and B sublattices of the monolayer graphene lattice, the momenta $\mathbf{q}_j$ are given by
\[
\mathbf{q}_1=k_\theta(0,-1)^T\ ,\qquad \mathbf{q}_2=k_\theta\left(\frac{\sqrt{3}}{2},\frac{1}{2}\right)^T\ ,\qquad \mathbf{q}_3=k_\theta\left(-\frac{\sqrt{3}}{2},\frac{1}{2}\right)^T\ ,
\]
and the interlayer hopping matrices
\[V_1=w_0\left(1_2+\sigma_x\right)\ ,\qquad V_2=w_0\left(1_2-\frac{1}{2}\sigma_x-\frac{\sqrt{3}}{2}\sigma_y\right)\ ,\qquad V_3=w_0\left(1_2-\frac{1}{2}\sigma_x+\frac{\sqrt{3}}{2}\sigma_y\right)\ ,\]
where $1_2$ stands for the $2\times2$ identity matrix. The parameters are given by $v_F\approx 610$meV$\cdot$nm, $w_0=110$meV, and $k_\theta=|\mathbf{q}_j|=(8\pi/3a_0)\sin(\theta/2)$, with the lattice constant $a_0=0.246$nm. In the example shown in main text Fig. 2, we take the twist angle $\theta=2.2^\circ$.

The reciprocal vectors of the TBG continuum model are given by $\bm{g}_1=\mathbf{q}_2-\mathbf{q}_3$ and $\bm{g}_2=\mathbf{q}_3-\mathbf{q}_1$. By Fourier transforming the zero field Hamiltonian (\ref{Seq-HtBLG}) into the momentum space, the Hamiltonian becomes a model in a honeycomb reciprocal lattice, where the orbitals of layer $1$ and layer $2$ are located at the two different sublattices $\mathbf{Q}_1\in\mathbf{q}_1+\bm{g}_1\mathbb{Z}+\bm{g}_2\mathbb{Z}$ and $\mathbf{Q}_2\in-\mathbf{q}_1+\bm{g}_1\mathbb{Z}+\bm{g}_2\mathbb{Z}$ of the honeycomb reciprocal lattice, respectively. Namely, the origin of the momentum in layer $\zeta=1,2$ is shifted by $(-1)^{\zeta-1}\mathbf{q}_1$, which is an example of choosing orbital-dependent momentum origins in Eq. (\ref{Seq-palpha}). Such a shift has the advantage of making the symmetries of the momentum space Hamiltonian more explicit, and thus is adopted in most literatures of TBG. 
Since the kinetic energy for sublattices $\{\mathbf{Q}_1\}$ and $\{\mathbf{Q}_2\}$ in Eq. (\ref{Seq-HtBLG}) are identical, we can use a single notation $\mathbf{Q}$ to denote both reciprocal sublattice sites $\{\mathbf{Q}_1,\mathbf{Q}_2\}$, i.e., the full honeycomb reciprocal lattice sites, and rewrite the Hamiltonian as
\begin{equation}\label{Seq-HQQTBG}
H_{\mathbf{Q}'\mathbf{Q}}(\mathbf{k})= v_F\bm{\sigma}^*\cdot(\mathbf{k}+\mathbf{Q}) \delta_{\mathbf{Q}'\mathbf{Q}}+  \sum_{j=1}^3 (V_{j}\delta_{\mathbf{Q}',\mathbf{Q}+\mathbf{q}_j}+h.c.)\ .
\end{equation}
One only needs to remember that the two different sublattices of the reciprocal lattice correspond to layers $1$ and $2$, respectively. The Hamiltonian under magnetic field in our basis is then given by the substitution $\mathbf{k}\rightarrow\hat{\bm{\kappa}}+\mathbf{k}_0$, with $\hat{\bm{\kappa}}= \frac{1}{{\sqrt{2}\ell}}(a+a^\dag, -ia+ia^\dag)$. In the calculation of main text Fig. 2, we set a twist angle $\theta=2.2^\circ$, take cutoffs $N_Q=37$ and $N_L=60$, and choose the central momentum $\mathbf{k}_0$ at the $\Gamma$ point of the first TBG BZ (A different choice of $\mathbf{k}_0$ only affects the spectrum at extremely small magnetic fluxes $|\varphi/2\pi|<1/N_L$, in which regime our calculation reduces to the LL calculation for the $\mathbf{k\cdot p}$ model expanded at momentum $\mathbf{k}_0$). The same spectrum with edge states present and with them deleted by the edge projection criteria of Eq. (\ref{Seq-Pb}) in a larger energy range is shown in Fig. \ref{FigS-TBG}(a)-(b). The edge states in Fig. \ref{FigS-TBG}(b) are deleted following the method described in Sec. \ref{Sec-numerical-cutoff}, where we used edge width $w=\min\{\ell^{-1},1.6\sqrt{\Omega_{BZ}}\}$ for the edge projector $P_{\kappa_b,w}$, and the edge projection threshold $P_c=0.5$. More examples of the TBG Hofstadter spectra calculated with our method can be found in Ref. \cite{lian2019b}.

\begin{figure}[tbp]
\begin{center}
\includegraphics[width=6.8in]{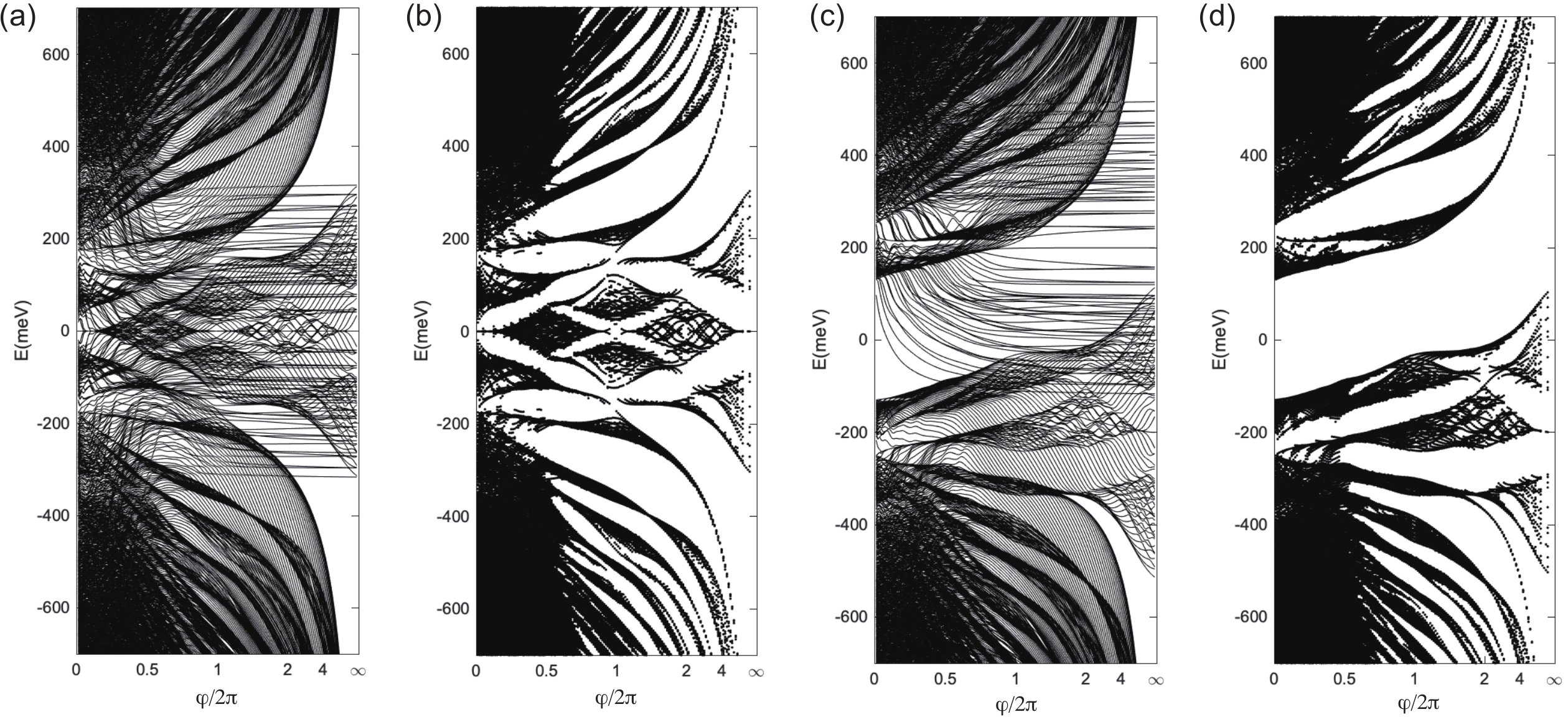}
\end{center}
\caption{Hofstadter butterfly and spectral flow of $\theta=2.2^\circ$ TBG with $N_Q=37$ and $N_L=60$ in a larger energy interval than that of main text Fig. 2, where (a)-(b) are calculated for a massless Dirac kinetic term, while (c)-(d) are calculated for a Dirac mass $m_D=200$meV. The horizontal axis linear coordinate is equal to $\varphi/2\pi$ in the range $[0,1]$, and is equal to $2-2\pi/\varphi$ in the range $[1,\infty]$. For convenience, we still label the horizontal axis by the values of $\varphi$. The edge states in (b) and (d) are deleted following the method described in Sec. \ref{Sec-numerical-cutoff}, where we used $w=\min\{\ell^{-1},1.6\sqrt{\Omega_{BZ}}\}$ for the edge projector $P_{\kappa_b,w}$, and the edge projection threshold $P_c=0.5$.
}
\label{FigS-TBG}
\end{figure}

As another example, we also calculated the spectrum of the TBG Hamiltonian with a Dirac mass term added (which can arise from hBN substrate alignment in TBG \cite{sharpe2019,serlin2020}), i.e., a model Hamiltonian
\begin{equation}\label{Seq-HQQTBGm}
H_{\mathbf{Q}'\mathbf{Q}}'(m_D,\mathbf{k})=[v_F\bm{\sigma}^*\cdot(\mathbf{k}+\mathbf{Q})+m_D\sigma_z] \delta_{\mathbf{Q}'\mathbf{Q}}+  \sum_{j=1}^3 (V_{j}\delta_{\mathbf{Q}',\mathbf{Q}+\mathbf{q}_j}+h.c.)\ .
\end{equation} 
The calculated Hofstadter butterfly with edge states present and deleted for $m_D=200$meV are shown in Fig. \ref{FigS-TBG}(c)-(d). We will comment more on this case of nonzero Dirac mass in Sec. \ref{Sec-spurious}.

\subsection{Spurious zero modes}\label{Sec-spurious}
In the main text, Fig. 2(a) (see also Fig. \ref{FigS-TBG}(a)) shows many nondispersive horizontal levels at large fluxes $\varphi$, which are roughly distributed in energies from $-0.3$eV to $0.3$eV. In Fig. \ref{FigS-TBG}(c) which has a nonzero Dirac mass $m_D=200$meV, these nondispersive horizontal levels are still present at large $\varphi$, but are distributed in an energy range $-0.1$eV to $0.5$eV. When $\varphi$ is small, these levels become dispersive with respect to $\varphi$ and tend to higher energies, merging either with the Hofstadter bulk bands or with the dispersive momentum edge states. These nondispersive horizontal levels at large fluxes are understood as spurious zero modes of Dirac fermions due to the LL cutoff $N_L$, which can be derived as follows. 

In the large $\varphi$ limit, the magnetic length $\ell\rightarrow0$, and the massless Dirac kinetic term of the TBG Hamiltonian (\ref{Seq-HQQTBG}) at momentum site $\mathbf{Q}$ will tend to

\begin{equation}\label{Seq-spuriousH}
v_F\bm{\sigma}^*\cdot(\hat{\bm{\kappa}}+\mathbf{k}_0+\mathbf{Q}) =\frac{\sqrt{2}v_F}{\ell}\left(\begin{array}{cc}
0&a-p_+\ell/\sqrt{2}\\
a^\dag-p_-\ell/\sqrt{2} &0
\end{array}
\right)\rightarrow
\frac{\sqrt{2}v_F}{\ell}\left(\begin{array}{cc}
0&a\\
a^\dag &0
\end{array}
\right)\ ,
\end{equation}
where $p_\pm=k_{0x}+Q_x\pm i(k_{0y}+Q_y)$. Therefore, up to error $\mathcal{O}(p_\pm\ell)$, the Dirac kinetic term on each site $\mathbf{Q}$ under LL cutoff $N_L$ has two zero energy modes:
\begin{equation}\label{Seq-spurious0}
\psi_0=\left(\begin{array}{c}0\\ |0\rangle\end{array}\right)\ ,\qquad\qquad \psi_s=\left(\begin{array}{c} |N_L\rangle\\ 0\end{array}\right)\ ,
\end{equation}
where we have used the fact that $a|0\rangle=0$ and $a^\dag|N_L\rangle=0$. Note that $a^\dag|N_L\rangle=0$ is only true because of the LL cutoff. The first mode $\psi_0$ is the physical Dirac zero mode, while the second mode $\psi_s$ is an unphysical spurious zero mode due to the LL cutoff $N_L$. With a reciprocal lattice cutoff of $N_Q$ BZs, we have $2N_Q$ spurious zero modes on the $2N_Q$ honeycomb reciprocal lattice (since there are two Dirac cones in each BZ). These spurious zero modes are located at radius $|\hat{\bm{\kappa}}|=\sqrt{\langle\hat{\bm{\kappa}}^2\rangle}\approx \frac{\sqrt{2N_L}}{\ell}$ in the momentum space (main text Fig. 1(b)). For large fluxes $\varphi>2\pi N_Q/N_L$, the momentum boundary is at radius $\kappa_b\approx\min\{\frac{\sqrt{2N_L}}{\ell}, \sqrt{\frac{N_Q\Omega_{BZ}}{\pi}}\}=\sqrt{\frac{N_Q\Omega_{BZ}}{\pi}}<\frac{\sqrt{2N_L}}{\ell}$, thus these spurious modes at radius $|\hat{\bm{\kappa}}|\approx \frac{\sqrt{2N_L}}{\ell}$ are outside the momentum space boundary and are not physical states of the Hofstadter butterfly. Moreover, these spurious modes on different reciprocal sites $\mathbf{Q}$ have zero Dirac kinetic energy, and only hop among nearest reciprocal sites with an amplitude $\approx \psi_s^\dag V_j\psi_s=w_0=110$ meV. Therefore, at large flux $\varphi$, they behave as a honeycomb tight-binding model in the reciprocal lattice with hopping amplitude $w_0$, which has $2N_Q$ energy levels independent of $\varphi$ distributed between $-3w_0$ and $3w_0$. These levels give the horizontal lines at large $\varphi$ in the main text Fig. 2(a) (as well as Fig. \ref{FigS-TBG}(a)).

When a Dirac mass term $m_D\sigma_z$ is added to the TBG Hamiltonian as given in Eq. (\ref{Seq-HQQTBGm}), one finds the spurious zero mode $\psi_s$ in Eq. (\ref{Seq-spurious0}) at each momentum site $\mathbf{Q}$ no longer has a Dirac zero kinetic energy; instead, it is shifted to energy $m_D$. Therefore, with the hopping $w_0$ among nearest momentum sites, one expects the $2N_Q$ spurious zero modes to be distributed in the energy range between $m_D-3w_0$ and $m_D+3w_0$. This is exactly the case in Fig. \ref{FigS-TBG}(c).

At small $\varphi$, one can no longer ignore the $p_\pm$ terms in Eq. (\ref{Seq-spuriousH}), thus the spurious modes $\psi_s$ in Eq. \ref{Seq-spurious0} is no longer a zero energy mode of the Dirac Hamiltonian in Eq. (\ref{Seq-spuriousH}). Therefore, one expect these spurious modes to disperse with respect to $\varphi$ at small $\varphi$, in agreement with the numerical result.

When counting the number of occupied states in a fixed $\lambda$ sector (in Eqs. (9) and (10) of the main text), these unphysical spurious modes at large magnetic fields should be excluded. Further, since these spurious modes are outside the momentum space boundary (when $\varphi/2\pi>N_Q/N_L$), they will be identified as edge states by the boundary projector $P_{\kappa_b,w}$ in Eq. (\ref{Seq-Pb}) (their expectation values $\langle P_{\kappa_b,w}\rangle$ are close to $1$), and thus will be removed in the edge-state removed spectrum (main text Fig. 2(b), and Fig. \ref{FigS-TBG}(b) and (d)).

\section{Basis Completeness and Matrix Elements for Tight-binding models}\label{sec-TB}

The open momentum space method can also be applied to the numerical calculation of the Hofstadter butterfly of tight-binding models, as we demonstrated in the main text Fig. 3. To understand why the method is also valid for tight-binding models, 
in this section, we construct the complete and orthonormal basis for tight binding models employed by our method. We prove that under the basis we construct, the tight-binding Hamiltonian under magnetic field (Peierls substitution) is block diagonalized into identical blocks, and each block has matrix elements given by the zero-magnetic-field momentum space Hamiltonian $H^{\alpha\beta}(\mathbf{k})$ with the simple substitution $\mathbf{k}\rightarrow\hat{\bm{\kappa}}+\mathbf{k}_0$, where $\mathbf{k}$ is the quasi-momentum, $\mathbf{k}_0$ is an arbitrary momentum vector, $\hat{\bm{\kappa}}= \frac{1}{{\sqrt{2}\ell}}(a+a^\dag, -ia+ia^\dag)$, and $a$ and $a^\dag$ are the Landau level raising and lowering operators.

\subsection{Standard Peierls substitution}\label{sec:standard-Peierls}

We denote the Wannier orbital $\alpha$ in the unit cell labeled by lattice vector $\bm{D}\in\bm{d}_1\mathbb{Z}+\bm{d}_2\mathbb{Z}$ in real space as $|\bm{D},\alpha\rangle$, and use $\bm{u}_\alpha$ to denote the position of orbital $\alpha$ in a unit cell. In the continuum space, we have
\begin{equation}\label{Seq-deltaWannier}
\langle \mathbf{r},\alpha|\bm{D},\alpha\rangle=W_\alpha(\mathbf{r}-\bm{D}-\bm{u}_\alpha)\ ,
\end{equation}
where $W_\alpha(\mathbf{r})$ is the Wannier function of orbital $\alpha$, and $|\mathbf{r},\alpha\rangle$ is the underlying continuum space basis at position $\mathbf{r}$ of intrinsic orbital $\alpha$ as defined by Eq. (\ref{Seq-r-basis}). For the discussion of tight-binding models here, without loss of generality, we shall choose the gauge where all $\mathbf{p}_\alpha=0$ in the definition of $|\mathbf{r},\alpha\rangle$ in Eq. (\ref{Seq-r-basis}) (recall that $\mathbf{p}_\alpha$ is the momentum space origin of orbital $\alpha$, which can be chosen freely), namely,
\begin{equation}\label{Seq-r-basis2}
|\mathbf{r},\alpha\rangle=c^\dag_\alpha(\mathbf{r})|0\rangle\ .
\end{equation}
This ensures that the Hamiltonian matrix in the real space basis is invariant under lattice vector translation $T_{\bm{d}_i}$ (instead of changing by a unitary transformation as shown in Eq. (\ref{Seq-Hr+d}) when $\mathbf{p}_\alpha\neq0$).

We first comment that the derivation of the standard Peierls substitution \cite{Luttinger1951} requires an approximation that the Wannier orbitals $|\bm{D},\alpha\rangle$ are infinitely localized, namely, $W_\alpha(\mathbf{r})=\delta^2(\mathbf{r})$. This is because the Peierls substitution only picks up the gauge phase factor connecting two points of Wannier positions, and is independent of the details (shapes, sizes, etc) of Wannier functions, which can be true only if each Wannier function is infinitely small and thus does not feel the magnetic field inside the orbital itself. This does not require the Wannier charge density to be localized on the site position $\bm{D}$, instead it can be a delta function localized at any position $\bm{D}+\bm{u}_\alpha$ away from the site position $\bm{D}$. 

However, in our paper here, we shall keep the Wannier function $W_\alpha(\mathbf{r})$ in Eq. (\ref{Seq-deltaWannier}) \emph{generic}, instead of assuming it is a delta function. This makes our proof of the method the most generic, which applies to nonstandard Peierls substitutions discussed in Sec. \ref{sec:nonstandard-Peierls}, too.

The tight-binding model in real space then generically takes the form
\begin{equation}\label{Seq-HTB}
H=\sum_{j,\alpha,\beta}t_{j}^{\alpha\beta} T^{\alpha\beta}_{\bm{D}_j+\bm{u}_\alpha- \bm{u}_\beta}\ ,
\end{equation}
where $t_j^{\alpha\beta}$ are the hopping amplitudes, and we have defined
\begin{equation}\label{Seq-trans}
T^{\alpha\beta}_{\bm{D}_j+\bm{u}_\alpha- \bm{u}_\beta}=\sum_{\bm{D}} e^{i\int_{c_{j,\alpha\beta}}\mathbf{A}(\bm{D}+\mathbf{r})\cdot \text{d}\mathbf{r}}|\bm{D}+\bm{D}_j,\alpha\rangle\langle \bm{D},\beta|
\end{equation}
as the translation operator from the Wannier orbital $|\bm{D},\beta\rangle$ at position $\bm{D}+\bm{u}_\beta$ to the Wannier orbital $|\bm{D}+\bm{D}_j,\alpha\rangle$ at position $\bm{D}+\bm{D}_j+\bm{u}_\alpha$ under the Peierls substitution of the gauge potential $\mathbf{A}(\mathbf{r})$ in the continuum space, with $\bm{D}_j\in\bm{d}_1\mathbb{Z}+\bm{d}_2\mathbb{Z}$, and $c_{j,\alpha\beta}$ being the straight line segment from $\bm{u}_\beta$ to $\bm{D}_j+\bm{u}_\alpha$. 

----- \emph{At zero magnetic field}, we can choose the gauge $\mathbf{A}(\mathbf{r})=0$, and the Hamiltonian can be written into the momentum space by Fourier transformation as
\begin{equation}\label{Seq-HTBk}
H^{\alpha\beta}(\mathbf{k})=\sum_{j}t_{j}^{\alpha\beta}e^{-i\mathbf{k}\cdot(\bm{D}_j+\bm{u}_\alpha- \bm{u}_\beta)}\ ,
\end{equation}
where $\mathbf{k}$ is the quasi-momentum which takes values in the Brillouin zone, and the basis is the Bloch basis $|\mathbf{k},\alpha\rangle=\frac{1}{N_\Omega}\sum_{\bm{D}}e^{i\mathbf{k}\cdot(\mathbf{D}+\bm{u}_\alpha)}|\bm{D},\alpha\rangle$, with $N_\Omega$ being the number of unit cells in real space.

----- \emph{At nonzero magnetic field}, the gauge potential $\mathbf{A}(\mathbf{r})$ satisfies $\partial_xA_y(\mathbf{r})-\partial_yA_x(\mathbf{r})=B$, where $B$ is a uniform magnetic field in the continuum space. We now proceed to define a complete orthonormal basis for Hamiltonian (\ref{Seq-HTB}) based on the continuum space. 

Before starting, we first note that the continuum space has a Hilbert space spanned by the real space basis $|\mathbf{r},\alpha\rangle$ of all positions $\mathbf{r}$, which is much larger than the Hilbert space of the tight binding model spanned by the Wannier basis $|\bm{D},\alpha\rangle$ defined in Eq. (\ref{Seq-deltaWannier}). In the following, we shall first define a complete basis for the continuum space; then we project the basis into the sub-Hilbert space of the tight-binding model spanned by $|\bm{D},\alpha\rangle$ using a projector, and prove that the resulting projected basis forms a complete orthonormal basis for the tight-binding model.

In the continuum space, we can define the kinematic momentum operator $\bm{\Pi}=-i\nabla-\mathbf{A}(\mathbf{r})$, and the guiding center operator $\mathbf{R}=\mathbf{r}-\frac{\ell^2}{\hbar}\hat{\mathbf{z}}\times\bm{\Pi}$, as we did in Sec. \ref{sec-algebra}. Similar to Sec. \ref{sec-algebra}, we define a guiding center $R_{\hat{\bm{\tau}}}$ along the $\hat{\bm{\tau}}$ direction, with $\frac{\hat{\bm{\tau}}\cdot(\hat{\mathbf{z}}\times\bm{g}_1)} {\hat{\bm{\tau}}\cdot(\hat{\mathbf{z}}\times\bm{g}_2)}$ irrational. Moreover, we denote the reciprocal lattice of the tight-binding model as $\mathbf{Q}\in \bm{g}_1\mathbb{Z}+\bm{g}_2\mathbb{Z}$. We can then define lowering and raising operators following the same procedure as we did in Sec. \ref{sec-CM}
\begin{equation}\label{Seq-aad2}
a_{\mathbf{Q}}=\frac{\ell}{\sqrt{2}}[\Pi_{x}-Q_{x}-k_{0,x}+ i(\Pi_y-Q_{y}-k_{0,y})] \ ,\qquad a_{\mathbf{Q}}^\dag=\frac{\ell}{\sqrt{2}}[\Pi_{x}-Q_{x}-k_{0,x}- i(\Pi_y-Q_{y}-k_{0,y})]\ ,
\end{equation}
where $\mathbf{k}_0$ is the center momentum which can be chosen freely. Afterwards, we can define a basis $|\lambda,\mathbf{Q},n,\alpha\rangle$ in the continuum space satisfying
\begin{equation}\label{Seq-Bbasis2}
R_{\hat{\bm{\tau}}}|\lambda,\mathbf{Q},n,\alpha \rangle=[\lambda-\ell^2\hat{\bm{\tau}}\cdot(\hat{\mathbf{z}}\times \mathbf{Q})]|\lambda,\mathbf{Q},n,\alpha\rangle\ , \qquad a_{\mathbf{Q}}^\dag a_{\mathbf{Q}}|\lambda,\mathbf{Q},n,\alpha \rangle= n|\lambda,\mathbf{Q},n,\alpha \rangle\ ,
\end{equation}
where $\lambda$ takes values in the quotient set defined in Eq. (\ref{Seq-lam}), namely, $\lambda\in\Lambda_{\hat{\bm{\tau}}}=\mathbb{R}/[\ell^2\hat{\bm{\tau}}\cdot(\hat{\mathbf{z}}\times\bm{g}_1)\mathbb{Z} +\ell^2\hat{\bm{\tau}}\cdot(\hat{\mathbf{z}}\times\bm{g}_2)\mathbb{Z}]$. 
The basis definition (\ref{Seq-Bbasis2}) follows exactly the same derivation of the basis $|\lambda,\mathbf{Q}_\alpha,n,\alpha\rangle$ in Eq. (\ref{Seq-Bbasis}) in Sec. \ref{sec-CM}, except that here we have chosen the gauge that the momentum origins of all orbitals $\alpha$ are at $\mathbf{p}_\alpha=\mathbf{0}$  (see Eq. (\ref{Seq-r-basis}) for the definition of $\mathbf{p}_\alpha$, and see Eq. (\ref{Seq-r-basis2}) for our gauge choice for tight binding models here), thus $\mathbf{Q}_\alpha=\mathbf{Q}$ for all $\alpha$ (recall the definition of $\mathbf{Q}_\alpha$ in Eq. (\ref{sec-CM})). 
As we have proved in Eq. (\ref{Seq-ortho}), the basis $|\lambda,\mathbf{Q},n,\alpha\rangle$ in Eq. (\ref{Seq-Bbasis2}) satisfies 
\begin{equation}\label{Seq-ortho2}
\langle\lambda',\mathbf{Q}',n',\beta|\lambda,\mathbf{Q},n,\alpha\rangle =\delta_{\beta\alpha} \delta_{\lambda\lambda'}  \delta_{\mathbf{Q}'\mathbf{Q}}\delta_{n'n}\ , 
\end{equation}
thus forms a complete orthonormal basis for the Hilbert space of the continuum space, where $\lambda\in\Lambda_{\hat{\bm{\tau}}}=\mathbb{R}/[\ell^2\hat{\bm{\tau}}\cdot(\hat{\mathbf{z}}\times\bm{g}_1)\mathbb{Z} +\ell^2\hat{\bm{\tau}}\cdot(\hat{\mathbf{z}}\times\bm{g}_2)\mathbb{Z}]$ as defined in Eq. (\ref{Seq-lam}), $\mathbf{Q}\in\bm{g}_1\mathbb{Z}+\bm{g}_2\mathbb{Z}$ are the reciprocal vectors, and $n\ge0$ denotes the LL number. We therefore have a completeness condition in the continuum space:
\begin{equation}\label{Seq-NQ1}
\frac{1}{N_Q}\sum_{\lambda,\mathbf{Q},n}|\lambda,\mathbf{Q},n,\alpha\rangle\langle \lambda,\mathbf{Q},n,\alpha| =\mathbf{1}_\alpha =\int \text{d}^2\mathbf{r}|\mathbf{r},\alpha\rangle\langle \mathbf{r},\alpha|\ ,
\end{equation}
where $N_Q$ is the number of reciprocal sites $\mathbf{Q}$ (which tends to infinity), and $\mathbf{1}_\alpha$ stands for the identity matrix in the orbital $\alpha$ subspace in the continuum space.

Next, based on the continuum space basis $|\lambda,\mathbf{Q},n,\alpha\rangle$, we would like to define a complete basis for the Hilbert space of the tight-binding model spanned by Wannier orbitals $|\bm{D},\alpha\rangle$. We define the basis for the tight-binding model as
\begin{equation}\label{Seq-tbbasis1}
\begin{split}
\overline{|\lambda,n,\alpha\rangle}&=\sum_{\bm{D}}|\bm{D},\alpha\rangle\langle \bm{D}+\bm{u}_\alpha,\alpha|\lambda,\mathbf{0},n,\alpha\rangle\ ,
\end{split}
\end{equation}
where $|\bm{D},\alpha\rangle$ is the Wannier orbital, $|\bm{D}+\bm{u}_\alpha,\alpha\rangle$ denotes the position eigenbasis at position $\mathbf{r}=\bm{D}+\bm{u}_\alpha$ (the center of the Wannier orbital), and $\lambda\in\Lambda_{\hat{\bm{\tau}}}=\mathbb{R}/[\ell^2\hat{\bm{\tau}}\cdot(\hat{\mathbf{z}}\times\bm{g}_1)\mathbb{Z} +\ell^2\hat{\bm{\tau}}\cdot(\hat{\mathbf{z}}\times\bm{g}_2)\mathbb{Z}]$ as defined in Eq. (\ref{Seq-lam}). One may wonder why only the continuum space states $|\lambda,\mathbf{0},n,\alpha\rangle$ at $\mathbf{Q=0}$ are used in defining the basis (\ref{Seq-tbbasis1}). In fact, by Eq. (\ref{Seq-Rq}), we have the following identity for any lattice vector $\bm{D}$ and reciprocal vector $\mathbf{Q}$ (in the equation below, $\mathbf{r}$ is understood as a number denoting the continuous space coordinates)
\begin{equation}\label{Seq-DQ}
\begin{split}
\langle\bm{D}+\bm{u}_\alpha,\alpha|\lambda,\mathbf{Q},n,\alpha\rangle&
=\int \text{d}^2\mathbf{r} \delta^2(\mathbf{r}-\bm{D}-\bm{u}_\alpha) \langle\mathbf{r},\alpha| e^{i\mathbf{Q\cdot r}} |\lambda,\mathbf{0},n,\alpha\rangle =e^{i\mathbf{Q}\cdot(\bm{D}+\bm{u}_\alpha)} \langle\bm{D}+\bm{u}_\alpha,\alpha |\lambda,\mathbf{0},n,\alpha\rangle \\
&=e^{i\mathbf{Q}\cdot\bm{u}_\alpha} \langle\bm{D}+\bm{u}_\alpha,\alpha |\lambda,\mathbf{0},n,\alpha\rangle\ .
\end{split}
\end{equation}
Thus, one can equivalently rewrite the basis definition as
\begin{equation}\label{Seq-tbbasis-q}
|\overline{\lambda,n,\alpha\rangle}=\frac{1}{\sqrt{N_Q}}\sum_{\mathbf{Q},\bm{D}}e^{-i\mathbf{Q}\cdot\bm{u}_\alpha} |\bm{D},\alpha\rangle\langle \bm{D}+\bm{u}_\alpha,\alpha |\lambda,\mathbf{Q},n,\alpha\rangle\ ,
\end{equation}
which is expressed using the continuum space basis of all reciprocal sites $\mathbf{Q}$.

By the orthonormal relation (\ref{Seq-ortho2}) and the basis expression in Eq. (\ref{Seq-tbbasis1}), it is easy to see the subset of basis $|\overline{\lambda,n,\alpha\rangle}$ is orthonormal:
\begin{equation}\label{seq-othornomal-D1}
\begin{split}
&\overline{\langle\lambda',n',\beta}|\overline{\lambda,n,\alpha\rangle}=\delta_{\beta\alpha}\sum_{\bm{D}} \langle \lambda',\mathbf{0},n',\alpha |\bm{D}+\bm{u}_\alpha,\alpha\rangle\langle \bm{D}+\bm{u}_\alpha,\alpha|\lambda,\mathbf{0},n,\alpha\rangle \\
&=\delta_{\beta\alpha}\langle\lambda',\mathbf{0},n',\alpha | \sum_{\mathbf{Q}} e^{i\mathbf{Q}\cdot(\mathbf{r}-\bm{u}_\alpha)}|\lambda,\mathbf{0},n,\alpha\rangle = \delta_{\beta\alpha}\sum_{\mathbf{Q}} e^{-i\mathbf{Q}\cdot\bm{u}_\alpha}\langle\lambda',\mathbf{0},n',\alpha |\lambda,\mathbf{Q},n,\alpha\rangle =\delta_{\lambda\lambda'}\delta_{n'n}\delta_{\beta\alpha}\ ,
\end{split}
\end{equation}
where in the 2nd line $\mathbf{r}$ is understood as the position operator (not number), and we have used Eq. (\ref{Seq-Rq}) which implies $|\lambda,\mathbf{Q},n,\alpha\rangle=e^{i\mathbf{Q\cdot r}}|\lambda,\mathbf{0},n,\alpha\rangle$. Besides, from the 1st line to the 2nd line we have used the following identity for each orbital $\alpha$:
\begin{equation}\label{Seq-tb-proj}
\begin{split}
&\sum_{\bm{D}}|\bm{D}+\bm{u}_\alpha,\alpha\rangle\langle \bm{D} +\bm{u}_\alpha,\alpha|=\int \text{d}^2\mathbf{r}|\mathbf{r},\alpha\rangle \langle\mathbf{r},\alpha|\sum_{\bm{D}}\delta^2(\mathbf{r}-\bm{D}-\bm{u}_\alpha) \\
&=\int \text{d}^2\mathbf{r}|\mathbf{r},\alpha\rangle \langle\mathbf{r},\alpha|\sum_{\mathbf{Q}}e^{i\mathbf{Q}\cdot(\mathbf{r}-\bm{u}_\alpha)} =\sum_{\mathbf{Q}}e^{i\mathbf{Q}\cdot(\mathbf{r}-\bm{u}_\alpha)}\ ,
\end{split}
\end{equation}
where $\mathbf{r}$ on the right-most hand side of the 2nd line is understood as the position operator (instead of a number). 

Furthermore, by Eq. (\ref{Seq-DQ}), we can prove the completeness of the basis (\ref{Seq-tbbasis1}) for the tight-binding model as follows:
\begin{equation}\label{seq-completeness-D1}
\begin{split}
\sum_{\lambda,n,\alpha}\overline{|\lambda,n,\alpha}\rangle\langle&\overline{\lambda,n,\alpha|} =\sum_{\lambda,n,\alpha} \sum_{\bm{D},\bm{D}'}|\bm{D},\alpha\rangle\langle \bm{D}+\bm{u}_\alpha,\alpha|\lambda,\mathbf{0},n,\alpha\rangle\langle \lambda,\mathbf{0},n,\alpha|\bm{D}'+\bm{u}_\alpha,\alpha\rangle\langle \bm{D}',\alpha| \\
=&\frac{1}{N_Q} \sum_{\lambda,\mathbf{Q},n,\alpha} \sum_{\bm{D},\bm{D}'}|\bm{D},\alpha\rangle e^{-i\mathbf{Q}\cdot\bm{u}_\alpha} \langle \bm{D}+\bm{u}_\alpha,\alpha|\lambda,\mathbf{Q},n,\alpha\rangle\langle \lambda,\mathbf{Q},n,\alpha|\bm{D}'+\bm{u}_\alpha,\alpha\rangle e^{i\mathbf{Q}\cdot\bm{u}_\alpha} \langle \bm{D}',\alpha|\\
=& \sum_{\alpha} \sum_{\bm{D},\bm{D}'}|\bm{D},\alpha\rangle\langle \bm{D}+\bm{u}_\alpha,\alpha|\left(\frac{1}{N_Q}\sum_{\lambda,\mathbf{Q},n}|\lambda,\mathbf{Q},n,\alpha\rangle\langle \lambda,\mathbf{Q},n,\alpha| \right)|\bm{D}'+\bm{u}_\alpha,\alpha\rangle\langle \bm{D}',\alpha|\\ 
=&\sum_{\bm{D},\bm{D}',\alpha}|\bm{D},\alpha\rangle\langle \bm{D}+\bm{u}_\alpha,\alpha| \left(\int \text{d}^2\mathbf{r}|\mathbf{r},\alpha\rangle\langle \mathbf{r},\alpha|\right) |\bm{D}'+\bm{u}_\alpha,\alpha\rangle\langle \bm{D}',\alpha|
=\sum_{\bm{D},\alpha}|\bm{D},\alpha\rangle\langle \bm{D},\alpha|\ .
\end{split}
\end{equation}
This proves that the basis $\overline{|\lambda,n,\alpha\rangle}$ forms a complete orthonormal basis for the Hilbert space of the tight-binding model spanned by Wannier orbitals $|\bm{D},\alpha\rangle$.

Now we discuss the translation operator in Eq. (\ref{Seq-trans}) under magnetic field. First, we prove the following identity of the position basis $|\mathbf{r},\alpha\rangle$ in the continuum space. Assume $c_{f0}$ is a path (not necessarily straight) from position $\mathbf{r}_0$ to position $\mathbf{r}_f$ in the continuum space. By path partitioning $c_{f0}$ into $N$ small segments $\delta\mathbf{r}_j=\mathbf{r}_j-\mathbf{r}_{j-1}$ ($1\le j\le N$), and note that $e^{-i\bm{\Pi}\cdot\delta\mathbf{r}}=e^{i\mathbf{A(r)}\cdot\delta\mathbf{r}}e^{-\delta\mathbf{r}\cdot\nabla}$ in the $\delta\mathbf{r}\rightarrow0$ limit, we can prove that
\begin{equation}\label{Seq-path}
\begin{split}
&\mathcal{P}e^{-i\int_{c_{f0}} \bm{\Pi}\cdot \text{d}\mathbf{r}}|\mathbf{r}_0,\alpha\rangle= \lim_{N\rightarrow\infty}\prod_{j=1}^Ne^{-i \bm{\Pi}\cdot \delta\mathbf{r}_j}|\mathbf{r}_0,\alpha\rangle \\ =&\lim_{N\rightarrow\infty}\int \text{d}\mathbf{r}_N|\mathbf{r}_N,\alpha\rangle \langle \mathbf{r}_N,\alpha| e^{-i \bm{\Pi}\cdot \delta\mathbf{r}_N} \int \text{d}\mathbf{r}_{N-1}|\mathbf{r}_{N-1},\alpha\rangle \langle \mathbf{r}_{N-1},\alpha| e^{-i \bm{\Pi}\cdot \delta\mathbf{r}_{N-1}} \cdots \int \text{d}\mathbf{r}_1|\mathbf{r}_1,\alpha\rangle \langle \mathbf{r}_1,\alpha| e^{-i \bm{\Pi}\cdot \delta\mathbf{r}_1}|\mathbf{r}_0,\alpha\rangle \\
=&\lim_{N\rightarrow\infty}\prod_{j=1}^N e^{i \mathbf{A}(\mathbf{r}_0+\sum_{i\le j}\delta\mathbf{r}_i)\cdot \delta\mathbf{r}_j} |\mathbf{r}_0+\sum_{j=1}^N\delta\mathbf{r}_j,\alpha\rangle\\
=&e^{i\int_{c_{f0}}\mathbf{A}(\mathbf{r})\cdot \text{d}\mathbf{r}}|\mathbf{r}_f,\alpha\rangle \ ,
\end{split}
\end{equation}
where $\mathcal{P}$ stands for path ordering, and we have used the fact that $e^{-\delta\mathbf{r}\cdot \nabla}|\mathbf{r},\alpha\rangle=|\mathbf{r}+\delta\mathbf{r},\alpha\rangle$. 
Therefore, note that the definition of the basis $\overline{|\lambda,n,\alpha\rangle}$ in Eq. (\ref{Seq-tbbasis1}) contains the position eigenstate $|\bm{D}+\bm{u}_\alpha,\alpha\rangle$ at position $\mathbf{r}=\bm{D}+\bm{u}_\alpha$, we can rewrite the action of the translation operator in Eq. (\ref{Seq-trans}) on the basis $\overline{|\lambda,n,\alpha\rangle}$ as
\begin{equation}\label{Seq-Td0}
\begin{split}
&T^{\alpha\beta}_{\bm{D}_j+\bm{u}_\alpha- \bm{u}_\beta}\overline{|\lambda,n,\beta'\rangle} 
= \delta_{\beta\beta'} \sum_{\bm{D}} e^{i\int_{c_{\alpha\beta}}\mathbf{A}(\bm{D}+\mathbf{r})\cdot \text{d}\mathbf{r}}|\bm{D}+\bm{D}_j, \alpha \rangle\langle \bm{D}+\bm{u}_\beta,\beta|\lambda,\mathbf{0},n,\beta\rangle\\
=& \delta_{\beta\beta'} \sum_{\bm{D}} |\bm{D}+\bm{D}_j,\alpha \rangle\langle \bm{D}+\bm{D}_j+\bm{u}_\alpha, \beta |\mathcal{P}e^{-i\int_{c_{j,\alpha\beta}}\bm{\Pi}\cdot \text{d}\mathbf{r}} |\lambda,\mathbf{0},n,\beta \rangle \ ,
\end{split}
\end{equation}
where notations of the form $|\bm{D}+\bm{u}_\alpha,\beta\rangle$ denotes the position basis of orbital $\beta$ in the continuum space at position $\bm{D}+\bm{u}_\alpha$ (note that $\beta$ need not be equal to $\alpha$). In contrast, $|\bm{D}+\bm{D}_j,\alpha \rangle$ denotes the Wannier orbital basis of the unit cell at $\bm{D}+\bm{D}_j$, as we have defined. We note that in Eq. (\ref{Seq-Td0}), from the 1st line to the 2nd line we have used the fact that the position basis in the continuum space satisfies $\langle \bm{D}+\bm{u}_\beta,\beta|=\langle \bm{D}+\bm{D}_j+\bm{u}_\alpha, \beta |\mathcal{P}e^{-i\int_{c_{j,\alpha\beta}}\bm{\Pi}\cdot \text{d}\mathbf{r}}$, which we have proved in Eq. (\ref{Seq-path}).

We then define lowering and raising operators $a=a_\mathbf{0}=\frac{\ell}{\sqrt{2}}[\Pi_{x}-k_{0,x}+ i(\Pi_y-k_{0,y})]$ and $a=a_\mathbf{0}^\dag$, where $a_\mathbf{0},a_\mathbf{0}^\dag$ are the lowering and raising operators in Eq. (\ref{Seq-aad2}) at reciprocal site $\mathbf{Q}=\mathbf{0}$. By further defining $\hat{\bm{\kappa}}=\frac{1}{{\sqrt{2}\ell}}(a+a^\dag, -ia+ia^\dag)$, we can rewrite the kinematic momentum as $\bm{\Pi}=\hat{\bm{\kappa}}+\mathbf{k}_0$, where $\mathbf{k}_0$ is the center momentum in Eq. (\ref{Seq-aad2}). We can then further simplify Eq. (\ref{Seq-Td0}) as
\begin{equation}\label{Seq-Td}
\begin{split}
&T^{\alpha\beta}_{\bm{D}_j+\bm{u}_\alpha- \bm{u}_\beta}\overline{|\lambda,n,\beta'\rangle} = \delta_{\beta\beta'} \sum_{\bm{D}} |\bm{D}+\bm{D}_j,\alpha \rangle\langle \bm{D}+\bm{D}_j+\bm{u}_\alpha, \beta |\mathcal{P}e^{-i\int_{c_{j,\alpha\beta}}(\hat{\bm{\kappa}}+\mathbf{k}_0)\cdot \text{d}\mathbf{r}} |\lambda,\mathbf{0},n,\beta \rangle \ ,\\
=& \delta_{\beta\beta'} \sum_{\bm{D}} |\bm{D}+\bm{D}_j,\alpha \rangle\langle \bm{D}+\bm{D}_j+\bm{u}_\alpha, \alpha |\mathcal{P}e^{-i\int_{c_{j,\alpha\beta}}(\hat{\bm{\kappa}}+\mathbf{k}_0)\cdot \text{d}\mathbf{r}} |\lambda,\mathbf{0},n,\alpha \rangle \ ,\\
=&\delta_{\beta\beta'}\sum_{\bm{D}} |\bm{D},\alpha \rangle\langle \bm{D} +\bm{u}_\alpha,\alpha| \mathcal{P}e^{-i\int_{c_{j,\alpha\beta}}(\hat{\bm{\kappa}}+\mathbf{k}_0)\cdot \text{d}\mathbf{r}} |\lambda,\mathbf{0},n,\alpha\rangle \\
=&\delta_{\beta\beta'} \mathcal{P}e^{-i\int_{c_{j,\alpha\beta}}(\hat{\bm{\kappa}}+\mathbf{k}_0)\cdot \text{d}\mathbf{r}} \overline{|\lambda,n,\alpha\rangle}\ ,
\end{split}
\end{equation}
where $\mathcal{P}$ stands for path ordering. From the 1st line to the 2nd line of Eq. (\ref{Seq-Td}), we have used the fact that the matrix elements of the operator $\mathcal{P}e^{-i\int_{c_{j,\alpha\beta}}(\hat{\bm{\kappa}}+\mathbf{k}_0)\cdot \text{d}\mathbf{r}}$ (acting from the right on the state $\langle \bm{D}+\bm{D}_j+\bm{u}_\alpha, \beta |$) only depend on the position $\mathbf{r}=\bm{D}+\bm{D}_j+\bm{u}_\alpha$ of the state (since $\hat{\bm{\kappa}}$ is defined in terms of $\bm{\Pi}$ which displaces $\mathbf{r}$), and are independent of the orbital index $\beta$. So we can simply change the orbital index $\beta$ in the 1st line into $\alpha$ in the 2nd line. In the derivation from the 3rd line to the 4-th line, we have also used the fact that the matrix elements of operator $\hat{\bm{\kappa}}$ only depends on the LL quantum number $n$, so in the last line of Eq. (\ref{Seq-Td}) one should understand the operator $\hat{\bm{\kappa}}$ as solely acting on the quantum number $n$ of basis $\overline{|\lambda,n,\alpha\rangle}$. More concretely, the LL lowering and raising operators in $\hat{\bm{\kappa}}$ act as $a\overline{|\lambda,n,\alpha\rangle}=\sqrt{n}\overline{|\lambda,n-1,\alpha\rangle}$ and $a^\dag\overline{|\lambda,n,\alpha\rangle}=\sqrt{n+1}\overline{|\lambda,n+1,\alpha\rangle}$.

For standard Peierls substitution, the path $c_{j,\alpha\beta}$ in Eq. (\ref{Seq-Td}) is a straight line segment (see the paragraph below Eq. (\ref{Seq-trans})), so the integral on the exponent of $\mathcal{P}e^{-i\int_{c_{j,\alpha\beta}}(\hat{\bm{\kappa}}+\mathbf{k}_0)\cdot \text{d}\mathbf{r}}$ is along the straight line segment from $\bm{D}+\bm{u}_\beta$ to $\bm{D}+\bm{D}_j+\bm{u}_\alpha$. Since the matrix elements of $\hat{\bm{\kappa}}$ only depend on $n$ when acting on $\overline{|\lambda,n,\alpha\rangle}$, the matrix $\hat{\bm{\kappa}}+\mathbf{k}_0$ should be a constant along the path of integration, and thus we can further simplify Eq. (\ref{Seq-Td}) into
\begin{equation}\label{Seq-Td1}
T^{\alpha\beta}_{\bm{D}_j+\bm{u}_\alpha- \bm{u}_\beta}\overline{|\lambda,n,\beta'\rangle}=\delta_{\beta\beta'} e^{-i(\hat{\bm{\kappa}}+\mathbf{k}_0)\cdot(\bm{D}_j+\bm{u}_\alpha- \bm{u}_\beta)} \overline{|\lambda,n,\alpha\rangle}\ .
\end{equation}
It is evident from Eq. (\ref{Seq-Td1}) that the matrix elements of $T_{\bm{D}_j+\bm{u}_\alpha- \bm{u}_\beta}$ is diagonal in quantum number $\lambda$ and independent of $\lambda$. Therefore, we can rewrite the translation operator in a fixed $\lambda$ subspace from basis $\overline{|\lambda,n,\beta\rangle}$ to basis $\overline{|\lambda,n',\alpha\rangle}$ ($n,n'\in\mathbb{Z})$ are nonnegative integers) as
\begin{equation}
T^{\lambda,\alpha\beta}_{\bm{D}_j+\bm{u}_\alpha- \bm{u}_\beta}=e^{-i(\hat{\bm{\kappa}}+\mathbf{k}_0)\cdot(\bm{D}_j+\bm{u}_\alpha- \bm{u}_\beta)}\ ,
\end{equation}
where $\hat{\bm{\kappa}}$ acting on the LL quantum number $n$ (recall that  $\bm{\Pi}=\hat{\bm{\kappa}}+\mathbf{k}_0$). 
Accordingly, the tight-binding Hamiltonian in magnetic field in a fixed $\lambda$ subspace from basis $\overline{|\lambda,n,\beta\rangle}$ to basis $\overline{|\lambda,n',\alpha\rangle}$ ($n,n'\in\mathbb{Z})$ takes the form
\begin{equation}\label{Seq-Htbb}
H^{\lambda,\alpha\beta}=\sum_{j}t_{j}^{\alpha\beta}T^{\lambda,\alpha\beta}_{\bm{D}_j+\bm{u}_\alpha- \bm{u}_\beta}=\sum_{j}t_{j}^{\alpha\beta}e^{-i(\hat{\bm{\kappa}}+\mathbf{k}_0)\cdot(\bm{D}_j+\bm{u}_\alpha- \bm{u}_\beta)}\ ,
\end{equation}
which is exactly the zero field momentum space Hamiltonian (\ref{Seq-HTBk}) with the substitution $\mathbf{k}\rightarrow\hat{\bm{\kappa}}+\mathbf{k}_0$.

\begin{figure}[tbp]
\begin{center}
\includegraphics[width=5in]{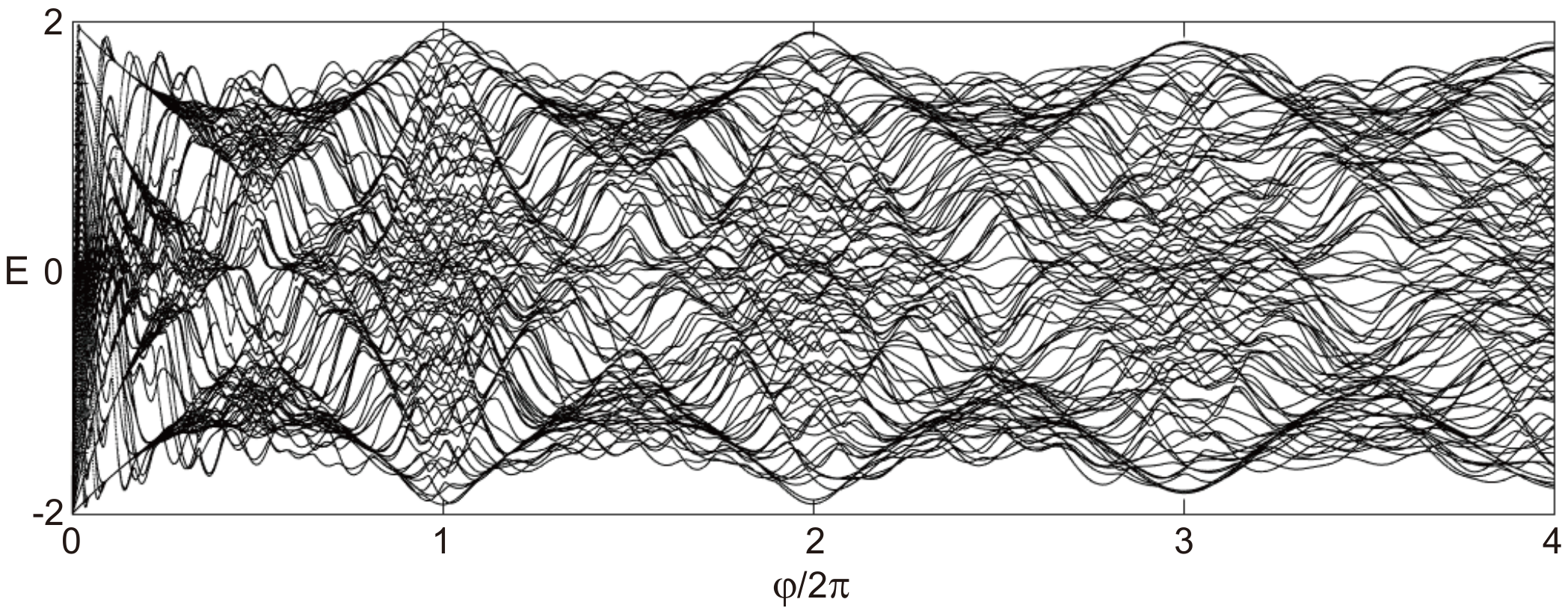}
\end{center}
\caption{The first 4 periods of Hofstadter butterfly of the tight-binding model $H(\mathbf{k})=-\cos k_x-\cos k_y$ in the square lattice calculated using our method of substitution $\mathbf{k}\rightarrow\hat{\bm{\kappa}}+\mathbf{k}_0$, where we take a LL cutoff $N_L=100$. The cutoff $N_L$ breaks the periodicity of the Hofstadter spectrum, and the larger the magnetic field $B$ is, the less clear the Hofstadter butterfly is. For sufficiently large LL cutoffs (e.g., $N_L\sim500$), the periodicity of Hofstadter spectrum will be well-preserved in the first few periods.
}
\label{tb4T}
\end{figure}

As an example of the above method, we numerically calculate the Hofstadter butterfly of the simplest square lattice model with one orbital per site and nearest hopping amplitude $-1/2$, which has a momentum space Hamiltonian $H(\mathbf{k})=-\cos k_x-\cos k_y$. The Hamiltonian matrix with LL cutoff $N_L$ is constructed by replacing $k_x$ by the Hermitian matrix $k_{0,x}\mathbf{1}_{N_L}+(a+a^\dag)/(\sqrt{2}\ell)$ and similarly for $k_y$, where $a$ is a matrix with matrix elements $[a]_{mn}=\langle m|a|n\rangle=\sqrt{n}\delta_{m,n-1}$ ($m,n$ are integers from $0$ to $N_L$). The cosine of a Hermitian matrix $W$ can be calculated by first diagonalizing the matrix into $W=U^\dag D_W U$ where $D_W$ is diagonal and $U$ is unitary. Then $\cos W$ can be calculated efficiently using the identity $\cos W=U^\dag(\cos D_W) U$ and the fact that $\cos D_W$ is simply the cosine of each element of the diagonal matrix $D_W$.

The spectra with cutoffs $N_L=100$ and $N_L=500$ are shown in the main text Fig. 3. Fig. 3 shows that the higher cutoff $N_L$ is, the clearer the Hofstadter butterfly is. Furthermore, the cutoff $N_L$ breaks the periodicity of the Hofstadter spectrum with respect to magnetic flux per unit cell $\varphi$. Fig. \ref{tb4T} shows the first 4 periods of the numerical Hofstadter butterfly with $N_L=100$. In particular, the larger $\varphi$ is, the less clear the Hofstadter butterfly is. This is because in Eq. (\ref{Seq-Htbb}) the operator $\hat{\bm{\kappa}}\propto\ell^{-1}(a,a^\dag)$; for a fixed cutoff $N_L$, the error in the LL operators $(a,a^\dag)$ is fixed, so the error in operator $\hat{\bm{\kappa}}$ is larger for larger $\varphi$ (which corresponds to smaller $\ell$).

Lastly, we note that throughout our derivation, we have not used the details of the Wannier function $W_\alpha(\mathbf{r})$ in Eq. (\ref{Seq-deltaWannier}). Therefore, our derivation holds generically as long as the Peierls substitution is valid, independent of the shape of the Wannier orbitals.


\subsection{Nonstandard Peierls substitution}\label{sec:nonstandard-Peierls}

In some models the Wannier orbitals cannot be approximated as localized at one position (but dominantly localizes at several positions respecting the symmetries of the lattice), so the standard Peierls approximations are no longer valid. However, in certain models, nonstandard Peierls substitution can be derived under certain approximations \cite{lian2019b}. Accordingly, the Peierls substitution might not be along the straight line segment path $c_{\alpha\beta}$, but may be along a nonstraight path or the sum of multiple nonstraight paths from $\bm{D}+\bm{u}_\beta$ to $\bm{D}+\bm{D}_j+\bm{u}_\alpha$. For instance, in the 4-band tight-binding model for TBG in Ref. \cite{songz2018}, the Wannier orbital is centered at AB and BA stackings of TBG but extends to the 3 closest AA stacking centers, and it is shown to have an approximate nonstandard Peierls substitution given by the summation of contributions of multiple broken-line paths \cite{lian2019b}.

For such a tight-binding model with nonstandard Peierls substitution, the tight-binding Hamiltonian is still of the form of Eq. (\ref{Seq-HTB}), except that the translation operator in Eq. (\ref{Seq-HTB}) now reads
\begin{equation}
T^{\alpha\beta}_{\bm{D}_j+\bm{u}_\alpha- \bm{u}_\beta}=\sum_{\bm{D},\mu} e^{i\int_{c^\mu_{j,\alpha\beta}}\mathbf{A}(\bm{D}+\mathbf{r})\cdot \text{d}\mathbf{r}}|\bm{D}+\bm{D}_j,\alpha\rangle\langle \bm{D},\beta|\ ,
\end{equation}
where the phase factor involves the summation of multiple different paths (e.g., broken-lines) $c^\mu_{j,\alpha\beta}$ (labeled by index $\mu=1,2,\cdots$) from position $\bm{u_\beta}$ to $\bm{D}_j+\bm{u}_\alpha$. For instance, in the 4-band tight-binding model for TBG studied in \cite{lian2019b}, the Peierls substitution between the nearest neighbors is given by the summation of the gauge phase factor of 2 different paths (i.e., $\mu=1,2$) from one AB site to another AB site via the nearest 2 AA sites.

Our method can still apply to such tight-binding models with nonstandard Peierls substitutions. To see this, we first note that throughout our derivations in Sec. \ref{sec:standard-Peierls} for the standard Peierls substitution case, we do not require at all the Wannier orbitals in Eq. (\ref{Seq-deltaWannier}) to be localized. Therefore, in the nonstandard Peierls substitution case here, we can still define the eigenbasis by Eq. (\ref{Seq-tbbasis1}), which still satisfies the orthogonality in Eq. (\ref{seq-othornomal-D1}) and the completeness in Eq. (\ref{seq-completeness-D1}).  

The action of the translation operator is still given by Eq. (\ref{Seq-Td}), except that one need to sum over all the paths $c^\mu_{j,\alpha\beta}$ (i.e., replace $c_{j,\alpha\beta}$ in Eq. (\ref{Seq-Td}) by $c^\mu_{j,\alpha\beta}$ and sum over the path index $\mu$). However, since $c^\mu_{j,\alpha\beta}$ are no longer straight paths, Eq. (\ref{Seq-Td}) cannot be further reduced to Eq. (\ref{Seq-Td1}). Therefore, the translation operator in a fixed $\lambda$ sector from basis $\overline{|\lambda,n,\beta\rangle}$ to $\overline{|\lambda,n',\alpha\rangle}$ can be expressed by Eq. (\ref{Seq-Td}) as
\begin{equation}
T^{\lambda,\alpha\beta}_{\bm{D}_j+\bm{u}_\alpha- \bm{u}_\beta}=\sum_\mu\mathcal{P}e^{-i\int_{c^\mu_{\alpha\beta}} (\hat{\bm{\kappa}}+\mathbf{k}_0)\cdot \text{d}\mathbf{r}}\ ,
\end{equation}
where $\hat{\bm{\kappa}}=\frac{1}{{\sqrt{2}\ell}}(a+a^\dag, -ia+ia^\dag)$ acts on the LL quantum number $n$, and $\mathcal{P}$ stands for path ordering. The Hamiltonian in a fixed $\lambda$ subspace then reads $H^{\lambda,\alpha\beta}=\sum_{j}t_{j}^{\alpha\beta}T^{\lambda,\alpha\beta}_{\bm{D}_j+\bm{u}_\alpha- \bm{u}_\beta}$, the matrix elements of which are independent of $\lambda$.

\section{Review of the Diophantine equation}\label{sec-Diophantine}
In this section, we briefly review the proof of the Diophantine equation.

For a lattice model with magnetic flux $\varphi=2\pi\frac{p}{q}$ per unit cell, where $p$ and $q$ are two coprime numbers, the energy spectrum forms a set of Hofstadter bands. In particular, each Hofstadter gap is characterized by two integers $t_\nu$ and $s_\nu$, which satisfy the Diophantine equation
\begin{equation}\label{Seq-diophantine1}
t_\nu p+s_\nu q=\nu\ ,
\end{equation}
where $\nu$ is an integer. In particular, $t_\nu$ is the Chern number of the Hofstadter gap.

Here we briefly review how the Diophantine equation is proved following Ref. \cite{dana1985} (See also Ref. \cite{bernevig2013book} Chapter 5.3 for a different proof). In fact, one can prove an equivalent statement, that each Hofstadter band satisfies a Diophantine equation
\begin{equation}\label{Seq-diophantine2}
\sigma p+m q=1\ ,
\end{equation}
where $\sigma$ is the Chern number of the Hofstadter band, and $m$ is another integer characterizing the band.

The proof is as follows. First, we note that a lattice Hamiltonian $H$ (either a continuum model in Eq. (\ref{Seq-Cmodel}) or a tight-binding model as in Eq. (\ref{Seq-HTB}) we considered) in a uniform magnetic field $\mathbf{B}=B\hat{\mathbf{z}}$ still has translation symmetries along the Bravais lattice vectors $\bm{d}_1$ and $\bm{d}_2$. However, the translation symmetry operators are not the simple translation operators 
\begin{equation}
T_{\bm{d}_j}=e^{-i\bm{\Pi}\cdot\bm{d}_j} \qquad (j=1,2) 
\end{equation}
in magnetic field in the continuum space, since the operator $T_{\bm{d}_j}$ do not commute with Hamiltonian $H$ as one can easily verify (since $H$ contains operator $\bm{\Pi}$, and $[\Pi_x,\Pi_y]\neq0$). Instead, the (magnetic) translation symmetry operators which commute with the Hamiltonian $H$ are given by (for simplicity here we choose all the $\mathbf{p}_\alpha=\mathbf{0}$ in Eq. (\ref{Seq-palpha})):
\begin{equation}\label{Seq-Ttilde}
\widetilde{T}_{\bm{d}_j}=T_{\bm{d}_j}e^{-i(\ell^{-2}\hat{\mathbf{z}}\times\mathbf{r})\cdot\bm{d}_j} =e^{-i(\bm{\Pi}+\ell^{-2}\hat{\mathbf{z}}\times\mathbf{r})\cdot\bm{d}_j} =e^{-i \ell^{-2}(\hat{\mathbf{z}}\times\mathbf{R})\cdot\bm{d}_j}\ ,\qquad  [\widetilde{T}_{\bm{d}_j},H]=0\ ,\qquad (j=1,2)
\end{equation}
where $\mathbf{R}$ is the guiding center operator. Namely, the translation operator $T_{\bm{d}_j}$ and the translation symmetry operator $\widetilde{T}_{\bm{d}_j}$ differ by a unitary transformation $e^{-i(\ell^{-2}\hat{\mathbf{z}}\times\mathbf{r})\cdot\bm{d}_j}$. At zero magnetic field, $T_{\bm{d}_j}$ and  $\widetilde{T}_{\bm{d}_j}$ become the same.

At rational flux $\varphi=2\pi\frac{p}{q}$ per unit cell, it is straightforward to show that the translation symmetry operators satisfy the commutation relation
\begin{equation}\label{Seq-transsym}
\widetilde{T}_{\bm{d}_1}\widetilde{T}_{\bm{d}_2} =e^{-i2\pi p/q} \widetilde{T}_{\bm{d}_2}\widetilde{T}_{\bm{d}_1}\ .
\end{equation}
Note that in contrast we have $T_{\bm{d}_1}T_{\bm{d}_2}=e^{i2\pi p/q}T_{\bm{d}_2}T_{\bm{d}_1}$. One can therefore define two magnetic translation symmetry operators $\widetilde{T}_{q\bm{d}_1}=(\widetilde{T}_{\bm{d}_1})^q$ and $\widetilde{T}_{\bm{d}_2}$ which commute with each other, i.e., $[\widetilde{T}_{q\bm{d}_1},\widetilde{T}_{\bm{d}_2}]=0$. Since they also commute with the Hamiltonian $H$, we can define the Bloch wave function eigenstates $|\psi_{n,\mathbf{k}}\rangle$ of a Hofstadter band $n$ of $H$, with quasimomentum $\mathbf{k}$ defined by
\begin{equation}\label{Seq-quasi}
\widetilde{T}_{q\bm{d}_1}|\psi_{n,\mathbf{k}}\rangle=e^{i q\bm{d}_1\cdot\mathbf{k}}|\psi_{n,\mathbf{k}}\rangle\ ,\qquad
\widetilde{T}_{\bm{d}_2}|\psi_{n,\mathbf{k}}\rangle=e^{i \bm{d}_2\cdot\mathbf{k}}|\psi_{n,\mathbf{k}}\rangle\ .
\end{equation}
The magnetic BZ is then a parallelogram spanned by momentum vectors $\bm{g}_1/q$ and $\bm{g}_2$, where $\bm{g}_i$ satisfies $\bm{g}_i\cdot\bm{d}_j=2\pi\delta_{ij}$ ($i,j=1,2$). For simplicity, we shall assume all the Hofstadter bands are nondegenerate in one magnetic BZ (namely, at each momentum $\mathbf{k}$ in the magnetic BZ, the energy eigenvalues are nondegenerate), which is generically true when there is no other symmetries (which may protect degeneracies) except for the translation symmetries. When the Hofstadter band $n$ has a Chern number $\sigma$, one can choose the Bloch wave function $|\psi_{n,\mathbf{k}}\rangle$ as a continuous function of $\mathbf{k}$ satisfying
\begin{equation}\label{Seq-bloch}
|\psi_{n,\mathbf{k}+\bm{g}_1/q}\rangle=|\psi_{n,\mathbf{k}}\rangle\ ,\qquad
|\psi_{n,\mathbf{k}+\bm{g}_2}\rangle=e^{i\sigma q\bm{d_1}\cdot\mathbf{k}}|\psi_{n,\mathbf{k}}\rangle\ ,
\end{equation}
as illustrated in Fig. \ref{FigS-diophantine}. Here we do not restrict $\mathbf{k}$ within the magnetic BZ.
One can easily verify that the above choice gives a Berry phase $\sigma q\bm{d_1}\cdot(\bm{g}_1/q)=2\pi\sigma$ circulating the boundary of the magnetic BZ (starting from momentum $(\mathbf{0,0})$ to $(\mathbf{0},\bm{g}_2)$ to $(\bm{g}_1/q,\bm{g}_2)$ to $(\bm{g}_1/q,\mathbf{0})$ and then back to $(\mathbf{0,0})$), which is required by the Chern number $\sigma$.

\begin{figure}[tbp]
\begin{center}
\includegraphics[width=1.8in]{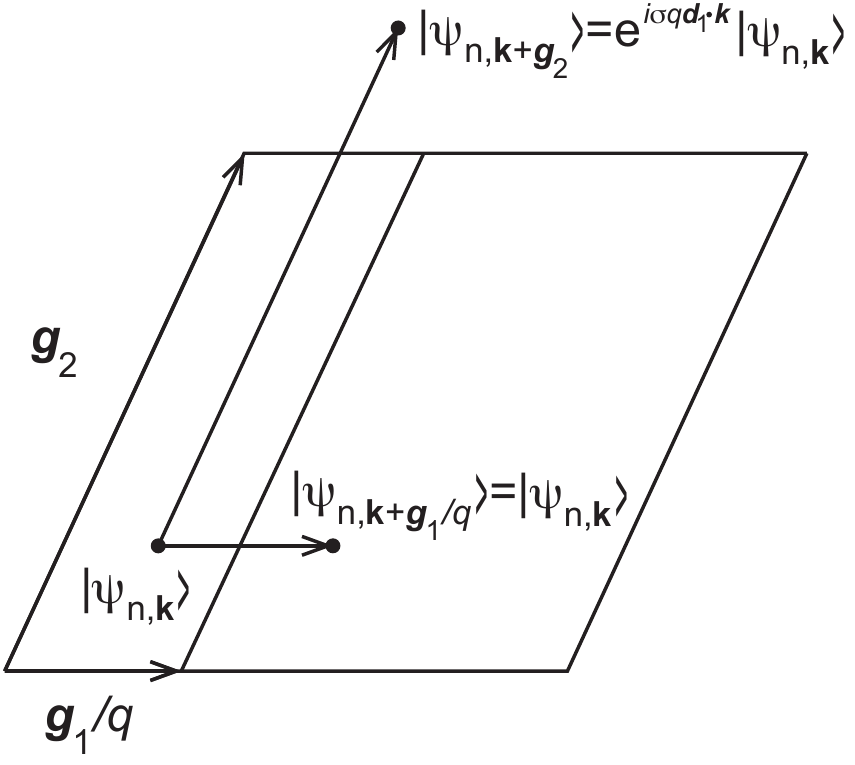}
\end{center}
\caption{Illustration of the gauge choices in Eq. (\ref{Seq-bloch}) for proving the Diophantine equation, where the magnetic flux is $\varphi/2\pi=p/q$.
}
\label{FigS-diophantine}
\end{figure}


By Eq. (\ref{Seq-transsym}) and the definition of quasimomentum $\mathbf{k}$ in Eq. (\ref{Seq-quasi}), the state $\widetilde{T}_{\bm{d}_1} |\psi_{n,\mathbf{k}}\rangle$ should be equal to the Bloch state $|\psi_{n,\mathbf{k}+p\bm{g}_2/q}\rangle$ up to a phase factor, and should be degenerate with the state $|\psi_{n,\mathbf{k}}\rangle$. In consistency with Eq. (\ref{Seq-bloch}), we can in general choose the phase such that
\begin{equation}\label{Seq-psi-Td1}
\widetilde{T}_{\bm{d}_1} |\psi_{n,\mathbf{k}}\rangle=e^{imq\bm{d}_1\cdot\mathbf{k}}|\psi_{n,\mathbf{k}+p\bm{g}_2/q}\rangle\ ,
\end{equation}
where $m$ is some coefficient to be determined. First, one can prove that $m\in\mathbb{Z}$ is an integer: this is because the first equation in (\ref{Seq-bloch}) requires $e^{imq\bm{d}_1\cdot\mathbf{k}}|\psi_{n,\mathbf{k}+p\bm{g}_2/q}\rangle=\widetilde{T}_{\bm{d}_1} |\psi_{n,\mathbf{k}}\rangle=\widetilde{T}_{\bm{d}_1} |\psi_{n,\mathbf{k}+\bm{g}_1/q}\rangle=e^{imq\bm{d}_1\cdot(\mathbf{k}+\bm{g}_1/q)}|\psi_{n,\mathbf{k}+\bm{g}_1/q+p\bm{g}_2/q}\rangle=e^{imq\bm{d}_1\cdot(\mathbf{k}+\bm{g}_1/q)}|\psi_{n,\mathbf{k}+p\bm{g}_2/q}\rangle$, namely, $e^{imq\bm{d}_1\cdot(\bm{g}_1/q)}=e^{i2\pi m}=1$. We can then apply $\widetilde{T}_{\bm{d}_1}$ by $q$ times to obtain
\begin{equation}
\widetilde{T}_{q\bm{d}_1}|\psi_{n,\mathbf{k}}\rangle =e^{imq^2\bm{d}_1\cdot\mathbf{k}}|\psi_{n,\mathbf{k}+p\bm{g}_2}\rangle =e^{i(\sigma p+mq)q\bm{d}_1\cdot\mathbf{k}}|\psi_{n,\mathbf{k}}\rangle =e^{i q\bm{d}_1\cdot\mathbf{k}}|\psi_{n,\mathbf{k}}\rangle\ .
\end{equation}
Since $\mathbf{k}$ can take any value, we conclude that the Diophantine equation \ref{Seq-diophantine2} for the band has to hold.

It is then straightforward to see the Diophantine equation \ref{Seq-diophantine1} holds in a Hofstadter gap: $\nu$ is simply the number of Hofstadter bands below the gap counted from some reference point of filling (see Sec. \ref{Seq-NK} for discussion of the reference point of filling). The Hofstadter band between the $\nu$-th gap and the $(\nu-1)$-th gap then has Chern number $\sigma=t_\nu-t_{\nu-1}$, and the other quantum number $m=s_\nu-s_{\nu-1}$.

Since $\nu$ has the physical meaning of number of occupied magnetic bands, and the magnetic unit cell has an area $q$ times of the original unit cell, we can define the number of electrons per original unit cell as $\rho=\nu/q$. One can then divide the Diophantine equation by $q$ to obtain our main text Eq. (7):
\begin{equation}\label{Seq-rho}
t_\nu\frac{\varphi}{2\pi}+s_\nu=\rho\ ,
\end{equation}
which holds even when the flux quanta $\varphi/2\pi$ is irrational. The derivative of this equation with respect to $\varphi$ gives the Streda formula $2\pi \frac{dn_r}{d\varphi}=t_r$ \cite{streda1982}.


\section{Quantized Lorentz Susceptibility from $s_\nu$}\label{sec-LS}
In this section, we show that the quantum number $s_\nu$, which can be viewed as a dual Chern number in the momentum space, yields a quantized Lorentz susceptibility $\gamma_{xy}=eBs_\nu$. This is analogous to the Hall conductance given by the Chern number $t_\nu$.

\subsection{Review of the quantized Hall conductance}

We first briefly review how the Chern number $t_\nu$ leads to a quantized Hall conductance $\sigma_{xy}=t_\nu \frac{e^2}{h}$ \cite{tknn1982}. We assume the flux per unit cell is $\varphi=2\pi p/q$, so that we can define the magnetic BZ quasimomentum $(k_x,k_y)$ by Eq. (\ref{Seq-quasi}). Using the Kubo formula, the Hall conductance under a uniform electric field $E_y$ can be written as
\begin{equation}\label{Seq-hallkubo}
\sigma_{xy}=i\frac{\partial}{\partial\omega} \int_{-\infty}^{\infty} \text{d}\omega'\langle G_{\omega+\omega'}\hat{j}_{x} G_{\omega'}\hat{j}_{y}\rangle\Big|_{\omega\rightarrow 0}
\end{equation}
where $\hat{\bm{j}}$ is the \emph{spatially uniform} current operator which is conjugate to a spatially uniform gauge field $\mathbf{A}$., while $G_\omega=(\omega-H)^{-1}$ is the Green's function in the magnetic field $B$ at energy $\omega$, and $H$ is the (first quantized) full single-particle Hamiltonian $H$ under magnetic field in the real space (under the position basis $|\mathbf{r},\alpha\rangle$). For example, for continuum models $H$ is as defined in Eq. (\ref{Seq-CmodelB}), while for tight-binding models $H$ is as given in Eq. (\ref{Seq-HTB}) embedded into the continuum space by the infinitely localized Wannier orbital assumption (\ref{Seq-deltaWannier}). The coefficient $\sigma_{xy}$ corresponds to a Hall response $j_x=-\sigma_{xy}dA_y/dt=\sigma_{xy}E_y$, where $E_y=-dA_y/dt$ is the uniform electric field in the $y$ direction.

We denote the Bloch eigenstates of $H$ in the $n$-th band as $|\psi_{n,\mathbf{k}}\rangle$ satisfying
\begin{equation}
 H|\psi_{n,\mathbf{k}}\rangle=\epsilon_{n,\mathbf{k}}|\psi_{n,\mathbf{k}}\rangle\ , 
\end{equation}
where the magnetic BZ quasimomentum $\mathbf{k}$ is defined by Eq. (\ref{Seq-quasi}) using the magnetic translation symmetry operators $\widetilde{T}_{q\bm{d}_1}=e^{-i q\ell^{-2}(\hat{\mathbf{z}}\times\mathbf{R})\cdot\bm{d}_1}$ and $\widetilde{T}_{\bm{d}_2}=e^{-i \ell^{-2}(\hat{\mathbf{z}}\times\mathbf{R})\cdot\bm{d}_2}$, and $\epsilon_{n,\mathbf{k}}$ is the energy of Bloch state $|\psi_{n,\mathbf{k}}\rangle$. 
Since $H$ is diagonal in the magnetic BZ quasimomentum $\mathbf{k}$, the Green's function (in the magnetic field $B$) is also diagonal in $\mathbf{k}$, and can be written as
\begin{equation}\label{Seq-Gomega}
G_{\omega}=\sum_{n,\mathbf{k}} \frac{|\psi_{n,\mathbf{k}}\rangle\langle\psi_{n,\mathbf{k}}|}{\omega-\epsilon_{n,\mathbf{k}}}\ .
\end{equation}
We comment that under the continuum space basis $|\lambda,\mathbf{Q}_\alpha,n,\alpha \rangle$ in Eq. (\ref{Seq-Bbasis}) employed by our Hofstadter method, the Bloch state $|\psi_{n,\mathbf{k}}\rangle$ does not live in a definite $\lambda$ sector. This is can be seen by noting that $|\psi_{n,\mathbf{k}}\rangle$ is the eigenstate of $\widetilde{T}_{q\bm{d}_1}$ and $\widetilde{T}_{\bm{d}_2}$, while we have $\widetilde{T}_{q\bm{d}_1}R_{\hat{\bm{\tau}}}\widetilde{T}_{q\bm{d}_1}^\dag=R_{\hat{\bm{\tau}}}+q\hat{\bm{\tau}}\cdot\bm{d}_1=R_{\hat{\bm{\tau}}}-p\ell^2\hat{\bm{\tau}}\cdot(\hat{z}\times\bm{g}_2)$ and $\widetilde{T}_{\bm{d}_2}R_{\hat{\bm{\tau}}}\widetilde{T}_{\bm{d}_2}^\dag=R_{\hat{\bm{\tau}}}+\hat{\bm{\tau}}\cdot\bm{d}_2=R_{\hat{\bm{\tau}}}+\frac{p}{q}\ell^2\hat{\bm{\tau}}\cdot(\hat{z}\times\bm{g}_1)$. Since the quantum number $\lambda \in \Lambda_{\hat{\bm{\tau}}}=\mathbb{R}/[\ell^2\hat{\bm{\tau}}\cdot(\hat{\mathbf{z}}\times\bm{g}_1)\mathbb{Z} +\ell^2\hat{\bm{\tau}}\cdot(\hat{\mathbf{z}}\times\bm{g}_2)\mathbb{Z}]$ as defined in Eq. (\ref{Seq-lam}), we find that the action of $\widetilde{T}_{q\bm{d}_1}$ preserves the coset of $\lambda$, while $\widetilde{T}_{\bm{d}_2}$ maps the sector of coset $\lambda$ to the sector of coset $\lambda+\frac{p}{q}\ell^2\hat{\bm{\tau}}\cdot(\hat{z}\times\bm{g}_1)$. Therefore, each Bloch state $|\psi_{n,\mathbf{k}}\rangle$ must be the superposition of the degenerate eigenstates in multiple of $q$ sectors labeled by quantum numbers $\lambda+m\frac{p}{q}\ell^2\hat{\bm{\tau}}\cdot(\hat{z}\times\bm{g}_1)$ ($\lambda$ belongs to a certain set, and $m=0,\cdots, q-1$). However, the wavefunction coefficients under the basis $|\lambda,\mathbf{Q}_\alpha,n,\alpha \rangle$ will be complicated, since the basis state $|\lambda,\mathbf{Q}_\alpha,n,\alpha \rangle$ with definite quantum numbers $\lambda,\mathbf{Q}_\alpha$, $n$ is not translationally invariant.

For any state preserving the translation symmetry, the expectation value of the current operator $\hat{\bm{j}}$ satisfies (recall we have set charge $e=1$)
\begin{equation}\label{Seq-jvalue}
\langle\hat{\bm{j}}\rangle=\langle \frac{\text{d}\mathbf{r}}{\text{d}t}\rangle=-i \langle [\mathbf{r},H]\rangle\ .
\end{equation}
The uniform current operator $\hat{\bm{j}}$ preserves the translation symmetry and is thus diagonal in magnetic BZ quasimomentum $\mathbf{k}$. In contrast, the position operator $\mathbf{r}$ does not commute with translation symmetry operators $\widetilde{T}_{q\bm{d}_1}$ and $\widetilde{T}_{\bm{d}_2}$, thus $\mathbf{r}$ is not fully diagonal in magnetic BZ quasimomentum $\mathbf{k}$, and accordingly $[\mathbf{r},H]$ is not fully diagonal in $\mathbf{k}$  (see Ref. \cite{karplus1954}). Therefore, Eq. (\ref{Seq-jvalue}) shows that the current operator $\hat{\bm{j}}$ is given by the part of $-i[\mathbf{r},H]$ that is diagonal in magnetic BZ quasimomentum $\mathbf{k}$, namely,
\begin{equation}\label{Seq-j}
\hat{\bm{j}}=\sum_{m,n,\mathbf{k}}\hat{\bm{j}}_{m\mathbf{k},n\mathbf{k}} |\psi_{m,\mathbf{k}}\rangle\langle\psi_{n,\mathbf{k}}|\ ,\qquad \hat{\bm{j}}_{m\mathbf{k},n\mathbf{k}}=-i\langle\psi_{m,\mathbf{k}}|[\mathbf{r},H]|\psi_{n,\mathbf{k}}\rangle\ .
\end{equation}

For a given magnetic BZ quasimomentum $\mathbf{k}$ (within the magnetic BZ spanned by vector $\bm{g}_1/q$ and $\bm{g}_2$), we can define a reduced magnetic BZ momentum space Hamiltonian $\hat{H}(\mathbf{k})$ (we label it by a hat, since it is distinguished from the full Hamiltonian $H$ in field in Eq. (\ref{Seq-CmodelB}) for continuum models or in Eq. (\ref{Seq-HTB}) for tight-binding models): 
\begin{equation}\label{Seq-Hhat}
\hat{H}(\mathbf{k})=P_\mathbf{0}(e^{-i\mathbf{k}\cdot\mathbf{r}}H e^{i\mathbf{k}\cdot\mathbf{r}})P_\mathbf{0}\ ,\qquad
P_\mathbf{0}=\frac{q}{N_\Omega}\sum_{j_1,j_2\in\mathbb{Z}} \widetilde{T}_{q j_1\bm{d}_1+j_2\bm{d}_2}\ ,
\end{equation}
where $\mathbf{k}$ is understood as a parameter, $\mathbf{r}$ is the position operator (understood as $\sum_\alpha\int \mathbf{r}|\mathbf{r},\alpha\rangle\langle\mathbf{r},\alpha|\text{d}^2\mathbf{r}$ when expanded in position basis, with $\alpha$ being the orbital), $P_\mathbf{0}$ is the projection operator into the magnetic translationally invariant subspace, $\widetilde{T}_{\bm{d}}=e^{-i \ell^{-2}(\hat{\mathbf{z}}\times\mathbf{R})\cdot\bm{d}}$ is the translation operator of displacement $\bm{d}$ by the guiding center operator $\mathbf{R}$ as defined in Eq. (\ref{Seq-Ttilde}), and $N_\Omega$ is the number of zero-magnetic-field unit cells (there are $N_\Omega/q$ magnetic unit cells at magnetic flux per unit cell $\varphi=2\pi p/q$). Note that any state with a nonzero momentum $\mathbf{k}\neq0$ in the magnetic BZ, for which the translation operator $\widetilde{T}_{q j_1\bm{d}_1+j_2\bm{d}_2}$ has an eigenvalue $e^{i(qj_1\bm{d}_1+j_2\bm{d}_2)\cdot\mathbf{k}}$, will be annihilated by $P_\mathbf{0}$ (since $\sum_{j_1,j_2}e^{i(qj_1\bm{d}_1+j_2\bm{d}_2)\cdot\mathbf{k}}=\frac{N_\Omega}{q}\delta_{\mathbf{k,0}}$).
Thus, unlike the full Hamiltonian $H$ in the real space which has a large Hilbert space dimension equal to the number of bands times the number of $\mathbf{k}$ in the magnetic BZ, the reduced Hamiltonian $\hat{H}(\mathbf{k})$ is defined at a fixed $\mathbf{k}$ and has a smaller Hilbert dimension equal to the number of bands, which can be understood as the full Hamiltonian $H$ projected into the sub-Hilbert space at magnetic BZ momentum $\mathbf{k}$. An explicit example of $\hat{H}(\mathbf{k})$ for a tight-binding model is given in Eqs. (\ref{Seq-Hhat-example0}) and (\ref{Seq-Hhat-example1}), where one can see $\hat{H}(\mathbf{k})$ is simply the magnetic BZ momentum space Hamiltonian under the Fourier transformed basis of the Wannier orbitals. The Hamiltonian $\hat{H}(\mathbf{k})$ has eigenstates
\begin{equation}\label{Seq-unk}
|u_{n,\mathbf{k}}\rangle=\sum_\alpha\int \text{d}^2\mathbf{r} e^{-i\mathbf{k}\cdot\mathbf{r}}|\mathbf{r},\alpha\rangle\langle\mathbf{r},\alpha|\psi_{n,\mathbf{k}}\rangle\ ,\qquad 
\hat{H}(\mathbf{k})|u_{n,\mathbf{k}}\rangle=\epsilon_{n,\mathbf{k}} |u_{n,\mathbf{k}}\rangle\ ,
\end{equation}
where $n$ runs over all the bands in the magnetic BZ. Note that the state $|u_{n,\mathbf{k}}\rangle$ has is magnetically translationally invariant, namely, $\widetilde{T}_{q\bm{d}_1}|u_{n,\mathbf{k}}\rangle =\widetilde{T}_{\bm{d}_2}|u_{n,\mathbf{k}}\rangle=|u_{n,\mathbf{k}}\rangle$. Therefore, $P_\mathbf{0} |u_{n,\mathbf{k}}\rangle=|u_{n,\mathbf{k}}\rangle$, and $|u_{n,\mathbf{k}}\rangle$ can be understood as the periodic part of the Bloch wave function $|\psi_{n,\mathbf{k}}\rangle$. 
In particular, from the definition of $\hat{H}(\mathbf{k})$ in Eq. (\ref{Seq-Hhat}) and the definition of current operator $\hat{\bm{j}}$ in Eq. (\ref{Seq-j}), we have (calculated by inserting the position basis)
\begin{equation}\label{Seq-dHdk}
\begin{split}
&\langle\psi_{m,\mathbf{k}}|\hat{\bm{j}}|\psi_{n,\mathbf{k}}\rangle =-i\sum_{\alpha,\alpha'}\int \text{d}^2\mathbf{r}d^2\mathbf{r}'\langle\psi_{m,\mathbf{k}}|\mathbf{r},\alpha\rangle\langle\mathbf{r},\alpha|(\mathbf{r}H-H\mathbf{r}')|\mathbf{r}',\alpha'\rangle \langle \mathbf{r}',\alpha'|\psi_{n,\mathbf{k}}\rangle \\
=&-i\sum_{\alpha,\alpha'}\int \text{d}^2\mathbf{r}d^2\mathbf{r}'\langle u_{m,\mathbf{k}}|\mathbf{r},\alpha\rangle\langle\mathbf{r},\alpha|e^{-i\mathbf{k\cdot r}}(\mathbf{r}H-H\mathbf{r}') e^{i\mathbf{k\cdot r}'}|\mathbf{r}',\alpha'\rangle \langle \mathbf{r}',\alpha'|u_{n,\mathbf{k}}\rangle \\
=& \langle u_{m,\mathbf{k}}|\frac{\partial}{\partial\mathbf{k}}\left(P_\mathbf{0}\int \text{d}^2\mathbf{r}d^2\mathbf{r}' |\mathbf{r},\alpha\rangle \langle \mathbf{r},\alpha|e^{-i\mathbf{k}\cdot\mathbf{r}} H e^{i\mathbf{k}\cdot\mathbf{r}'}|\mathbf{r}',\alpha'\rangle \langle \mathbf{r}',\alpha'| P_\mathbf{0}\right)|u_{n,\mathbf{k}}\rangle \\
=&\langle u_{m,\mathbf{k}}|\frac{\partial \hat{H}(\mathbf{k})}{\partial\mathbf{k}}|u_{n,\mathbf{k}}\rangle\ ,
\end{split}
\end{equation}
where we have used the first equation in (\ref{Seq-unk}) and the fact that $P_\mathbf{0} |u_{n,\mathbf{k}}\rangle=|u_{n,\mathbf{k}}\rangle$. 
Therefore, by carrying out the $\omega'$ integral in Eq. (\ref{Seq-hallkubo}), we arrive at
\begin{equation}\label{Seq-kuboH}
\begin{split}
\sigma_{xy}&=i\frac{e^2}{\hbar}\sum_{n,m}\int_{\text{MBZ}} \frac{\text{d}^2\mathbf{k}}{4\pi^2} \frac{n_F(\epsilon_{n,\mathbf{k}})-n_F(\epsilon_{m,\mathbf{k}})} {(\epsilon_{m,\mathbf{k}}-\epsilon_{n,\mathbf{k}})^2} \langle \psi_{n,\mathbf{k}}|\hat{j}_{x}|\psi_{m,\mathbf{k}}\rangle \langle \psi_{m,\mathbf{k}}|\hat{j}_{y}|\psi_{n,\mathbf{k}}\rangle \\
&=i\frac{e^2}{\hbar}\sum_{n,m}\int_{\text{MBZ}} \frac{\text{d}^2\mathbf{k}}{4\pi^2} \frac{n_F(\epsilon_{n,\mathbf{k}})-n_F(\epsilon_{m,\mathbf{k}})} {(\epsilon_{m,\mathbf{k}}-\epsilon_{n,\mathbf{k}})^2} \langle u_{n,\mathbf{k}}|\frac{\partial \hat{H}(\mathbf{k})}{\partial k_x}|u_{m,\mathbf{k}}\rangle \langle u_{m,\mathbf{k}}|\frac{\partial \hat{H}(\mathbf{k})}{\partial k_y}|u_{n,\mathbf{k}}\rangle \\
&=i\frac{e^2}{\hbar}\sum_{n\in\text{occ},m\not{\in}\text{occ}}\int_{\text{MBZ}} \frac{\text{d}^2\mathbf{k}}{4\pi^2} \frac{\langle u_{n,\mathbf{k}}|\frac{\partial \hat{H}(\mathbf{k})}{\partial k_x}|u_{m,\mathbf{k}}\rangle \langle u_{m,\mathbf{k}}|\frac{\partial \hat{H}(\mathbf{k})}{\partial k_y}|u_{n,\mathbf{k}}\rangle -\langle u_{n,\mathbf{k}}|\frac{\partial \hat{H}(\mathbf{k})}{\partial k_y}|u_{m,\mathbf{k}}\rangle \langle u_{m,\mathbf{k}}|\frac{\partial \hat{H}(\mathbf{k})}{\partial k_x}|u_{n,\mathbf{k}}\rangle} {(\epsilon_{m,\mathbf{k}}-\epsilon_{n,\mathbf{k}})^2}  \ ,
\end{split}
\end{equation}
where the integral of $\mathbf{k}$ runs over one magnetic BZ (MBZ), $n_F(\epsilon)=\frac{1-\text{sgn}(\epsilon)}{2}$ is the zero temperature Fermi-Dirac distribution function, $n\in\text{occ}$ means $n$ runs over all the occupied bands, and $m\not\in\text{occ}$ means $m$ runs over all the empty bands. Since we are in a Hofstadter gap, $n_F(\epsilon_{n,\mathbf{k}})$ only depends on the band index $n$. Then, using the fact that
\begin{equation}\label{Seq-dHdk1}
\begin{split}
&\langle u_{m,\mathbf{k}}|\frac{\partial \hat{H}(\mathbf{k})}{\partial k_i}|u_{n,\mathbf{k}}\rangle= \partial_{k_i}(\langle u_{m,\mathbf{k}}|\hat{H}(\mathbf{k})|u_{n,\mathbf{k}}\rangle)-\langle \partial_{k_i}u_{m,\mathbf{k}}|\hat{H}(\mathbf{k})|u_{n,\mathbf{k}}\rangle -\langle u_{m,\mathbf{k}}|\hat{H}(\mathbf{k})|\partial_{k_i}u_{n,\mathbf{k}}\rangle \\
&= 0- \epsilon_{n,\mathbf{k}} \langle \partial_{k_i}u_{m,\mathbf{k}}|u_{n,\mathbf{k}}\rangle - \epsilon_{m,\mathbf{k}}\langle u_{m,\mathbf{k}}|\partial_{k_i}u_{n,\mathbf{k}}\rangle = (\epsilon_{n,\mathbf{k}}-\epsilon_{m,\mathbf{k}})\langle u_{m,\mathbf{k}}|\partial_{k_i}u_{n,\mathbf{k}}\rangle =(\epsilon_{m,\mathbf{k}}-\epsilon_{n,\mathbf{k}})\langle \partial_{k_i} u_{m,\mathbf{k}}|u_{n,\mathbf{k}}\rangle\ ,
\end{split}
\end{equation}
we can rewrite the Hall conductance as the TKNN formula \cite{tknn1982}
\begin{equation}
\begin{split}
&\sigma_{xy}=i\frac{e^2}{\hbar}\sum_{n\in\text{occ}}\sum_{m\not\in\text{occ}} \int_{\text{MBZ}} \frac{\text{d}^2\mathbf{k}}{4\pi^2} \left(\langle\partial_{k_x}u_{n,\mathbf{k}}|u_{m,\mathbf{k}}\rangle \langle u_{m,\mathbf{k}}|\partial_{k_y}u_{n,\mathbf{k}}\rangle -\langle\partial_{k_y}u_{n,\mathbf{k}}|u_{m,\mathbf{k}}\rangle \langle u_{m,\mathbf{k}}|\partial_{k_x}u_{n,\mathbf{k}}\rangle\right)\\
&=i\frac{e^2}{\hbar}\sum_{n\in\text{occ}}\sum_m \int_{\text{MBZ}} \frac{\text{d}^2\mathbf{k}}{4\pi^2} \left(\langle\partial_{k_x}u_{n,\mathbf{k}}|u_{m,\mathbf{k}}\rangle \langle u_{m,\mathbf{k}}|\partial_{k_y}u_{n,\mathbf{k}}\rangle -\langle\partial_{k_y}u_{n,\mathbf{k}}|u_{m,\mathbf{k}}\rangle \langle u_{m,\mathbf{k}}|\partial_{k_x}u_{n,\mathbf{k}}\rangle\right)\\
&=i\frac{e^2}{\hbar}\sum_{n\in\text{occ}} \int_{\text{MBZ}} \frac{\text{d}^2\mathbf{k}}{4\pi^2} \left(\langle\partial_{k_x}u_{n,\mathbf{k}}|\partial_{k_y}u_{n,\mathbf{k}}\rangle -\langle\partial_{k_y}u_{n,\mathbf{k}}|\partial_{k_x}u_{n,\mathbf{k}}\rangle\right)\\
&=\frac{e^2}{\hbar}\sum_{n\in\text{occ}} \int_{\text{MBZ}} \frac{\text{d}^2\mathbf{k}}{4\pi^2} \mathcal{F}^n_{xy}(\mathbf{k}) = t_\nu\frac{e^2}{h}\ ,
\end{split}
\end{equation}
where we have used the fact that $t_\nu$ in the Streda formula (Eq. (\ref{Seq-rho})) is the total Chern number of occupied bands, $\sum_m |u_{m,\mathbf{k}}\rangle \langle u_{m,\mathbf{k}}|$ is the identity matrix for the hatted Hamiltonian $\hat{H}(\mathbf{k})$ at magnetic BZ quasimomentum $\mathbf{k}$, and $\mathcal{F}^n_{xy}(\mathbf{k})$ is the Berry curvature of the $n$-th band. The corresponding U(1) Berry gauge field is 
\begin{equation}\label{Seq-berry}
\mathcal{A}^n(\mathbf{k})=i\langle u_{n,\mathbf{k}}|\partial_{\mathbf{k}} u_{n,\mathbf{k}}\rangle\ ,\qquad \mathcal{F}^n_{xy}(\mathbf{k})=\partial_x \mathcal{A}^n_y(\mathbf{k})-\partial_y \mathcal{A}^n_x(\mathbf{k})\ .
\end{equation}

\subsection{Quantized Lorentz susceptibility from the Kubo formula}

We now try to find a quantity that can be viewed as the momentum space dual of the Hall conductance. The Hall conductance gives a current density $j_x=\partial H/\partial_{k_x}$ in response to a uniform electric field $E_y=\text{d}k_y/\text{d}t$ where $\mathbf{k}$ is the canonical momentum. Therefore, a natural dual response can be defined by exchanging the roles of the real space position $\mathbf{r}$ and the momentum $\mathbf{k}$. Such a position-momentum dual response (dual to the Hall conductance response) corresponds to a force density (force per unit cell) $F_x=-\partial H/\partial x$ acting on the lattice in response to a position pumping, i.e., velocity $v_y=\text{d}y/\text{d}t$ of the system, which is nothing but the Lorentz force. 
Therefore, we name the coefficient $\gamma_{xy}$ in the response 
\begin{equation}\label{Seq-Lorentz-response}
F_x=\gamma_{xy}v_y
\end{equation}
as the Lorentz susceptibility.

We first identify the spatially uniform force operator $\hat{\bm{F}}$ representing the Lorentz force felt by the electrons as the lattice moves rigidly. Recall that an electron in a magnetic field is centered at the guiding center $\mathbf{R}$. In a translationally invariant state, if the guiding centers $\mathbf{R}$ of the electrons drift at velocity $\langle\frac{\text{d}\mathbf{R}}{\text{d}t}\rangle$, the electrons will feel a Lorentz force
\begin{equation}\label{Seq-Fvalue}
\langle \hat{\bm{F}}\rangle =B\hat{\mathbf{z}}\times\langle\frac{\text{d}\mathbf{R}}{\text{d}t}\rangle= -iB\hat{\mathbf{z}}\times \langle [\mathbf{R},H]\rangle =-i\langle [\bm{\Pi},H]\rangle -iB\hat{\mathbf{z}}\times\langle [\mathbf{r}, H]\rangle\ ,
\end{equation}
where in the last equality we have used the definition of guiding center $\mathbf{R}=\mathbf{r}-\ell^2\hat{\mathbf{z}}\times \bm{\Pi}$. For adiabatic processes, the electrons are in equilibrium, so the Lorentz force $\langle \hat{\bm{F}}\rangle$ the electrons felt has to be balanced by a force $-\langle \hat{\bm{F}}\rangle$ the lattice exerts on the electrons. According to Newton's third law, the electrons will then exert a force $\langle \hat{\bm{F}}\rangle$ on the lattice. Therefore, experimentally one could measure the force felt by the lattice to find the Lorentz force $\langle \hat{\bm{F}}\rangle$ felt by the electrons.

Since the force operator $\hat{\bm{F}}$ is spatially uniform and preserves the translation symmetry, it is diagonal in magnetic BZ quasimomentum $\mathbf{k}$. On the right hand side of Eq. (\ref{Seq-Fvalue}), the operator $[\bm{\Pi},H]$ is readily diagonal in magnetic BZ quasimomentum $\mathbf{k}$, since $\bm{\Pi}$ and hence $H$ both commute with the translation symmetry operators $\widetilde{T}_{q\bm{d}_1}=e^{-i q\ell^{-2}(\hat{\mathbf{z}}\times\mathbf{R})\cdot\bm{d}_1}$ and $\widetilde{T}_{\bm{d}_2}=e^{-i \ell^{-2}(\hat{\mathbf{z}}\times\mathbf{R})\cdot\bm{d}_2}$. The other operator $[\mathbf{r},H]$ is not diagonal in $\mathbf{k}$ (Ref. \cite{karplus1954}), but its diagonal part is simply the current operator $\hat{\bm{j}}$ as we discussed in Sec. \ref{sec-LS}A. Therefore, we conclude the force operator is given by
\begin{equation}\label{Seq-F}
\hat{\bm{F}}= -i[\bm{\Pi},H]+B\hat{\mathbf{z}}\times\hat{\bm{j}}\ ,
\end{equation}
which is diagonal in magnetic BZ quasimomentum $\mathbf{k}$. Note that here $H$ is the full Hamiltonian defined in Eq. (\ref{Seq-CmodelB}) for continuum models or in Eq. (\ref{Seq-HTB}) for tight-binding models (embedded in the continuum space).

In particular, if there is no periodic lattice potential for the electrons at all, $H$ will be solely a function of $\bm{\Pi}$ (i.e., the kinetic term of electrons in the free space), and one will have $\langle\hat{\bm{F}}\rangle =-iB\hat{\mathbf{z}}\times \langle[\mathbf{R},H]\rangle\equiv\mathbf{0}$ since $[\mathbf{R},\bm{\Pi}]=0$. Physically, this is because without a lattice potential, the electrons cannot feel a force in respond to the movement of the lattice and thus cannot exert a force on the lattice.

Similar to $\sigma_{xy}$ which is given by the Kubo formula of the spatially uniform current operator $\hat{\bm{j}}$, we can define the Lorentz susceptibility in Eq. (\ref{Seq-Lorentz-response}) by the Kubo formula of the spatially uniform force operator $\hat{\bm{F}}$ as
\begin{equation}\label{Seq-kuboLS}
\gamma_{xy}=-i\frac{\partial}{\partial\omega} \int_{-\infty}^{\infty} \text{d}\omega'\langle G_{\omega+\omega'}\hat{F}_{x} G_{\omega'}\hat{F}_{y}\rangle \ ,
\end{equation}
where $G_\omega$ is the Green's function defined in Eq. (\ref{Seq-Gomega}).

In the below, we first calculate the quantized value of the Lorentz susceptibility in Sec. \ref{Sec-lorentz-diophantine}, and then derive a dual formula for the Lorentz susceptibility analogous to the TKNN formula for Hall conductance in Sec. \ref{Sec-lorentz-TKNN}.

\subsubsection{Calculation of the Lorentz susceptibility from the Diophantine equation}\label{Sec-lorentz-diophantine}

By the expression of $\hat{\bm{F}}$ in Eq. (\ref{Seq-F}), we can rewrite the Lorentz susceptibility defined in Eq. (\ref{Seq-kuboLS}) as
\begin{equation}\label{Seq-kuboLS2}
\begin{split}
\gamma_{xy}=&-i e\Omega\sum_{n\in\text{occ},m\not\in\text{occ}}\int_{\text{MBZ}} \frac{\text{d}^2\mathbf{k}}{4\pi^2} \frac{1} {(\epsilon_{m,\mathbf{k}}-\epsilon_{n,\mathbf{k}})^2} \left[\langle \psi_{n,\mathbf{k}}|\hat{F}_{x}|\psi_{m,\mathbf{k}}\rangle \langle \psi_{m,\mathbf{k}}|\hat{F}_{y}|\psi_{n,\mathbf{k}}\rangle -(x\leftrightarrow y)\right] \\
=&-i e\Omega \sum_{n\in\text{occ},m\not\in\text{occ}} \frac{1} {(\epsilon_{m,\mathbf{k}}-\epsilon_{n,\mathbf{k}})^2} \int_{\text{MBZ}} \frac{\text{d}^2\mathbf{k}}{4\pi^2} \Big[-B^2 \Big(\langle \psi_{n,\mathbf{k}}|\hat{j}_{y}|\psi_{m,\mathbf{k}}\rangle \langle \psi_{m,\mathbf{k}}|\hat{j}_{x}|\psi_{n,\mathbf{k}}\rangle-(x\leftrightarrow y)\Big) \\
&-iB \Big(\langle \psi_{n,\mathbf{k}}|[\Pi_x,H]|\psi_{m,\mathbf{k}}\rangle \langle \psi_{m,\mathbf{k}}|\hat{j}_{x}|\psi_{n,\mathbf{k}}\rangle - \langle \psi_{n,\mathbf{k}}|\hat{j}_{x}|\psi_{m,\mathbf{k}}\rangle \langle \psi_{m,\mathbf{k}}|[\Pi_x,H]|\psi_{n,\mathbf{k}}\rangle \\
&-\langle \psi_{n,\mathbf{k}}|[\Pi_y,H]|\psi_{m,\mathbf{k}}\rangle \langle \psi_{m,\mathbf{k}}|\hat{j}_{y}|\psi_{n,\mathbf{k}}\rangle + \langle \psi_{n,\mathbf{k}}|\hat{j}_{y}|\psi_{m,\mathbf{k}}\rangle \langle \psi_{m,\mathbf{k}}|[\Pi_y,H]|\psi_{n,\mathbf{k}}\rangle\Big)\\
&- \langle \psi_{n,\mathbf{k}}|[\Pi_x,H]|\psi_{m,\mathbf{k}}\rangle \langle \psi_{m,\mathbf{k}}|[\Pi_y,H]|\psi_{n,\mathbf{k}}\rangle +
\langle \psi_{n,\mathbf{k}}|[\Pi_x,H]|\psi_{m,\mathbf{k}}\rangle \langle \psi_{m,\mathbf{k}}|[\Pi_y,H]|\psi_{n,\mathbf{k}}\rangle \Big]\ , \\
\end{split}
\end{equation}
where the integral of $\mathbf{k}$ runs over one magnetic BZ (MBZ). Note that so far everything is expressed in the eigenbasis $|\psi_{m,\mathbf{k}}\rangle$ of the full Hamiltonian $H$. The first two terms of correlations of $\hat{j}_x$ and $\hat{j}_y$ in Eq. (\ref{Seq-kuboLS2}) simply yield the Berry curvature, which can be calculated by transforming from the eigenbasis $|\psi_{m,\mathbf{k}}\rangle$ of the full Hamiltonian $H$ into the reduced basis $|u_{m,\mathbf{k}}\rangle$ of the reduced magnetic BZ momentum space Hamiltonian $\hat{H}(\mathbf{k})$ as we have shown in Sec. \ref{sec-LS}A. We now examine the remaining terms of Eq. (\ref{Seq-kuboLS2}) (which can be calculated simply in the eigenbasis $|\psi_{m,\mathbf{k}}\rangle$ of the full Hamiltonian $H$). Since $[\bm{\Pi},H]$ is diagonal in $\mathbf{k}$, and $\hat{\bm{j}}$ is the part of operator $-i[\mathbf{r},H]$ that is diagonal in $\mathbf{k}$, we can rewrite the terms of correlations of $[\Pi_x,H]$ and $\hat{j}_{x}$ as
\begin{equation}
\begin{split}
&\sum_{n\in\text{occ},m\not\in\text{occ}}\frac{\langle \psi_{n,\mathbf{k}}|[\Pi_x,H]|\psi_{m,\mathbf{k}}\rangle \langle \psi_{m,\mathbf{k}}|\hat{j}_{x}|\psi_{n,\mathbf{k}}\rangle } {(\epsilon_{m,\mathbf{k}}-\epsilon_{n,\mathbf{k}})^2}- \Big([\Pi_x,H]\leftrightarrow \hat{j}_{x}\Big) \\
&=\sum_{n\in\text{occ}}\sum_{m\not\in\text{occ},\mathbf{k}'}\frac{\langle \psi_{n,\mathbf{k}}|[\Pi_x,H]|\psi_{m,\mathbf{k}'}\rangle \langle \psi_{m,\mathbf{k}'}|(-i[x,H])|\psi_{n,\mathbf{k}}\rangle} {(\epsilon_{m,\mathbf{k}'}-\epsilon_{n,\mathbf{k}})^2} - \Big([\Pi_x,H]\leftrightarrow -i[x,H]\Big)\\
&=\sum_{n\in\text{occ}}\sum_{m\not\in\text{occ},\mathbf{k}'}\frac{ i(\epsilon_{m,\mathbf{k}'}-\epsilon_{n,\mathbf{k}})^2 \langle \psi_{n,\mathbf{k}}|\Pi_x|\psi_{m,\mathbf{k}'}\rangle \langle \psi_{m,\mathbf{k}'}|x|\psi_{n,\mathbf{k}}\rangle} {(\epsilon_{m,\mathbf{k}'}-\epsilon_{n,\mathbf{k}})^2} -\Big(\Pi_x\leftrightarrow x\Big) \\
&=\sum_{n\in\text{occ}}\sum_{m,\mathbf{k}'}i \langle \psi_{n,\mathbf{k}}|\Pi_x|\psi_{m,\mathbf{k}'}\rangle \langle \psi_{m,\mathbf{k}'}|x|\psi_{n,\mathbf{k}}\rangle -\Big(\Pi_x\leftrightarrow x\Big) \\
&=\sum_{n\in\text{occ}}i\langle \psi_{n,\mathbf{k}}|[\Pi_x, x]|\psi_{n,\mathbf{k}}\rangle\ ,
\end{split}
\end{equation}
where we have used the fact that $\sum_{m,\mathbf{k}'} |\psi_{m,\mathbf{k}'}\rangle \langle \psi_{m,\mathbf{k}'}|$ is identity for the entire Hilbert space. A similar equality holds for the terms of correlations of $[\Pi_y,H]$ and $\hat{j}_{y}$. Lastly, by the same derivation technique, we have
\begin{equation}
\sum_{n\in\text{occ},m\not\in\text{occ}}\frac{\langle \psi_{n,\mathbf{k}}|[\Pi_x,H]|\psi_{m,\mathbf{k}}\rangle \langle \psi_{m,\mathbf{k}}|[\Pi_y,H]|\psi_{n,\mathbf{k}}\rangle} {(\epsilon_{m,\mathbf{k}}-\epsilon_{n,\mathbf{k}})^2} =- \sum_{n\in\text{occ}} \langle \psi_{n,\mathbf{k}}|[\Pi_x, \Pi_y]|\psi_{n,\mathbf{k}}\rangle\ .
\end{equation}
Therefore, we can rewrite Eq. (\ref{Seq-kuboLS2}) as
\begin{equation}\label{Seq-gamma-xy-laststep}
\begin{split}
\gamma_{xy}=&e\Omega\sum_{n\in\text{occ}}\int_{\text{MBZ}} \frac{\text{d}^2\mathbf{k}}{4\pi^2} \Big(-B^2\mathcal{F}^n_{xy}(\mathbf{k}) -i B\langle \psi_{n,\mathbf{k}}|[\Pi_x, x]|\psi_{n,\mathbf{k}}\rangle +i B\langle \psi_{n,\mathbf{k}}|[\Pi_y, y]|\psi_{n,\mathbf{k}}\rangle -i \langle \psi_{n,\mathbf{k}}|[\Pi_x, \Pi_y]|\psi_{n,\mathbf{k}}\rangle \Big)\\
=&e\Omega\sum_{n\in\text{occ}}\int_{\text{MBZ}} \frac{\text{d}^2\mathbf{k}}{4\pi^2} \Big(-B^2\mathcal{F}^n_{xy}(\mathbf{k}) +B\Big)=eB \left(\rho-t_\nu\frac{\varphi}{2\pi}\right)=eB s_\nu\ ,
\end{split}
\end{equation}
where we have used commutation relations $[\Pi_x, x]=[\Pi_y, y]=-i$, $[\Pi_x, \Pi_y]=i\ell^{-2}=iB$, and the Diophantine equation rewritten in the form of Eq. (\ref{Seq-rho}). As we discussed in both the main text and the supplementary Sec. \ref{sec-Diophantine}, $\rho=\nu/q$ is the number of electrons per zero-magnetic-field unit cell. This proves that the Lorentz susceptibility is quantized in terms of $s_\nu$, and corresponds to a Lorentz force per original (zero field) unit cell
\begin{equation}
F_x=\gamma_{xy}v_y=eBs_\nu v_y\ .
\end{equation}

\subsubsection{A formula for Lorentz susceptibility similar to the TKNN formula for Hall conductance}\label{Sec-lorentz-TKNN}
In Eq. (\ref{Seq-dHdk}), we know the matrix elements of the spatially uniform current operator $\hat{\bm{j}}$ can be calculated by $\frac{\partial \hat{H}(\mathbf{k})}{\partial\mathbf{k}}$, where $\hat{H}(\mathbf{k})$ is the reduced Hamiltonian defined in Eq. (\ref{Seq-Hhat}). Here for the force operator $\hat{\bm{F}}$, we can find a similar expression as follows. First, we define another reduced Hamiltonian
\begin{equation}\label{Seq-Htildehat}
\hat{\widetilde{H}}(\bm{d})=P_\mathbf{0}( \widetilde{T}_{-\bm{d}} H \widetilde{T}_{\bm{d}})P_\mathbf{0}=P_\mathbf{0}( e^{i \ell^{-2}(\hat{\mathbf{z}}\times\mathbf{R})\cdot\bm{d}} H e^{-i \ell^{-2}(\hat{\mathbf{z}}\times\mathbf{R})\cdot\bm{d}})P_\mathbf{0}\ ,\qquad
P_\mathbf{0}=\frac{q}{N_\Omega}\sum_{j_1,j_2\in\mathbb{Z}} \widetilde{T}_{q j_1\bm{d}_1+j_2\bm{d}_2}\ ,
\end{equation}
where $\widetilde{T}_{\bm{d}}=\widetilde{T}_{-\bm{d}}^\dag=e^{-i \ell^{-2}(\hat{\mathbf{z}}\times\mathbf{R})\cdot\bm{d}}$ (similar to $e^{i\mathbf{k\cdot r}}$) is the translation operator generated by guiding center $\mathbf{R}$, and $P_{\mathbf{0}}$ is still the projection operator into the magnetic translationally invariant subspace as defined in Eq. (\ref{Seq-Hhat}). Using the Bloch eigenstates $|\psi_{n,\mathbf{k}}\rangle$ of the full Hamiltonian $H$, we can define a set of states
\begin{equation}\label{Seq-define-w}
|w_{n,\bm{d}}\rangle=\widetilde{T}_{-\bm{d}}|\psi_{n,\ell^{-2}\hat{\mathbf{z}}\times\bm{d}}\rangle=e^{i \ell^{-2}(\hat{\mathbf{z}}\times\mathbf{R})\cdot\bm{d}}|\psi_{n,\ell^{-2}\hat{\mathbf{z}}\times\bm{d}}\rangle\ ,
\end{equation}
where $|\psi_{n,\ell^{-2}\hat{\mathbf{z}}\times\bm{d}}\rangle$ is the Bloch wave function at momentum $\mathbf{k}=\ell^{-2}\hat{\mathbf{z}}\times\bm{d}$. More discussions and an explicit example of the Hamiltonian $\hat{\widetilde{H}}(\bm{d})$ and wavefunction $|w_{n,\bm{d}}\rangle$ are given in Sec. \ref{sec:w-example}. 
Making use of the fact that $\widetilde{T}_{\bm{d}}\widetilde{T}_{\bm{d}'} =e^{-i\ell^{-2}\hat{\mathbf{z}}\cdot(\bm{d}\times\bm{d}')} \widetilde{T}_{\bm{d}'}\widetilde{T}_{\bm{d}}=e^{-i\ell^{-2}\hat{\mathbf{z}}\cdot(\bm{d}\times\bm{d}')/2}\widetilde{T}_{\bm{d}+\bm{d}'}$ (since $[R_x,R_y]=-i\ell^2$), we have
\begin{equation}
\begin{split}
\widetilde{T}_{q\bm{d}_1}|w_{n,\bm{d}}\rangle&=\widetilde{T}_{q\bm{d}_1} \widetilde{T}_{-\bm{d}}|\psi_{n,\ell^{-2}\hat{\mathbf{z}}\times\bm{d}}\rangle=e^{i\ell^{-2}\hat{\mathbf{z}}\cdot(q\bm{d}_1\times\bm{d})} \widetilde{T}_{-\bm{d}} \widetilde{T}_{q\bm{d}_1}|\psi_{n,\ell^{-2}\hat{\mathbf{z}}\times\bm{d}}\rangle \\
&=e^{-iq\bm{d}_1\cdot (\ell^{-2}\hat{\mathbf{z}}\times\bm{d})}  \widetilde{T}_{-\bm{d}} e^{iq\bm{d}_1\cdot (\ell^{-2}\hat{\mathbf{z}}\times\bm{d})}|\psi_{n,\ell^{-2}\hat{\mathbf{z}}\times\bm{d}}\rangle=|w_{n,\bm{d}}\rangle\ , \\
\widetilde{T}_{\bm{d}_2}|w_{n,\bm{d}}\rangle&=\widetilde{T}_{\bm{d}_2} \widetilde{T}_{-\bm{d}}|\psi_{n,\ell^{-2}\hat{\mathbf{z}}\times\bm{d}}\rangle=e^{i\ell^{-2}\hat{\mathbf{z}}\cdot(\bm{d}_2\times\bm{d})} \widetilde{T}_{-\bm{d}} \widetilde{T}_{\bm{d}_2}|\psi_{n,\ell^{-2}\hat{\mathbf{z}}\times\bm{d}}\rangle \\
&=e^{-i\bm{d}_2\cdot (\ell^{-2}\hat{\mathbf{z}}\times\bm{d})}  \widetilde{T}_{-\bm{d}} e^{i\bm{d}_2\cdot (\ell^{-2}\hat{\mathbf{z}}\times\bm{d})}|\psi_{n,\ell^{-2}\hat{\mathbf{z}}\times\bm{d}}\rangle=|w_{n,\bm{d}}\rangle\ , \\
\end{split}
\end{equation}
where we have used the definition of Bloch momentum in Eq. (\ref{Seq-quasi}). Therefore, we find the states $|w_{n,\bm{d}}\rangle$ are translationally invariant, namely, 
\begin{equation}
P_{\mathbf{0}}|w_{n,\bm{d}}\rangle=|w_{n,\bm{d}}\rangle\ .
\end{equation}
It is then easy to see that the eigenstates of the reduced Hamiltonian $\hat{\widetilde{H}}(\bm{d})$ are given by $|w_{n,\bm{d}}\rangle$, namely,
\begin{equation}
\begin{split}
&\hat{\widetilde{H}}(\bm{d})|w_{n,\bm{d}}\rangle=P_\mathbf{0}\widetilde{T}_{-\bm{d}} H \widetilde{T}_{\bm{d}} |w_{n,\bm{d}}\rangle =P_\mathbf{0}\widetilde{T}_{-\bm{d}} H |\psi_{n,\ell^{-2}\hat{\mathbf{z}}\times\bm{d}}\rangle \\
=&P_\mathbf{0}\widetilde{T}_{-\bm{d}}\epsilon_{n,\ell^{-2}\hat{\mathbf{z}}\times\bm{d}}|\psi_{n,\ell^{-2}\hat{\mathbf{z}}\times\bm{d}}\rangle =\epsilon_{n,\ell^{-2}\hat{\mathbf{z}}\times\bm{d}}|w_{n,\ell^{-2}\hat{\mathbf{z}}\times\bm{d}}\rangle\ ,
\end{split}
\end{equation}
where $\epsilon_{n,\ell^{-2}\hat{\mathbf{z}}\times\bm{d}}$ are the eigenenergies at momentum $\mathbf{k}=\ell^{-2}\hat{\mathbf{z}}\times\bm{d}$. 
In particular, we can further prove that the force operator satisfies (recall Eq. (\ref{Seq-Fvalue}))
\begin{equation}\label{Seq-dHdd}
\begin{split}
&\langle \psi_{n,\ell^{-2}\hat{\mathbf{z}}\times\bm{d}}|\hat{\bm{F}}|\psi_{n,\ell^{-2}\hat{\mathbf{z}}\times\bm{d}}\rangle=-i\langle \psi_{n,\ell^{-2}\hat{\mathbf{z}}\times\bm{d}}|B[\hat{\mathbf{z}}\times\mathbf{R},H]|\psi_{n,\ell^{-2}\hat{\mathbf{z}}\times\bm{d}}\rangle \\
&= -i\langle w_{n,\bm{d}}|\ell^{-2}e^{i \ell^{-2}(\hat{\mathbf{z}}\times\mathbf{R})\cdot\bm{d}}[\hat{\mathbf{z}}\times\mathbf{R},H]e^{-i \ell^{-2}(\hat{\mathbf{z}}\times\mathbf{R})\cdot\bm{d}}|w_{n,\bm{d}}\rangle \\
&= -\langle w_{n,\bm{d}}|\frac{\partial}{\partial \bm{d}}\left(P_\mathbf{0} e^{i \ell^{-2}(\hat{\mathbf{z}}\times\mathbf{R})\cdot\bm{d}}He^{-i \ell^{-2}(\hat{\mathbf{z}}\times\mathbf{R})\cdot\bm{d}}P_0\right)|w_{n,\bm{d}}\rangle \\
&= -\langle w_{n,\bm{d}}|\frac{\partial \hat{\widetilde{H}}(\bm{d})}{\partial \bm{d}}  |w_{n,\bm{d}}\rangle .\\
\end{split}
\end{equation}
Eq. (\ref{Seq-dHdd}) is then the dual analog to Eq. (\ref{Seq-dHdk}) (see Sec. \ref{sec:w-example} for an explicit example). 

We then consider the periodicity of the wave function $|w_{n,\bm{d}}\rangle$ in the space of parameter $\bm{d}$. We first note that by the definition of the projection operator $P_{\mathbf{0}}$ in Eq. (\ref{Seq-Htildehat}), one has
\begin{equation}
P_\mathbf{0}\widetilde{T}_{\bm{d}_2/p}=\frac{q}{N_\Omega} \widetilde{T}_{\bm{d}_2/p} \sum_{j_1,j_2\in\mathbb{Z}} e^{-i\ell^{-2}\hat{\mathbf{z}}\cdot[(qj_1\bm{d}_1+j_2\bm{d}_2)\times(\bm{d}_2/p)]}  \widetilde{T}_{q j_1\bm{d}_1+j_2\bm{d}_2}=\widetilde{T}_{\bm{d}_2/p} P_\mathbf{0}\ ,
\end{equation}
where we have used the fact that $\ell^{-2}\hat{\mathbf{z}}\cdot(\bm{d}_1\times\bm{d}_2)=2\pi p/q$. Therefore, we find 
\begin{equation}
\begin{split}
&\hat{\widetilde{H}}(\bm{d}+\bm{d}_2/p)=P_\mathbf{0}( \widetilde{T}_{-\bm{d}-\bm{d}_2/p} H \widetilde{T}_{\bm{d}+\bm{d}_2/p})P_\mathbf{0}=P_\mathbf{0} \widetilde{T}_{-\bm{d}_2/p} \widetilde{T}_{-\bm{d}}  H \widetilde{T}_{\bm{d}} \widetilde{T}_{\bm{d}_2/p} P_\mathbf{0}= \widetilde{T}_{-\bm{d}_2/p} (P_\mathbf{0} \widetilde{T}_{-\bm{d}}  H \widetilde{T}_{\bm{d}} P_\mathbf{0}) \widetilde{T}_{\bm{d}_2/p} \\
&= \widetilde{T}_{-\bm{d}_2/p} \hat{\widetilde{H}}(\bm{d})  \widetilde{T}_{\bm{d}_2/p}\ ,
\end{split}
\end{equation}
namely, the reduced Hamiltonian $\hat{\widetilde{H}}(\bm{d})$ and $\hat{\widetilde{H}}(\bm{d}+\bm{d}_2/p)$ differ by a $\bm{d}$-independent unitary transformation $\widetilde{T}_{\bm{d}_2/p}$. Moreover, the lattice translation symmetry tells us that 
\begin{equation}
\hat{\widetilde{H}}(\bm{d}+\bm{d}_1) = P_\mathbf{0}( \widetilde{T}_{-\bm{d}-\bm{d}_1} H \widetilde{T}_{\bm{d}+\bm{d}_1})P_\mathbf{0}=P_\mathbf{0}  \widetilde{T}_{-\bm{d}}\widetilde{T}_{-\bm{d}_1}  H \widetilde{T}_{\bm{d}_1} \widetilde{T}_{\bm{d}}  P_\mathbf{0}= P_\mathbf{0} \widetilde{T}_{-\bm{d}}  H \widetilde{T}_{\bm{d}} P_\mathbf{0} =\hat{\widetilde{H}}(\bm{d}), 
\end{equation}
where we have used the fact that the full Hamiltonian satisfies $[\widetilde{T}_{\bm{d}_1}, H]=0$ (translational symmetry), as given in Eq. (\ref{Seq-Ttilde}). 
Therefore, we conclude that the eigenstates of $\hat{\widetilde{H}}(\bm{d})$ satisfies
\begin{equation}
|w_{n,\bm{d}+\bm{d}_2/p}\rangle=e^{i\varphi_2(\bm{d})} \widetilde{T}_{-\bm{d}_2/p} |w_{n,\bm{d}}\rangle\ ,\qquad |w_{n,\bm{d}+\bm{d}_1}\rangle=e^{i\varphi_1(\bm{d})} |w_{n,\bm{d}}\rangle\ ,
\end{equation}
where $\varphi_i(\bm{d})$ are some $\bm{d}$-dependent phases. This effectively defines a dual "Brillouin zone" $\Omega_M$ in the $\bm{d}$ space with periods $\bm{d}_1$ and $\bm{d}_2/p$ (which is a torus). The unitary transformation $\widetilde{T}_{-\bm{d}_2/p}$ serves as an embedding matrix for mapping state $|w_{n,\bm{d}}\rangle$ to $|w_{n,\bm{d}+\bm{d}_2/p}\rangle$. Note that $\Omega_M$ is simple $1/p$ fraction of the zero magnetic field unit cell. Furthermore, this allows us to define a U(1) Berry gauge field and its field strength for band $n$ on the dual "Brillouin zone" $\Omega_m$ (which is a closed manifold):
\begin{equation}\label{Seq-new-berry}
\widetilde{\bm{a}}^n(\bm{d})=-i\langle w_{n,\bm{d}}|\partial_{\bm{d}}w_{n,\bm{d}}\rangle\ , \qquad  \widetilde{f}^n_{xy}(\bm{d})=\partial_x \widetilde{\bm{a}}^n_y(\bm{d})-\partial_y \widetilde{\bm{a}}^n_x(\bm{d})\ .
\end{equation}
In the following, we will show that the Lorentz susceptibility is given by a formula similar to the TKNN formula \cite{tknn1982}, but with the usual Berry gauge field in Eq. (\ref{Seq-berry}) replaced by the new gauge field we defined in Eq. (\ref{Seq-new-berry}).

We first note that the Lorentz susceptibility in Eq. (\ref{Seq-kuboLS2}) can be re-expressed in the $\bm{d}$ space (by variable substitution $\mathbf{k}=\ell^{-2}\hat{\mathbf{z}}\times\bm{d}$) as
\begin{equation}\label{Seq-kuboLS3}
\begin{split}
&\gamma_{xy}=-i e\Omega\sum_{n\in\text{occ},m\not\in\text{occ}}\int_{\text{MBZ}} \frac{\text{d}^2\mathbf{k}}{4\pi^2} \frac{1} {(\epsilon_{m,\mathbf{k}}-\epsilon_{n,\mathbf{k}})^2} \left[\langle \psi_{n,\mathbf{k}}|\hat{F}_{x}|\psi_{m,\mathbf{k}}\rangle \langle \psi_{m,\mathbf{k}}|\hat{F}_{y}|\psi_{n,\mathbf{k}}\rangle -(x\leftrightarrow y)\right] \\
=&-i \frac{eB p}{2\pi q}\sum_{n\in\text{occ},m\not\in\text{occ}}\int_{\bm{d}\in\mathcal{M}_{\{q\bm{d}_1/p,\bm{d}_2/p\}}} \text{d}^2\bm{d} \frac{\left[\langle \psi_{n,\ell^{-2}\hat{\mathbf{z}}\times\bm{d}}|\hat{F}_{x}|\psi_{m,\ell^{-2}\hat{\mathbf{z}}\times\bm{d}}\rangle \langle \psi_{m,\ell^{-2}\hat{\mathbf{z}}\times\bm{d}}|\hat{F}_{y}|\psi_{n,\ell^{-2}\hat{\mathbf{z}}\times\bm{d}}\rangle -(x\leftrightarrow y)\right]} {(\epsilon_{m,\ell^{-2}\hat{\mathbf{z}}\times\bm{d}}-\epsilon_{n,\ell^{-2}\hat{\mathbf{z}}\times\bm{d}})^2} , \\
\end{split}
\end{equation}
where $\Omega$ is the zero magnetic field unit cell area, MBZ stands for the magnetic BZ spanned by $\bm{g}_1/q$ and $\bm{g}_2$, and $\mathcal{M}_{\{q\bm{d}_1/p,\bm{d}_2/p\}}$ denotes a torus with periods $q\bm{d}_1/p$ and $\bm{d}_2/p$, which is nothing but the MBZ rotated by $\pi/2$ and size scaled by a factor $\ell^2$ (this can be seen by noting that $\bm{d}_2=\frac{\Omega}{2\pi}\hat{\mathbf{z}}\times\bm{g}_1$, and $\bm{d}_1=-\frac{\Omega}{2\pi}\hat{\mathbf{z}}\times\bm{g}_2$, and $B\Omega=\ell^{-2}\Omega=2\pi p/q$). 

Then, by Eq. (\ref{Seq-psi-Td1}) we know that $\widetilde{T}_{\bm{d}_1} |\psi_{n,\mathbf{k}}\rangle$ is the same as $|\psi_{n,\mathbf{k}-p\bm{g}_2/q}\rangle$ up to a phase factor. Since $p$ and $q$ are coprime, there exists an integer $j$ so that $jp\equiv -1(\text{mod}\ q)$, and thus $\widetilde{T}_{j\bm{d}_1} |\psi_{n,\mathbf{k}}\rangle=e^{i\alpha_\mathbf{k}} |\psi_{n,\mathbf{k}+jp\bm{g}_2/q}\rangle=e^{i\alpha_\mathbf{k}'}|\psi_{n,\mathbf{k}-\bm{g}_2/q}\rangle=e^{i\alpha_\mathbf{k}'}|\psi_{n,\mathbf{k}+\ell^{-2}\hat{\mathbf{z}}\times\bm{d}_1/p}\rangle$, where $\alpha_\mathbf{k}$ and $\alpha_\mathbf{k}'$ are some phase factors. Making use of the fact that $[\hat{\bm{F}}, \widetilde{T}_{\bm{d}_1}]=0$ (translationally invariant), we then have
\begin{equation}
\langle \psi_{n,\ell^{-2}\hat{\mathbf{z}}\times\bm{d}}|\hat{\bm{F}}|\psi_{m,\ell^{-2}\hat{\mathbf{z}}\times\bm{d}}\rangle = \langle \psi_{n,\ell^{-2}\hat{\mathbf{z}}\times\bm{d}}| \widetilde{T}_{-j\bm{d}_1}\hat{\bm{F}}\widetilde{T}_{j\bm{d}_1} |\psi_{m,\ell^{-2}\hat{\mathbf{z}}\times\bm{d}}\rangle =\langle \psi_{n,\ell^{-2}\hat{\mathbf{z}}\times(\bm{d}+\bm{d}_1/p)}| \hat{\bm{F}} |\psi_{m,\ell^{-2}\hat{\mathbf{z}}\times(\bm{d}+\bm{d}_1/p)}\rangle\ ,
\end{equation}
namely, the quantity $\langle \psi_{n,\ell^{-2}\hat{\mathbf{z}}\times\bm{d}}|\hat{\bm{F}}|\psi_{m,\ell^{-2}\hat{\mathbf{z}}\times\bm{d}}\rangle$ invariant under displacement $\bm{d}\rightarrow \bm{d}+\bm{d}_1/p$. Therefore, the integral in region $\mathcal{M}_{\{q\bm{d}_1/p,\bm{d}_2/p\}}$ in Eq. (\ref{Seq-kuboLS3}) is equal to the same integral in a region $\mathcal{M}_{\{\bm{d}_1,\bm{d}_2/p\}}$ times a global factor $q/p$. Note that region $\mathcal{M}_{\{\bm{d}_1,\bm{d}_2/p\}}$ is nothing but the torus dual "Brillouin zone" $\Omega_M$ on which we defined the U(1) Berry gauge field in Eq. (\ref{Seq-new-berry}). We can then rewrite the Lorentz susceptibility in Eq. (\ref{Seq-kuboLS3}) as
\begin{equation}\label{Seq-kuboHf}
\begin{split}
\gamma_{xy}=&-i \frac{eB}{2\pi}\sum_{n\in\text{occ},m\not\in\text{occ}}\int_{\bm{d}\in\Omega_M} \text{d}^2\bm{d} \frac{\left[\langle \psi_{n,\ell^{-2}\hat{\mathbf{z}}\times\bm{d}}|\hat{F}_{x}|\psi_{m,\ell^{-2}\hat{\mathbf{z}}\times\bm{d}}\rangle \langle \psi_{m,\ell^{-2}\hat{\mathbf{z}}\times\bm{d}}|\hat{F}_{y}|\psi_{n,\ell^{-2}\hat{\mathbf{z}}\times\bm{d}}\rangle -(x\leftrightarrow y)\right]} {(\epsilon_{m,\ell^{-2}\hat{\mathbf{z}}\times\bm{d}}-\epsilon_{n,\ell^{-2}\hat{\mathbf{z}}\times\bm{d}})^2} \\
=&-i \frac{eB}{2\pi}\sum_{n\in\text{occ},m\not\in\text{occ}}\int_{\bm{d}\in\Omega_M} \text{d}^2\bm{d} \frac{\left[\langle \psi_{n,\ell^{-2}\hat{\mathbf{z}}\times\bm{d}}|\frac{\partial \hat{\widetilde{H}}(\bm{d})}{\partial d_x}|\psi_{m,\ell^{-2}\hat{\mathbf{z}}\times\bm{d}}\rangle \langle \psi_{m,\ell^{-2}\hat{\mathbf{z}}\times\bm{d}}|\frac{\partial \hat{\widetilde{H}}(\bm{d})}{\partial d_y}|\psi_{n,\ell^{-2}\hat{\mathbf{z}}\times\bm{d}}\rangle -(x\leftrightarrow y)\right]} {(\epsilon_{m,\ell^{-2}\hat{\mathbf{z}}\times\bm{d}}-\epsilon_{n,\ell^{-2}\hat{\mathbf{z}}\times\bm{d}})^2} ,
\end{split}
\end{equation}
where we have used Eq. (\ref{Seq-dHdd}). Then, by noting that 
\begin{equation}\label{Seq-dHdd1}
\begin{split}
&\langle w_{m,\bm{d}}|\frac{\partial \hat{\widetilde{H}}(\bm{d})}{\partial d_i}|w_{n,\bm{d}}\rangle= \partial_{d_i}(\langle w_{m,\bm{d}}|\hat{\widetilde{H}}(\bm{d})|w_{n,\bm{d}}\rangle)-\langle \partial_{d_i}w_{m,\bm{d}}|\hat{\widetilde{H}}(\bm{d})|w_{n,\bm{d}}\rangle -\langle w_{m,\bm{d}}|\hat{\widetilde{H}}(\bm{d})|\partial_{d_i}w_{n,\bm{d}}\rangle \\
&= 0- \epsilon_{n,\ell^{-2}\hat{\mathbf{z}}\times\bm{d}} \langle \partial_{d_i}w_{m,d}|w_{n,\bm{d}}\rangle - \epsilon_{m,\ell^{-2}\hat{\mathbf{z}}\times\bm{d}}\langle w_{m,\bm{d}}|\partial_{d_i}w_{n,\bm{d}}\rangle \\
&= (\epsilon_{n,\ell^{-2}\hat{\mathbf{z}}\times\bm{d}}-\epsilon_{m,\ell^{-2}\hat{\mathbf{z}}\times\bm{d}})\langle w_{m,\bm{d}}|\partial_{d_i}w_{n,\bm{d}}\rangle 
=(\epsilon_{m,\ell^{-2}\hat{\mathbf{z}}\times\bm{d}}-\epsilon_{n,\ell^{-2}\hat{\mathbf{z}}\times\bm{d}})\langle \partial_{k_i} w_{m,\bm{d}}|w_{n,\bm{d}}\rangle\ ,
\end{split}
\end{equation}
we can then derive the following:
\begin{equation}\label{Seq-kuboHf1}
\begin{split}
\gamma_{xy}=&-i \frac{eB}{2\pi}\sum_{n\in\text{occ},m\not\in\text{occ}}\int_{\bm{d}\in\Omega_M} \text{d}^2\bm{d} \left(\langle\partial_{d_x}w_{n,\bm{d}}|w_{m,\bm{d}}\rangle \langle w_{m,\bm{d}}|\partial_{d_y}w_{n,\bm{d}}\rangle -\langle\partial_{d_y}w_{n,\bm{d}}|w_{m,\bm{d}}\rangle \langle w_{m,\bm{d}}|\partial_{k_x}w_{n,\bm{d}}\rangle\right) \\
=&-i \frac{eB}{2\pi}\sum_{n\in\text{occ},m}\int_{\bm{d}\in\Omega_M} \text{d}^2\bm{d} \left(\langle\partial_{d_x}w_{n,\bm{d}}|w_{m,\bm{d}}\rangle \langle w_{m,\bm{d}}|\partial_{d_y}w_{n,\bm{d}}\rangle -\langle\partial_{d_y}w_{n,\bm{d}}|w_{m,\bm{d}}\rangle \langle w_{m,\bm{d}}|\partial_{d_x}w_{n,\bm{d}}\rangle\right) \\
=&-i \frac{eB}{2\pi}\sum_{n\in\text{occ}}\int_{\bm{d}\in\Omega_M} \text{d}^2\bm{d} \left(\langle\partial_{d_x}w_{n,\bm{d}}|\partial_{d_y}w_{n,\bm{d}}\rangle -\langle\partial_{d_y}w_{n,\bm{d}}|\partial_{d_x}w_{n,\bm{d}}\rangle\right) \\
=& eB\sum_{n\in\text{occ}}\int_{\bm{d}\in\Omega_M} \frac{\text{d}^2\bm{d}}{2\pi} \widetilde{f}^n_{xy}(\bm{d})\ ,\\
\end{split}
\end{equation}
where $\widetilde{f}^n_{xy}(\bm{d})$ is the new Berry curvature in the $\bm{d}$ space we defined in Eq. (\ref{Seq-new-berry}). This is then a dual analogy to the TKNN formula. In particular, our results for $\gamma_{xy}$ in Sec. \ref{Sec-lorentz-diophantine} implies
\begin{equation}\label{Seq-dual-Chern}
\sum_{n\in\text{occ}}\int_{\bm{d}\in\Omega_M} \frac{\text{d}^2\bm{d} }{2\pi}\widetilde{f}^n_{xy}(\bm{d})=s_\nu\ .
\end{equation}
Therefore, $s_\nu$ can be viewed as a dual Chern number, where the integral is within a torus ``dual magnetic Brillouin zone" $\Omega_M$ with periods $\bm{d}_1$ and $\bm{d}_2/p$.

--- Eq. (\ref{Seq-dual-Chern}) can also be proved by the gauge choices of the Bloch wave functions in Eqs.  (\ref{Seq-bloch}) and (\ref{Seq-psi-Td1}) (which we employed to prove the Diophantine equation), namely, one can fix $|\psi_{n,\mathbf{k}+\bm{g}_1/q}\rangle=|\psi_{n,\mathbf{k}}\rangle$, 
$|\psi_{n,\mathbf{k}+\bm{g}_2}\rangle=e^{-i\sigma q\bm{d_1}\cdot\mathbf{k}}|\psi_{n,\mathbf{k}}\rangle$ and $\widetilde{T}_{\bm{d}_1} |\psi_{n,\mathbf{k}}\rangle=e^{imq\bm{d}_1\cdot\mathbf{k}}|\psi_{n,\mathbf{k}+p\bm{g}_2/q}\rangle$ for a Hofstadter band $n$ between the $(\nu-1)$-th gap and the $\nu$-th gap, where $\sigma=t_\nu-t_{\nu-1}$ and $m=s_\nu-s_{\nu-1}$. Then by Eq. (\ref{Seq-define-w}), we find
\begin{equation}
\begin{split}
&|w_{n,\bm{d}+\bm{d}_2/p}\rangle=\widetilde{T}_{-\bm{d}-\bm{d}_2/p}|\psi_{n,\ell^{-2}\hat{\mathbf{z}}\times(\bm{d}+\bm{d}_2/p)}\rangle= e^{i\ell^{-2}\hat{\mathbf{z}}\cdot(\bm{d}_2\times\bm{d})/2p}\widetilde{T}_{-\bm{d}_2/p} \widetilde{T}_{-\bm{d}} |\psi_{n,\ell^{-2}\hat{\mathbf{z}}\times\bm{d}-\bm{g}_1/q}\rangle \\
=& e^{i\ell^{-2}\hat{\mathbf{z}}\cdot(\bm{d}_2\times\bm{d})/2p}\widetilde{T}_{-\bm{d}_2/p} \widetilde{T}_{-\bm{d}} |\psi_{n,\ell^{-2}\hat{\mathbf{z}}\times\bm{d}}\rangle=e^{i\ell^{-2}\hat{\mathbf{z}}\cdot(\bm{d}_2\times\bm{d})/2p} \widetilde{T}_{-\bm{d}_2/p} |w_{n,\bm{d}}\rangle \ , \\
&|w_{n,\bm{d}+\bm{d}_1}\rangle= \widetilde{T}_{-\bm{d}-\bm{d}_1}|\psi_{n,\ell^{-2}\hat{\mathbf{z}}\times(\bm{d}+\bm{d}_1)}\rangle =e^{-i\ell^{-2}\hat{\mathbf{z}}\cdot(\bm{d}_1\times\bm{d})/2}\widetilde{T}_{-\bm{d}} \widetilde{T}_{-\bm{d}_1} |\psi_{n,\ell^{-2}\hat{\mathbf{z}}\times\bm{d}+p\bm{g}_2/q}\rangle \\
=&e^{-i\ell^{-2}\hat{\mathbf{z}}\cdot(\bm{d}_1\times\bm{d})/2-imq\bm{d}_1\cdot(\ell^{-2}\hat{\mathbf{z}}\times\bm{d})}\widetilde{T}_{-\bm{d}} |\psi_{n,\ell^{-2}\hat{\mathbf{z}}\times\bm{d}}\rangle = e^{-i\ell^{-2}\hat{\mathbf{z}}\cdot(\bm{d}_1\times\bm{d})/2+imq\ell^{-2}\hat{\mathbf{z}}\cdot(\bm{d}_1\times\bm{d})} |w_{n,\bm{d}}\rangle\ .
\end{split}
\end{equation}
Therefore, for band $n$ between the $(\nu-1)$-th gap and the $\nu$-th gap, we find the dual Berry gauge field in Eq. (\ref{Seq-new-berry}) satisfies
\begin{equation}
\begin{split}
&\int_{\bm{d}\in\Omega_M} \frac{\text{d}^2\bm{d} }{2\pi}\widetilde{f}^n_{xy}(\bm{d})=\frac{1}{2\pi}\oint_{\partial\Omega_M}\widetilde{\bm{a}}^n(\bm{d})\cdot\text{d}\bm{d} \\
&= \frac{1}{2\pi}[mq\ell^{-2}\hat{\mathbf{z}}\cdot(\bm{d}_1\times\bm{d}_2/p)-\ell^{-2}\hat{\mathbf{z}}\cdot(\bm{d}_1\times\bm{d}_2/p)/2+ \ell^{-2}\hat{\mathbf{z}}\cdot(\bm{d}_1\times\bm{d}_2)/2p] =\frac{mq}{2\pi p}\ell^{-2}\Omega \\
&=m=s_\nu-s_{\nu-1}\ .
\end{split}
\end{equation}
Summing over all occupied bands then proves Eq. (\ref{Seq-dual-Chern}).

\subsection{Relation between the dual wave functions, and a model example}\label{sec:w-example}

This subsection is devoted for a better understanding of the relation between the two wave functions $|u_{n,\mathbf{k}}\rangle$ defined in Eq. (\ref{Seq-unk}) and $|w_{n,\bm{d}}\rangle$ defined in Eq. (\ref{Seq-define-w}). To do this, we examine their form in the real space basis $|\mathbf{r}\rangle$. If we denote the Bloch wave function of momentum $\mathbf{k}$ as $|\psi_{n,\mathbf{k}}\rangle$, by the definition of the periodic wavefunction $|u_{n,\mathbf{k}}\rangle$ in Eq. (\ref{Seq-unk}), we have
\begin{equation}
\psi_{n,\mathbf{k}}(\mathbf{r})=\langle \mathbf{r}|\psi_{n,\mathbf{k}}\rangle\ ,\qquad  u_{n,\mathbf{k}}(\mathbf{r})=\langle \mathbf{r}|u_{n,\mathbf{k}}\rangle=e^{-i\mathbf{k\cdot r}}\psi_{n,\mathbf{k}}(\mathbf{r})\ .
\end{equation}
To find the real space wave function of $|w_{n,\bm{d}}\rangle=\widetilde{T}_{-\bm{d}}|\psi_{n,\ell^{-2}\hat{\mathbf{z}}\times\bm{d}}\rangle$ as defined in Eq. (\ref{Seq-define-w}), we first note that
\begin{equation}\label{seq-Td-example}
\begin{split}
\widetilde{T}_{\bm{d}}|\mathbf{r}\rangle&=e^{-i \ell^{-2}(\hat{\mathbf{z}}\times\mathbf{R})\cdot\bm{d}}|\mathbf{r}\rangle= e^{-i \ell^{-2}(\hat{\mathbf{z}}\times\mathbf{r})\cdot\bm{d}-i\bm{\Pi}\cdot\bm{d}}|\mathbf{r}\rangle=e^{i\int_\mathbf{r}^{\mathbf{r}+\bm{d}}[\mathbf{A}(\mathbf{r}')-\ell^{-2}\hat{\mathbf{z}}\times\mathbf{r}']\cdot\text{d}\mathbf{r}'}|\mathbf{r}+\bm{d}\rangle\\
& = e^{i \ell^{-2}(\hat{\mathbf{z}}\times\bm{d})\cdot (\mathbf{r}+\bm{d})}e^{i\int_\mathbf{r}^{\mathbf{r}+\bm{d}}\mathbf{A}(\mathbf{r}')\cdot\text{d}\mathbf{r}'}|\mathbf{r}+\bm{d}\rangle\ ,
\end{split}
\end{equation}
which can be seen by following a derivation similar to Eq. (\ref{Seq-path}), where the integral on the exponent is along the straight line segment from $\mathbf{r}$ to $\mathbf{r}+\bm{d}$. Therefore, we find
\begin{equation}\label{Seq-w-u-relation1}
\begin{split}
&w_{n,\bm{d}}(\mathbf{r})=\langle \mathbf{r}|w_{n,\bm{d}}\rangle =\langle \mathbf{r}|\widetilde{T}_{-\bm{d}}|\psi_{n,\ell^{-2}\hat{\mathbf{z}}\times\bm{d}}\rangle = e^{-i \ell^{-2}(\hat{\mathbf{z}}\times\bm{d})\cdot (\mathbf{r}+\bm{d})}e^{-i\int_\mathbf{r}^{\mathbf{r}+\bm{d}}\mathbf{A}(\mathbf{r}')\cdot\text{d}\mathbf{r}'} \langle \mathbf{r}+\bm{d}|\psi_{n,\ell^{-2}\hat{\mathbf{z}}\times\bm{d}}\rangle\\
& =e^{-i\int_\mathbf{r}^{\mathbf{r}+\bm{d}}\mathbf{A}(\mathbf{r}')\cdot\text{d}\mathbf{r}'} u_{n,\ell^{-2}\hat{\mathbf{z}}\times\bm{d}}(\mathbf{r}+\bm{d})\ .
\end{split}
\end{equation}
This is the explicit form of the wave function $w_{n,\bm{d}}(\mathbf{r})$.

Without going into the real space position basis, we can also derive that (here $\mathbf{r}$ below denotes the position operator)
\begin{equation}\label{Seq-w-u-relation2}
|w_{n,\bm{d}}\rangle=\widetilde{T}_{-\bm{d}}|\psi_{n,\ell^{-2}\hat{\mathbf{z}}\times\bm{d}}\rangle = e^{i \ell^{-2}(\hat{\mathbf{z}}\times\mathbf{R})\cdot\bm{d}} e^{i \ell^{-2}(\hat{\mathbf{z}}\times\bm{d})\cdot\mathbf{r}}|u_{n,\ell^{-2}\hat{\mathbf{z}}\times\bm{d}}\rangle = e^{i\bm{\Pi}\cdot \bm{d}}|u_{n,\ell^{-2}\hat{\mathbf{z}}\times\bm{d}}\rangle\ ,
\end{equation}
where we have used the fact that $[(\hat{\mathbf{z}}\times\mathbf{R})\cdot\bm{d},(\hat{\mathbf{z}}\times\mathbf{r})\cdot\bm{d}]=0$, and $\mathbf{R-r}=-\ell^2\hat{\mathbf{z}}\times\bm{\Pi}$. By this relation, the dual Berry curvature $\widetilde{f}^n_{xy}(\bm{d})$ defined in Eq. (\ref{Seq-new-berry}) can be explicitly related to the Berry curvature $\mathcal{F}^n_{xy}(\mathbf{k})$ in Eq. (\ref{Seq-berry}). We have
\begin{equation}
|\partial_{d_i}w_{n,\bm{d}}\rangle =(\partial_{d_i} e^{i\bm{\Pi}\cdot \bm{d}})|u_{n,\ell^{-2}\hat{\mathbf{z}}\times\bm{d}}\rangle+ e^{i\bm{\Pi}\cdot \bm{d}} |\partial_{d_i} u_{n,\ell^{-2}\hat{\mathbf{z}}\times\bm{d}}\rangle = e^{i\bm{\Pi}\cdot \bm{d}}[i(\Pi_i+\epsilon_{ij}d_j/2)|u_{n,\ell^{-2}\hat{\mathbf{z}}\times\bm{d}}\rangle +|\partial_{d_i} u_{n,\ell^{-2}\hat{\mathbf{z}}\times\bm{d}}\rangle]  \ ,
\end{equation}
where $\epsilon_{ji}$ ($i,j=x,y$) is the Levi-Civita tensor ($\epsilon_{xy}=-\epsilon_{yx}=1$). Therefore, we find the dual Berry curvature
\begin{equation}\label{Seq-2Berry-relation}
\begin{split}
&\widetilde{f}^n_{xy}(\bm{d})=-i\langle \partial_{d_x}w_{n,\bm{d}}|\partial_{d_y}w_{n,\bm{d}}\rangle+ i \langle \partial_{d_y}w_{n,\bm{d}}|\partial_{d_x}w_{n,\bm{d}}\rangle \\
&=-i\langle u_{n,\ell^{-2}\hat{\mathbf{z}}\times\bm{d}}|[\Pi_x,\Pi_y] |u_{n,\ell^{-2}\hat{\mathbf{z}}\times\bm{d}}\rangle -i\langle \partial_{d_x}u_{n,\ell^{-2}\hat{\mathbf{z}}\times\bm{d}}|\partial_{d_y}u_{n,\ell^{-2}\hat{\mathbf{z}}\times\bm{d}}\rangle+ i \langle \partial_{d_y}u_{n,\ell^{-2}\hat{\mathbf{z}}\times\bm{d}}|\partial_{d_x}u_{n,\ell^{-2}\hat{\mathbf{z}}\times\bm{d}}\rangle \\
& \qquad -\langle \partial_{d_y}u_{n,\ell^{-2}\hat{\mathbf{z}}\times\bm{d}}| \Pi_x | u_{n,\ell^{-2}\hat{\mathbf{z}}\times\bm{d}}\rangle- \langle u_{n,\ell^{-2}\hat{\mathbf{z}}\times\bm{d}}| \Pi_x | \partial_{d_y}u_{n,\ell^{-2}\hat{\mathbf{z}}\times\bm{d}}\rangle \\
&\qquad + \langle \partial_{d_x}u_{n,\ell^{-2}\hat{\mathbf{z}}\times\bm{d}}| \Pi_y | u_{n,\ell^{-2}\hat{\mathbf{z}}\times\bm{d}}\rangle+ \langle u_{n,\ell^{-2}\hat{\mathbf{z}}\times\bm{d}}| \Pi_y | \partial_{d_x}u_{n,\ell^{-2}\hat{\mathbf{z}}\times\bm{d}}\rangle \\
&=\ell^{-2}-\ell^{-4}\mathcal{F}^n_{xy}(\ell^{-2}\hat{\mathbf{z}}\times\bm{d}) +\partial_{d_x} (\langle u_{n,\ell^{-2}\hat{\mathbf{z}}\times\bm{d}}| \Pi_y | u_{n,\ell^{-2}\hat{\mathbf{z}}\times\bm{d}}\rangle) -\partial_{d_y} (\langle u_{n,\ell^{-2}\hat{\mathbf{z}}\times\bm{d}}| \Pi_x | u_{n,\ell^{-2}\hat{\mathbf{z}}\times\bm{d}}\rangle)  \\
&=B-B^2\mathcal{F}^n_{xy}(\ell^{-2}\hat{\mathbf{z}}\times\bm{d}) +\partial_{d_x} (\langle w_{n,\bm{d}}| \Pi_y | w_{n,\bm{d}}\rangle) -\partial_{d_y} (\langle w_{n,\bm{d}}| \Pi_x | w_{n,\bm{d}}\rangle)\ .
\end{split}
\end{equation}
This gives the relationship between the dual Berry curvature $\widetilde{f}^n_{xy}(\bm{d})$ and the Berry curvature $\mathcal{F}^n_{xy}(\mathbf{k})$. In particular, we note that the two terms $\partial_{d_x} (\langle w_{n,\bm{d}}| \Pi_y | w_{n,\bm{d}}\rangle)-\partial_{d_y} (\langle w_{n,\bm{d}}| \Pi_x | w_{n,\bm{d}}\rangle)$ in the last line of Eq. (\ref{Seq-2Berry-relation}) are total derivatives which do not contribute to the dual Chern number $s_\nu$, since $\langle w_{n,\bm{d}}| \Pi_i | w_{n,\bm{d}}\rangle$ are well-defined periodic functions (physical quantities) in the dual Brillouin zone $\Omega_M$ spanned by $\bm{d}_1$ and $\bm{d}_2/p$. 

\subsubsection{An example}



For simplicity, we consider a single-orbital tight-binding model on a square lattice with a magnetic flux per unit cell $\varphi=2\pi/3$. Assume the lattice vectors are $\bm{d}_1=(1,0)$ and $\bm{d}_2=(0,1)$, and the Wannier orbitals $|\bm{D}_{n_1,n_2}\rangle$ are delta functions at lattice sites $\bm{D}_{n_1,n_2}=(n_1,n_2)\in \bm{d}_1 \mathbb{Z}+ \bm{d}_2 \mathbb{Z}$ in the continuous space. At zero magnetic field, we assume all the nearest bonds have hopping $-t$ (with $t$ being a real number), while all the longer range hoppings are zero. We then add a uniform magnetic field $B=2\pi p/q=4\pi/3$ (so that the flux $\varphi=B|\bm{d}_1\times \bm{d}_2|=4\pi/3$), and we adopt the Landau gauge $\mathbf{A}(\mathbf{r})=(0,Bx)$. This correspond to $p=2$ and $q=3$. Accordingly, the magnetic length $\ell$ satisfies $\ell^{-2}=4\pi/3$.

The tight-binding Hamiltonian in magnetic field $B$ can then be written as
\begin{equation}
H=-t\sum_{n_1,n_2\in\mathbb{Z}} \left(|\bm{D}_{n_1+1,n_2}\rangle \langle \bm{D}_{n_1,n_2}| + e^{i 4\pi n_1/3}|\bm{D}_{n_1,n_2+1}\rangle \langle \bm{D}_{n_1,n_2}| + h.c. \right)\ .
\end{equation}
Therefore, we have
\begin{equation}
e^{-i\mathbf{k\cdot r}}He^{i\mathbf{k\cdot r}}=-t\sum_{n_1,n_2\in\mathbb{Z}} \left(e^{-i k_x}|\bm{D}_{n_1+1,n_2}\rangle \langle \bm{D}_{n_1,n_2}| + e^{-ik_y+i 4\pi n_1/3}|\bm{D}_{n_1,n_2+1}\rangle \langle \bm{D}_{n_1,n_2}| + h.c. \right)\ ,
\end{equation}
where $\mathbf{r}$ is the position operator (instead of a parameter), while $\mathbf{k}=(k_x,k_y)$ is the momentum parameter. By Eq. (\ref{seq-Td-example}), the projector $P_\mathbf{0}$ in Eq. (\ref{Seq-Htildehat}) in the Hilbert space of Wannier basis $|\bm{D}_{n_1,n_2}\rangle$ can be rewritten as
\begin{equation}\label{Seq-Hhat-example0}
P_\mathbf{0}=\frac{3}{N_\Omega}\sum_{j_1,j_2\in\mathbb{Z}} \widetilde{T}_{3 j_1\bm{d}_1+j_2\bm{d}_2}=\sum_{j=1}^3 |j\rangle\langle j|\ ,\qquad |j\rangle =\sqrt{\frac{3}{N_\Omega}}\sum_{n_1,n_2\in\mathbb{Z}} |\bm{D}_{3n_1+j-1,n_2}\rangle\ ,\quad (j=1,2,3).
\end{equation}
The three basis $|j\rangle$ ($j=1,2,3$) correspond exactly to the three orbitals in a magnetic unit cell. Therefore, under the three basis $|j\rangle$, we find the reduced momentum space Hamiltonian $\hat{H}(\mathbf{k})$ is given by
\begin{equation}\label{Seq-Hhat-example1}
\hat{H}(\mathbf{k})=P_\mathbf{0}e^{-i\mathbf{k\cdot r}}He^{i\mathbf{k\cdot r}}P_\mathbf{0} =
-t\left(\begin{array}{ccc}
2\cos(k_y) & e^{ik_x} & e^{-ik_x} \\
e^{-ik_x} & 2\cos(k_y+2\pi/3) & e^{ik_x} \\
e^{ik_x} & e^{-ik_x} & 2\cos(k_y-2\pi/3) \\
\end{array}\right)\ .
\end{equation}
This is exactly the momentum space Hamiltonian we are familiar with.

We now consider the other reduced Hamiltonian $\hat{\widetilde{H}}(\bm{d})$. By Eq. (\ref{seq-Td-example}), for $\bm{d}=(d_x,d_y)$ we have
\begin{equation}
\widetilde{T}_{-\bm{d}}H\widetilde{T}_{\bm{d}}=-t\sum_{n_1,n_2\in\mathbb{Z}} \left(|\bm{D}_{n_1+1,n_2}-\bm{d}\rangle \langle \bm{D}_{n_1,n_2}-\bm{d}| + e^{-i4\pi d_x/3+ i 4\pi n_1/3}|\bm{D}_{n_1,n_2+1}-\bm{d}\rangle \langle \bm{D}_{n_1,n_2}-\bm{d}| + h.c. \right),
\end{equation}
where we have used $\mathbf{A}(\mathbf{r})=(0,Bx)$ in the continuous space. We note that the Hamiltonian is transformed into a shifted Wannier basis $|\bm{D}_{n_1+1,n_2}-\bm{d}\rangle$. Note that by Eq. (\ref{seq-Td-example}), the lattice translation operators act as
\begin{equation}
\widetilde{T}_{3\bm{d}_1}|\bm{D}_{n_1,n_2}-\bm{d} \rangle = e^{-i4\pi d_y} |\bm{D}_{n_1+3,n_2}-\bm{d} \rangle\ ,\qquad \widetilde{T}_{\bm{d}_2}|\bm{D}_{n_1,n_2}-\bm{d} \rangle = |\bm{D}_{n_1,n_2+1}-\bm{d} \rangle\ .
\end{equation}
Therefore, the projector $P_\mathbf{0}$ in the Hilbert space of Wannier basis $|\bm{D}_{n_1+1,n_2}-\bm{d}\rangle$ is
\begin{equation}
P_\mathbf{0}=\frac{3}{N_\Omega}\sum_{j_1,j_2\in\mathbb{Z}} \widetilde{T}_{3 j_1\bm{d}_1+j_2\bm{d}_2}=\sum_{j=1}^3 |j,\bm{d}\rangle\langle j,\bm{d}|\ ,\qquad |j,\bm{d}\rangle =\sqrt{\frac{3}{N_\Omega}}\sum_{n_1,n_2\in\mathbb{Z}} e^{-i4\pi (3n_1+j)d_y/3}|\bm{D}_{3n_1+j-1,n_2}-\bm{d}\rangle,
\end{equation}
where $j=1,2,3$. Under the basis $|j,\bm{d}\rangle$, we find
\begin{equation}
\hat{\widetilde{H}}(\bm{d})=P_\mathbf{0}\widetilde{T}_{-\bm{d}}H\widetilde{T}_{\bm{d}} P_\mathbf{0} = -t\left(\begin{array}{ccc}
2\cos(4\pi d_x/3) & e^{-i 4\pi d_y/3} & e^{i 4\pi d_y/3} \\
e^{i 4\pi d_y/3} & 2\cos(4\pi d_x/3+2\pi/3) & e^{-i 4\pi d_y/3} \\
e^{-i 4\pi d_y/3} & e^{i 4\pi d_y/3} & 2\cos(4\pi d_x/3-2\pi/3) \\
\end{array}\right)\ .
\end{equation}
Therefore, under the Landau gauge, we find $\hat{\widetilde{H}}(\bm{d})$ takes a form analogous to $\hat{H}(\mathbf{k})$ with $\mathbf{k}=(4\pi/3)\hat{\mathbf{z}}\times \bm{d}$, except that the basis $|j,\bm{d}\rangle$ is $\bm{d}$ dependent. Therefore, if the eigenstates of $\hat{H}(\mathbf{k})$ are $|u_{n,\mathbf{k}}\rangle=\sum_{j=1}^3 u_{n,j}(\mathbf{k})|j\rangle$, the eigenstates of $\hat{\widetilde{H}}(\bm{d})$ will be $|w_{n,\bm{d}}\rangle =\sum_{j=1}^3 u_{n,j}(\frac{4\pi}{3}\hat{\mathbf{z}}\times \bm{d})|j,\bm{d}\rangle$, in agreement with Eq. (\ref{Seq-w-u-relation1}). 

The Chern numbers of the 3 bands of $\hat{H}(\mathbf{k})$ can be calculated to be $\sigma_n=\{-1,2,-1\}$ (from the lowest band to the highest band). According to Eq. (\ref{Seq-2Berry-relation}), we find the dual Chern numbers of the 3 bands of $\hat{\widetilde{H}}(\bm{d})$ are given by $m_n=\frac{1}{3}-\frac{2}{3}\sigma_n=\{1,-1,1\}$ (from the lowest band to the highest band).

\begin{figure}[tbp]
\begin{center}
\includegraphics[width=2.4in]{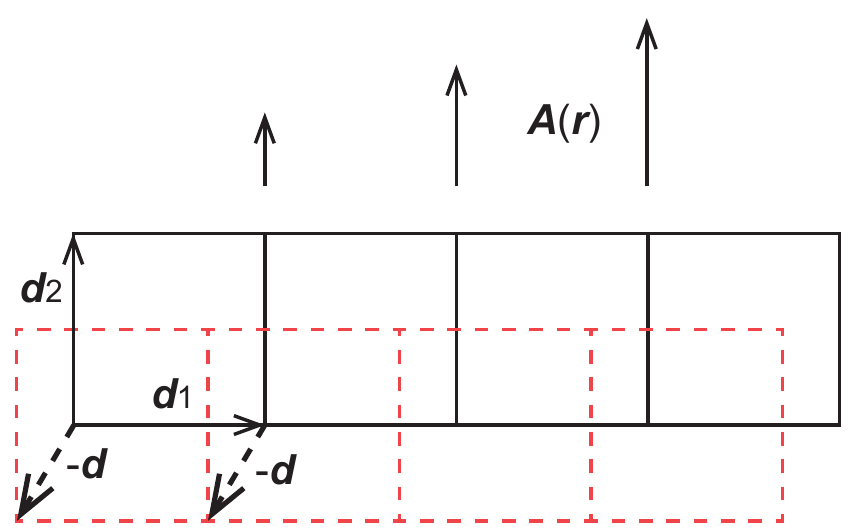}
\end{center}
\caption{An example of tight binding model on a square lattice in a magnetic field, where Landau gauge is chosen.
}
\label{FigS-example}
\end{figure}

\subsection{Physical understanding of the quantized Lorentz susceptibility}
The quantized Lorentz susceptibility $\gamma_{xy}=eBs_\nu$ can be easily understood from the following physical picture. Consider a lattice model in the $x$-$y$ plane moving with a velocity $\mathbf{v}=v_y\hat{\mathbf{y}}$ in the $\hat{\mathbf{z}}$ direction magnetic field $B$, and the Fermi level is in a Hofstadter gap $\nu$. In the rest frame of the lattice system, the magnetic field $B$ will move with a velocity $-\mathbf{v}=-v_y\hat{\mathbf{y}}$, which produces an electric field $\mathbf{E}=v_yB\mathbf{\hat{\mathbf{x}}}$. This electric field then produces a Hall current density $\widetilde{j}_y=-\sigma_{xy}v_yB$ in the rest frame of the lattice. If we go back to the laboratory frame where the system moves with velocity $\mathbf{v}=v_y\hat{\mathbf{y}}$, we would find a total current density
\[j_y=\frac{e\rho v_y}{\Omega}+\widetilde{j}_y=\left(\frac{e\rho}{\Omega}-\sigma_{xy}B\right)v_y\ ,\]
where $\rho$ is the number of occupied electron per unit cell, and $\Omega$ is the unit cell area. Therefore, the Lorentz force per unit cell is given by
\begin{equation}
F_x=j_y\Omega B=\left(e\rho-\sigma_{xy}\Omega B\right)Bv_y=\left(e\rho-t_\nu\frac{e^2}{h}\Omega B\right)Bv_y =eB\left(\rho-t_\nu\frac{\varphi}{2\pi}\right)v_y=eBs_\nu v_y\ ,
\end{equation}
where we have used the Diophantine equation. Therefore, we find the Lorentz susceptibility is $\gamma_{xy}=eBs_\nu$.

\end{widetext}

\end{document}